\documentclass[fleqn]{article}
\usepackage{xeCJK}
\setCJKsansfont[BoldFont=FandolHei-Bold.otf]{FandolHei-Regular.otf}
\setCJKmonofont{FandolFang-Regular.otf}
\xeCJKsetup{AutoFakeBold=3}
\xeCJKDeclareCharClass{CJK}{"0370->"03FF}
\xeCJKDeclareCharClass{CJK}{"2070->"209F}
\xeCJKDeclareCharClass{CJK}{"2190->"21FF}
\xeCJKDeclareCharClass{CJK}{"2200->"22FF}
\xeCJKDeclareCharClass{CJK}{"2460->"24FF}
\xeCJKDeclareCharClass{CJK}{"2500->"257F}
\xeCJKDeclareCharClass{CJK}{"25A0->"25FF}
\xeCJKDeclareCharClass{CJK}{"2713->"2713}
\xeCJKDeclareCharClass{CJK}{"2032->"2032}
\usepackage{pifont}
\usepackage{newunicodechar}
\newunicodechar{ᵀ}{\ensuremath{^{\mathsf{T}}}}
\newunicodechar{⟨}{\ensuremath{\langle}}
\newunicodechar{⟩}{\ensuremath{\rangle}}
\newunicodechar{∈}{\ensuremath{\in}}
\newunicodechar{∀}{\ensuremath{\forall}}
\newunicodechar{∩}{\ensuremath{\cap}}
\newunicodechar{⊙}{\ensuremath{\odot}}
\newunicodechar{′}{\ensuremath{{}^{\prime}}}
\newunicodechar{✓}{\ding{51}}
\newunicodechar{►}{\ensuremath{\blacktriangleright}}
\newunicodechar{◄}{\ensuremath{\blacktriangleleft}}
\newunicodechar{⟺}{\ensuremath{\Longleftrightarrow}}
\newunicodechar{≲}{\ensuremath{\lesssim}}
\newunicodechar{₀}{\ensuremath{_{0}}}
\newunicodechar{₁}{\ensuremath{_{1}}}
\newunicodechar{₂}{\ensuremath{_{2}}}
\newunicodechar{⁵}{\ensuremath{^{5}}}
\newunicodechar{⁽}{\ensuremath{^{(}}}
\newunicodechar{⁾}{\ensuremath{^{)}}}
\usepackage[nobottomtitles*]{titlesec}

\usepackage{float}
\usepackage{needspace}
\usepackage{indentfirst}
\usepackage{tikz}
\usetikzlibrary{positioning,arrows.meta,calc}
\newcommand{\fitwidth}[1]{\resizebox{\ifdim\width>\linewidth\linewidth\else\width\fi}{!}{#1}}
\newcommand{\fittab}[1]{\resizebox{\ifdim\width>\dimexpr\linewidth-2em\relax\dimexpr\linewidth-2em\relax\else\width\fi}{!}{#1}}
\newcommand\mwmparagraph[1]{\par\addvspace{8pt plus 2pt minus 1pt}%
  {\leftskip=0pt\relax\noindent\normalfont\bfseries #1\par}%
  \leftskip=2em\relax%
  \nobreak\addvspace{3pt plus 1pt}\noindent\ignorespaces}
\AtBeginDocument{%
  \setlength{\parindent}{0pt}%
  \setlength{\leftskip}{2em}%
  \setlength{\parskip}{3pt plus 1.2pt minus 0.8pt}%
  \linespread{1.2}\selectfont%
  \clubpenalty=10000 \widowpenalty=10000%
  \interlinepenalty=200%
  \raggedbottom%
  \setlength{\mathindent}{2em}%
  \titlespacing*{\section}{0pt}{16pt plus 4pt minus 2pt}{6pt plus 2pt minus 1pt}%
  \titlespacing*{\subsection}{0pt}{13pt plus 3pt minus 2pt}{4pt plus 2pt minus 1pt}%
  \titlespacing*{\subsubsection}{0pt}{10pt plus 3pt minus 2pt}{3pt plus 1pt minus 1pt}%
  \titlespacing*{\paragraph}{0pt}{8pt plus 2pt minus 1pt}{3pt plus 1pt}%
  \let\mwmsection\section%
  \renewcommand{\section}{\clearpage\mwmsection}%
  \let\paragraph\mwmparagraph%
}
\renewenvironment{quote}%
  {\par\addvspace{0.5\baselineskip}\small\setlength{\leftskip}{2em}\setlength{\rightskip}{0pt}\noindent\ignorespaces}%
  {\par\addvspace{0.5\baselineskip}}
\makeatletter
\g@addto@macro\normalsize{%
  \setlength{\parindent}{0pt}%
  \setlength{\leftskip}{2em}%
  \setlength\abovedisplayskip{7pt plus 2pt minus 2pt}%
  \setlength\belowdisplayskip{7pt plus 2pt minus 2pt}%
  \setlength\abovedisplayshortskip{7pt plus 2pt minus 2pt}%
  \setlength\belowdisplayshortskip{7pt plus 2pt minus 2pt}%
}
\makeatother
\usepackage{etoolbox}
\makeatletter
\AtEndPreamble{\apptocmd{\@afterheading}{\leftskip=2em\relax\parindent=0pt\relax}{}{}}
\makeatother

\usepackage{preprint}
\usepackage{hyperref}
\usepackage[numbers,square]{natbib}

\usepackage{amsmath}
\usepackage{booktabs}
\usepackage{graphicx}
\usepackage{url}

\hypersetup{colorlinks=true, linkcolor=purple, urlcolor=blue, citecolor=cyan, anchorcolor=black}

\usepackage{xcolor}
\usepackage{lineno}					
\usepackage{tikz} 					

\usepackage{newfloat}
\DeclareFloatingEnvironment[name={Supplementary Figure}]{suppfigure}
\usepackage{sidecap}
\sidecaptionvpos{figure}{c}

\usepackage{titlesec}
\titlespacing\section{0pt}{12pt plus 3pt minus 3pt}{1pt plus 1pt minus 1pt}
\titlespacing\subsection{0pt}{10pt plus 3pt minus 3pt}{1pt plus 1pt minus 1pt}
\titlespacing\subsubsection{0pt}{8pt plus 3pt minus 3pt}{1pt plus 1pt minus 1pt}

\definecolor{lime}{HTML}{A6CE39}

\title{From MWM to iSLIP: A Linear-Algebraic Tutorial on Input-Queued Switch Scheduling}

\usepackage{authblk}

\author[1\thanks{\texttt{miaomiao\_tong@outlook.com}}]{Xiaotong Yuan (袁晓彤)}
\author[1]{An Guo (郭安)}
\affil[1]{Enrigin Technology Co., Ltd.}

\begin{document}

\maketitle

\begin{abstract}
\linespread{1.5}\selectfont\setlength{\leftskip}{3em}\setlength{\rightskip}{3em}\setlength{\parindent}{0pt}%
This paper uses three objects, namely the queue matrix Q, the matching matrix P, and the Lyapunov energy function V = ‖Q‖\textsuperscript{2}, as a shared mathematical language to explain, within a single framework, the scheduling objective of maximum weight matching (MWM), queue stability under admissible traffic (per-port loads strictly below 1), and the mechanics of iSLIP's Grant--Accept row-column decoupling together with the long-run average service matrix P̄. The setting throughout is an N$\times$N SoC crossbar, where each clock cycle permits at most one cell transfer per input--output port pair. For the experimental comparison, we built a C++ discrete-event simulator and used exact MWM (solved by the Hungarian algorithm) as the performance reference. All three approximate algorithms are given a fixed iteration budget: r = 3 rounds per cycle for iSLIP and for spectral scheduling, and r\_sink = 10 Sinkhorn normalization rounds for entropy-regularized optimal transport (OT). Throughput and average cell delay are measured across four traffic patterns. Spectral scheduling and entropy-regularized OT track MWM closely in both throughput and delay across most tested conditions. iSLIP, by contrast, hits a throughput ceiling of roughly 80\% under non-uniform admissible traffic at high load (unbalanced pattern w = 0.5, $\rho$\_load $\geq$ 0.9), with bottleneck queues growing without bound and delays reaching two orders of magnitude above MWM. Under uniform traffic this breakdown does not occur: at $\rho$\_load = 0.99 iSLIP delay is about 3.7$\times$ that of MWM. The performance gains of spectral scheduling and OT come at an additional per-cycle compute cost on the order of O(r·N\textsuperscript{2}) multiply-accumulate or exponential operations; whether this overhead is feasible in real hardware, in terms of die area, power, and timing closure, remains to be evaluated.\\
\end{abstract}


\vspace{0.5cm}


\mwmsection*{Document Note (Language Versions)}

\textbf{Part I. English (this section first):} The English text was produced with LLM assistance (Claude Sonnet 4.6) from the original Chinese manuscript and reviewed by the authors.

\textbf{Part II. Chinese / 中文版 (follows after a page break):} The Chinese text is the original manuscript.

For arXiv submission metadata: original language Chinese; English translation included per arXiv multilingual policy.

\section{Introduction}

\subsection{Background and Motivation}

In on-chip interconnect architectures, an N$\times$N centralized crossbar must complete high-speed arbitration and path matching within a single clock cycle, making it the core component for on-chip data transfer. The central task of arbitration is to select valid input--output matching pairs; within each cycle, at most one flit of data is allowed to pass between any given pair of ports. The overall queue state of the hardware can be represented by a queue matrix Q(t), and the arbitration decision for each cycle can be represented as a binary (0/1) matching matrix P(t). Constrained by the small iteration budgets that hardware imposes, existing industrial iterative algorithms are all approximate solutions to the optimal matching problem; complex exact solvers such as the Hungarian algorithm cannot be deployed. The core research motivation of this paper is therefore: under fair comparison conditions with a fixed iteration count, to quantify how closely various approximate scheduling algorithms approach the optimal performance of MWM, and to characterize the limitations of each.

The standard model for IQ switch scheduling theory has been studied extensively; within a time slot, the scheduling problem is essentially that of finding an optimal matching in a bipartite graph. Traditional single-input FIFO queues suffer from severe head-of-line (HOL) blocking [1],[13], while the virtual output queue (VOQ) architecture eliminates HOL at the architectural level by maintaining a separate queue for each port pair (i,j) [4],[14]. Classical theory shows that MWM scheduling on top of a VOQ architecture can achieve 100\% throughput [4],[7], but the algorithmic complexity is prohibitively high for real-time arbitration.

To balance hardware cost and scheduling performance, low-complexity iterative algorithms such as iSLIP [3] have become mainstream in industry. These algorithms rely only on a 1-bit queue-nonempty signal to carry out round-robin iterations; they have low hardware cost and good timing convergence, and can guarantee queue stability under uniform admissible traffic. Under certain non-uniform traffic patterns their performance degrades. When compared with other approximate scheduling schemes, the performance boundaries need to be quantified in controlled experiments.

\subsection{Three Research Threads}

The literature on crossbar arbitration and IQ scheduling is often split into three largely separate research threads, with few unified cross-domain treatments.

\textbf{Theory thread:} Existing work based on queueing theory and Lyapunov stability focuses on steady-state and throughput analysis, systematically characterizing the global optimality and throughput upper bound of MWM [7],[8],[9]. However, purely theoretical analyses generally ignore the practical constraints imposed by the limited iteration counts of hardware cycles.

\textbf{Engineering implementation thread:} Iterative algorithms such as iSLIP focus on RTL implementability and represent a mature hardware architecture [3], but they typically rely on simulation for performance evaluation and lack a mathematical explanation of their approximation mechanisms and performance boundaries.

\textbf{SoC/NoC research thread:} The SoC/NoC field focuses on flit transfer, crossbar architecture, and buffer depth design. Its terminology is inconsistent with the data communications field, and it similarly lacks mathematical explanations of MWM optimality theory and iterative approximation algorithms.

In summary, purely mathematical tutorials typically do not emphasize the hardware constraint of a small number of iterations per clock cycle in a crossbar, while purely SoC-oriented documents rarely include a mathematical treatment of MWM stability. Bringing iSLIP or spectral analysis into an arbiter design requires establishing a set of comparable mathematical descriptions that bridge these three threads.

\subsection{Main Contributions}

\begin{itemize}
\item \textbf{A unified mathematical language:} Using the queue matrix Q, the matching matrix P, the inner product ⟨Q,P⟩, and the Lyapunov energy V=‖Q‖\textsuperscript{2} as a shared vocabulary, this paper places MWM's scheduling objective and stability, continuous relaxations (Sinkhorn iteration/BvN decomposition, spectral scheduling), and iSLIP's 0/1 iterations all within the same algebraic framework for comparison, bridging the three separate research threads mentioned above.
\item \textbf{A derivation chain that can be worked out by hand:} Starting from the MWM scheduling objective and the stability condition under admissible traffic, proceeding to iSLIP's Grant--Accept mechanism and the cross-cycle average service matrix P̄, the entire development is accompanied by worked examples that can be reproduced with pencil and paper.
\item \textbf{A unified comparison of easily confused concepts:} The various ``throughput'' percentages that appear throughout the paper (the HOL architecture upper bound, the approximation ratio lower bound for maximal matching, the 100\% in the sense of admissible-traffic stability) are collected and defined side by side so the reader can distinguish them.
\item \textbf{A quantitative comparison and cost analysis under a fair iteration budget:} With iteration counts fixed at r=3 and r\_sink=10, and MWM as the performance reference, this paper compares iSLIP, spectral scheduling, and entropy-regularized OT in terms of throughput and delay, and carries out an analysis-level Pareto comparison of arithmetic operation counts and queue-read bit widths.
\end{itemize}

\subsection{Paper Structure}

The paper is organized along three main threads, namely linear-algebraic modeling, graph-theoretic principles, and the engineering implementation of iterative algorithms, with the following specific structure:

\bigskip\noindent
\par\leftskip=2em\relax\noindent \fittab{\begin{tabular}{p{\dimexpr 0.333\linewidth-2\tabcolsep}p{\dimexpr 0.333\linewidth-2\tabcolsep}p{\dimexpr 0.333\linewidth-2\tabcolsep}}
\toprule
Thread & Core question & Corresponding sections \\
\midrule
I. Linear algebra & System modeling, MWM theory and stability, continuous relaxations (scaling and spectral) projected onto discrete matchings & §2, §3, §4.1, §4.2 \\
II. Graph theory & Graph-theoretic formulation of exact MWM, connection to industrial implementation & §4.3, Appendix B \\
III. iSLIP & Principles of 0/1 iterative approximation and its stability & §5 \\
IV. Taxonomy, experiments, and conclusions & Algorithm taxonomy, metric comparison, comparative experiments, conclusions and future directions & §6--§8 (non-square extension in Appendix C) \\
\bottomrule
\end{tabular}}\par\leftskip=2em\relax{}

\bigskip

\subsection{Prerequisites and Notation}

Readers are assumed to have an introductory background in linear algebra (vectors, matrix multiplication, inner products, eigenvalues) and basic probability theory.

\textbf{Core notation:}

\bigskip\noindent
\par\leftskip=2em\relax\noindent \fittab{\begin{tabular}{p{\dimexpr 0.500\linewidth-2\tabcolsep}p{\dimexpr 0.500\linewidth-2\tabcolsep}}
\toprule
Symbol & Meaning \\
\midrule
Q(t) & N$\times$N matrix; Q\_ij(t) = backlog at (i,j) at time t (packets/cells awaiting forwarding; see next row); on-chip this is the VOQ / egress queue depth \\
backlog & Packets/cells in the (i,j) direction that are queued and awaiting a crossbar grant, counted by Q\_ij. Q\_ij \textgreater  0 means there is backlog and a Request may be sent \\
P(t) & Binary grant matrix for the current cycle; at most one 1 per row and per column \\
P̄ & Time-averaged service matrix; P̄\_ij = E[P\_ij] \\
$\Lambda$ & Traffic / arrival rate matrix; $\lambda$\_ij is the mean arrival rate in direction (i,j) \\
$\lambda$ & Arrival vector ($\Lambda$ flattened) \\
$\rho$($\Lambda$) & Spectral radius of the traffic matrix $\Lambda$ \\
$\rho$\_load & Normalized load factor (per-input-port utilization, swept in experiments); distinct from spectral radius $\rho$($\Lambda$) \\
MWM & Maximum weight matching; each step solves max ⟨Q,P⟩ \\
MSM & Maximal matching schedulers: iSLIP and similar algorithms that seek only a maximal matching per cycle rather than the globally optimal MWM \\
IQ / VOQ & Input-queued switch; virtual output queue \\
iSLIP & McKeown's iterative round-robin matching algorithm [3] \\
HOL & Head-of-line blocking \\
R--G--A & Request--Grant--Accept \\
admissible & Admissible traffic \\
\bottomrule
\end{tabular}}\par\leftskip=2em\relax{}

\bigskip

\subsection{Terminology Alignment}

The table below aligns concepts from on-chip networks and data communications:

\bigskip\noindent
\par\leftskip=2em\relax\noindent \fittab{\begin{tabular}{p{\dimexpr 0.333\linewidth-2\tabcolsep}p{\dimexpr 0.333\linewidth-2\tabcolsep}p{\dimexpr 0.333\linewidth-2\tabcolsep}}
\toprule
On-chip / NoC & Datacom / IQ switch (paper and references) & Notes \\
\midrule
flit (burst=1) & cell / packet & At most one basic unit per cycle \\
master/slave IP; initiator/target & input/output port & N$\times$N crossbar assumption \\
crossbar matrix; centralized crossbar & switch fabric; IQ crossbar & Not a mesh NoC \\
egress queue per (i,j); Q & VOQ length Q & N/A \\
grant / arbitration / scheduler & switch scheduling; matching & iSLIP still called scheduling \\
link bandwidth; number of arbitration rounds per cycle & port nominal bandwidth; iterations per slot & Experiments: aligned at r rounds \\
saturation throughput; backpressure & throughput; admissible $\Lambda$; admissible & N/A \\
\bottomrule
\end{tabular}}\par\leftskip=2em\relax{}

\bigskip

Hardware-related terms such as per-cycle, weight-unaware, bit-plane matching, and row/column masking are explained in Appendix A.

\subsection{Related Work}

Existing IQ switch scheduling research can be classified by queue-read resolution and per-cycle iteration count; see §6.2 and §6.4 Table 2 for details.

\bigskip\noindent
\par\leftskip=2em\relax\noindent \fittab{\begin{tabular}{p{\dimexpr 0.250\linewidth-2\tabcolsep}p{\dimexpr 0.250\linewidth-2\tabcolsep}p{\dimexpr 0.250\linewidth-2\tabcolsep}p{\dimexpr 0.250\linewidth-2\tabcolsep}}
\toprule
Topic & Representative work & Core idea & Position in this paper \\
\midrule
Max-weight / Lyapunov stability & Tassiulas--Ephremides [7]; McKeown et al. 100\% throughput [4] & Queue-inner-product-based maximum weight optimization and Lyapunov stability; VOQ architecture can achieve full throughput under admissible traffic & Restated in §3.2--§3.4, connected to BvN relaxation theory [8],[9] \\
Iterative matching scheduling (industrial mainstream) & iSLIP [3]; McKeown PhD thesis [6]; comparative study [16] & Discrete matching via R-G-A iteration and round-robin pointers, trading low overhead for implementability & Core theory in §5; simulation comparison baseline in §7 \\
iSLIP family variants & DRRM [17]; FIRM [19]; SRR/DRDSRR [20],[21]; starvation-free MSM [15] & Improved delay and fairness via dual pointers, fairness constraints, etc. & §5.7.1 algorithm variants \\
Global-queue randomized scheduling & Tassiulas [22], APSARA/SERENA [23],[24],[25]; randomized evaluation [26] & Low-complexity weight matching using global queue information and randomization & §6.2, §8 \\
Distributed progressively optimal matching & $\epsilon$-Auction / min-sum [41] & Multi-round iteration converging to exact MWM & §6.2, §8 \\
Continuous-relaxation discrete matching & Sinkhorn [12], BvN scheduling [31]--[35] & Doubly stochastic matrix relaxation and BvN permutation decomposition & §4.1 theory; §7 entropy-regularized OT exp(Q/$\epsilon$) baseline with r\_sink rounds \\
Switch architecture extensions & CIOQ [27],[28],[29], CICQ/LIPS [36],[38],[39] & Internal queue architecture modifications to relieve global matching pressure & §6.2, §8 \\
Programmable queue scheduling & PIFO [40] & Programmable dataplane queue scheduling & §8 future research directions \\
\bottomrule
\end{tabular}}\par\leftskip=2em\relax{}

\bigskip

{\small\textbf{Table 1.} Related work summary (selected)}

\section{Problem Background and System Model}

\subsection{HOL Blocking, the 58.6\% Upper Bound, and the VOQ Motivation}

In classical space-division switching, if each input port maintains only a single FIFO (with no per-output queue structure), a packet at the head of the queue that is destined for an already-occupied output will block all subsequent packets on the same input port; this is head-of-line (HOL) blocking [1],[13]. Karol et al. showed that, in the large-N limit, the saturation throughput of this input-FIFO architecture is approximately 2-√2 $\approx$ 58.6\% (an architectural upper bound, not a property of the scheduler algorithm itself). Output-queued switches can approach 100\% bandwidth utilization, but in the worst case N inputs simultaneously target the same output port, so each output buffer requires write bandwidth of N $\times$ port rate, making the implementation cost prohibitive.

VOQ [4],[14] maintains a separate Q\_ij for each (i,j) pair, eliminating HOL at the architectural level; the scheduling problem reduces to choosing P(t) from within the matching set each cycle. Under admissible traffic, MWM using Q\_ij as weights can achieve 100\% throughput [4],[7]. Throughout this paper we assume an N$\times$N IQ + VOQ + single crossbar structure, which is isomorphic to the (i,j) egress queue setup of an SoC.

\subsection{System Model}

Consider an N$\times$N IQ switch: each input port i to output port j maintains an independent VOQ [4], with Q\_ij(t) being the backlog at (i,j). The scheduler selects a valid matching P(t) (a binary grant matrix with at most one 1 per row and per column) each clock cycle t, determining which cells are forwarded in this cycle (one flit when burst=1).

The rest of the paper proceeds along three main threads: (I) linear-algebraic modeling, MWM derivation, and stability; (II) graph-theoretic exact algorithms and hardware constraints; (III) the industrial iSLIP approximation and its statistical properties in the long-run average sense.

\section{Algebraic Principles of MWM and Stability}

This chapter establishes the algebraic and stability foundations for the rest of the paper, answering two questions: why MWM is a mathematically justified scheduling objective, and under what traffic conditions the system can remain stable. §3.1 gives the linear-algebraic model of the queueing system (queue matrix Q, traffic $\Lambda$, matching P, and the evolution equation); §3.2 defines MWM and its weight variants; §3.3 uses a Lyapunov energy function to explain the connection between MWM and the decrease in system energy, which is the algebraic source of the stability analysis; §3.4 uses eigenvalues and the Birkhoff--von Neumann theorem to characterize the range of admissible traffic, and explains that ``within the admissible region $\neq$ any scheduler is stable.'' Doubly stochastic matrices, the Birkhoff polytope, and the BvN theorem (§3.4) also form the mathematical foundation for the continuous-relaxation algorithms in §4.

\subsection{Linear-Algebraic Modeling of the Queueing System}

\subsubsection{State Vector}

Flattening the N$\times$N queue lengths (or keeping them in matrix form) gives the state:
\begin{equation}
Q(t)\in\mathbb{R}^{N^2}\ (\text{flattened as a vector})\quad\text{or}\quad Q(t)\in\mathbb{R}^{N\times N}\ (\text{kept as a matrix}).
\end{equation}

Q(t) changes over time: it increases when new packets arrive and decreases when packets are forwarded.

\begin{quote}
\textbf{Inner-product and energy notation:} For brevity, the vector notation $Q^{\mathsf{T}}P$ is used for matrices Q and P throughout to denote their scalar inner product: it means the inner product of Q and P after both are flattened into $\mathbb{R}^{N^2}$ vectors, which equals the Frobenius inner product $\langle Q,P\rangle=\operatorname{tr}(Q^{\mathsf{T}}P)=\sum_{i,j}Q_{ij}P_{ij}$. Similarly, the energy function $V=Q^{\mathsf{T}}Q$ means $\langle Q,Q\rangle=\operatorname{tr}(Q^{\mathsf{T}}Q)=\lVert Q\rVert_F^2$ (i.e., $\lVert Q\rVert^2$ throughout). When Q and P are treated as matrices, $Q^{\mathsf{T}}P$ and $Q^{\mathsf{T}}Q$ are both understood as scalars under this convention, not as matrix products.
\end{quote}

\subsubsection{Arrival Rate and Service}

\begin{itemize}
\item Arrival rate $\lambda$ (vector) or $\Lambda$ (matrix): the average number of packets arriving per unit time in each direction (i,j).
\item Service matrix P(t): the matching matrix actually executed in this cycle. In hardware, each selected (i,j) pair forwards exactly 1 cell/flit, so \texttt{P\_ij(t) ∈ \{0,1\}}, with at most one 1 per row and per column.
\end{itemize}

\subsubsection{State Evolution Equation}

In discrete time, the queue update within one clock cycle is [4],[7] (McKeown's IQ/VOQ cycle model; Tassiulas--Ephremides max-weight theory):
\begin{equation}
Q(t+1) = Q(t) + \lambda - P(t)
\end{equation}

In a full stochastic model this should be $Q(t+1)=\max\bigl(Q(t)+A(t) -P(t),\,0\bigr)$, where $A(t)$ is the random arrival quantity and $\mathbb{E}[A(t)]=\lambda$; throughout this paper we use the deterministic $\lambda$ in place of $\mathbb{E}[A(t)]$ to simplify worked examples and derivations, and temporarily ignore queue underflow (length going negative).

\textbf{Why subtraction rather than M(t)·Q(t)?}

\begin{itemize}
\item Matrix multiplication M·Q means draining a proportion of queue contents (e.g., processing 20\% each time).
\item Switch hardware is discrete: each cycle each port forwards at most 1 fixed-size cell, behavior that is independent of the absolute queue length (the matching selection depends on queue lengths, but that is a separate step).
\item Therefore vector subtraction \texttt{Q - P} more accurately captures the physical behavior of ``subtracting a fixed number per cycle.''
\end{itemize}

\subsection{Maximum Weight Matching Algorithm}

\subsubsection{Weight and Optimization Objective}

The scheduling algorithm makes only a 0/1 ``transmit or not'' decision each cycle, but it can introduce weight preferences when selecting the matching. MWM (Maximum Weight Matching) [4] uses the queue lengths themselves as weights:
\begin{equation}
\max_{P\in M}\ \sum_{i,j} Q_{ij}(t)\, P_{ij}(t) \;=\; \max_{P\in M}\ \langle Q(t),\, P(t)\rangle
\end{equation}

where M is the set of all valid matching matrices (permutation matrices or their subsets). \texttt{⟨Q, P⟩} denotes the inner product of Q and P (element-wise multiply then sum).

\textbf{Physical intuition:} The longer the queue in direction (i,j), the higher the priority of that direction receiving a forwarding slot in the current cycle.

\subsubsection{Introductory Example (2 Queues / Single Server, Scalar Illustration)}

\begin{quote}
\textbf{Notation note:} This example uses a simplified ``2 queues competing for 1 server'' scalar illustration (only 1 packet forwarded per cycle), not the N$\times$N queue matrix and permutation matrices defined in §3.1.1.
\end{quote}

Current queue vector (input 1 severely congested, input 2 idle):
\begin{equation}
Q = \begin{bmatrix} 10 \\ 2 \end{bmatrix}
\end{equation}

Only 1 packet can be forwarded per cycle. Two options:

\bigskip\noindent
\par\leftskip=2em\relax\noindent \fittab{\begin{tabular}{p{\dimexpr 0.250\linewidth-2\tabcolsep}p{\dimexpr 0.250\linewidth-2\tabcolsep}p{\dimexpr 0.250\linewidth-2\tabcolsep}p{\dimexpr 0.250\linewidth-2\tabcolsep}}
\toprule
Option & Service vector P & Next-step Q & Energy V = ‖Q‖\textsuperscript{2} \\
\midrule
A: prioritize queue 1 & \texttt{[1, 0]ᵀ} & \texttt{[9, 2]ᵀ} & 85 \\
B: prioritize queue 2 & \texttt{[0, 1]ᵀ} & \texttt{[10, 1]ᵀ} & 101 \\
\bottomrule
\end{tabular}}\par\leftskip=2em\relax{}

\bigskip

Initial energy:
\begin{equation}
V(0) = 10^2 + 2^2 = 104
\end{equation}

MWM selects option A (inner product \texttt{10$\times$1 + 2$\times$0 = 10 \textgreater  2}), with energy decrease \texttt{104 - 85 = 19}; option B decreases it by only 3.

\textbf{Conclusion:} MWM, in physical terms, means that every step chooses the direction along which the system's ``total energy'' decreases most rapidly.

\subsubsection{MWM Weight Variants: LQF, OCF, and LPF}

McKeown et al. [4] compared several max-weight definitions in the VOQ setting; all belong to the MWM family (each step solves max ⟨Q, P⟩, but with a different weight function):

\bigskip\noindent
\par\leftskip=2em\relax\noindent \fittab{\begin{tabular}{p{\dimexpr 0.250\linewidth-2\tabcolsep}p{\dimexpr 0.250\linewidth-2\tabcolsep}p{\dimexpr 0.250\linewidth-2\tabcolsep}p{\dimexpr 0.250\linewidth-2\tabcolsep}}
\toprule
Weight & Definition & 100\% throughput & Fairness / hardware \\
\midrule
LQF & Q\_ij itself (longest queue first) & ✓ [4] & May starve short queues \\
OCF & Oldest cell first & ✓ [4] & Better fairness \\
LPF & Largest port weight within maximal matching set [5] & ✓ [5] & Easier to implement in hardware than LQF \\
\bottomrule
\end{tabular}}\par\leftskip=2em\relax{}

\bigskip

\subsection{Lyapunov Stability and the Algebraic Origin of MWM}

We want the queues to remain bounded, so that ‖Q(t)‖ does not grow without bound. The standard device is to define an ``energy'' that measures total backlog and study when it can decrease; this turns the stability question into an algebraic condition on the per-cycle change $\Delta$V. The rest of this section derives $\Delta$V first without arrivals (§3.3.2) and then with them (§3.3.3), and reads off the scheduling rule that the condition implies.

\subsubsection{Energy Function (Lyapunov Function)}

Define the total system energy as the squared backlog:
\begin{equation}
V(t) = Q(t)^{\mathsf{T}} Q(t) = \lVert Q(t)\rVert^2
\end{equation}

a standard quadratic form. ``Energy decreasing'' is exactly ``queues draining.''

\subsubsection{Derivation Without Arrivals ($\lambda$ = 0)}

The state equation simplifies to \texttt{Q(t+1) = Q(t) - P(t)}.

Expanding V(t+1):
\begin{align}
V(t+1) &= (Q - P)^{\mathsf{T}} (Q - P) \nonumber \\
&= Q^{\mathsf{T}}Q - 2\,Q^{\mathsf{T}}P + P^{\mathsf{T}}P
\end{align}

Energy change:
\begin{equation}
\Delta V = V(t+1) - V(t) = -2\,Q^{\mathsf{T}}P + \lVert P\rVert^2
\end{equation}

Analysis of the two terms:

\bigskip\noindent
\par\leftskip=2em\relax\noindent \fittab{\begin{tabular}{p{\dimexpr 0.500\linewidth-2\tabcolsep}p{\dimexpr 0.500\linewidth-2\tabcolsep}}
\toprule
Term & Property \\
\midrule
‖P‖\textsuperscript{2} & Assuming a perfect matching is achieved each cycle (N edges), this term equals the constant N \\
-2·QᵀP & The only controllable term; to make $\Delta$V as negative as possible, we must maximize QᵀP \\
\bottomrule
\end{tabular}}\par\leftskip=2em\relax{}

\bigskip

\begin{quote}
\textbf{Note:} The premise \texttt{‖P‖\textsuperscript{2} = N} requires a perfect matching every cycle. If some ports are idle, \texttt{‖P‖\textsuperscript{2} \textless  N}, so the constant term is smaller and the conclusion that $\Delta$V $\leq$ 0 is not affected (a smaller ‖P‖\textsuperscript{2} is more favorable for $\Delta$V $\leq$ 0). A more complete Lyapunov derivation handles \texttt{‖P‖\textsuperscript{2}} using a fixed upper bound N. The relevant treatment is in §3.3--§3.4; iSLIP stability under uniform i.i.d. traffic is discussed in McKeown et al. [3].
\end{quote}

Therefore:
\begin{equation}
\max_{P\in M} Q^{\mathsf{T}}P \iff \text{MWM}
\end{equation}

\textbf{Conclusion:} MWM is not an arbitrary design choice; it is derived from the algebraic condition that makes the quadratic energy decrease as fast as possible [4],[7].

\subsubsection{Full Formula With Arrivals}

Substituting the full state equation \texttt{Q(t+1) = Q(t) + $\lambda$ - P(t)}:
\begin{equation}
\Delta V = 2\,Q^{\mathsf{T}}(\lambda - P) + \lVert \lambda - P\rVert^2
\end{equation}

\begin{itemize}
\item Qᵀ$\lambda$ \textgreater  0 (when arrivals occur on directions that have backlog): new traffic injects energy, growing the queues.
\item -QᵀP \textless  0: the scheduling algorithm removes energy, shrinking the queues.
\end{itemize}

The basic condition for system stability is that the energy removed by scheduling (long-run average) must exceed the energy injected.

\subsubsection{Worked Example: $\Delta$V With Arrivals and the Injection--Service Trade-off}

\begin{quote}
\textbf{Notation note:} This example follows the simplified ``2 queues competing for 1 server'' scalar illustration from §3.2.2 (Q, $\lambda$, P are 2-dimensional vectors), not the N$\times$N queue matrix and permutation matrices from §3.1.1. A complete 2$\times$2 switch worked example is given in §3.4.6.
\end{quote}

Let N=2, with the following queue and traffic vectors in a given cycle:
\begin{equation}
Q = \begin{bmatrix} 100 \\ 50 \end{bmatrix}, \quad
\lambda = \begin{bmatrix} 1 \\ 1 \end{bmatrix}, \quad
P = \begin{bmatrix} 1 \\ 0 \end{bmatrix}
\end{equation}

Meaning: both queues are long; 1 packet arrives on each path this cycle; the scheduler serves only input 1 (which is what MWM would choose).

\textbf{Step 1. Substitute into the $\Delta$V formula}
\begin{equation}
\Delta V = 2\,Q^{\mathsf{T}}(\lambda - P) + \lVert \lambda - P\rVert^2
\end{equation}

\textbf{Step 2. Compute the inner-product term Qᵀ($\lambda$ - P)}
\begin{align}
&\lambda - P = [0,\ 1]^{\mathsf{T}} \nonumber \\
&Q^{\mathsf{T}}(\lambda - P) = 100\times 0 + 50\times 1 = 50 \ > 0
\end{align}

\ensuremath{\rightarrow} The net effect this cycle is to inject energy (queue 2 received a packet but was not served).

\textbf{Step 3. Compute the constant term}
\begin{equation}
\lambda - P = [0,\ 1]^{\mathsf{T}} \ \to\ \lVert \lambda - P\rVert^2 = 0^2 + 1^2 = 1
\end{equation}

\textbf{Step 4. Combine}
\begin{equation}
\Delta V = 2\times 50 + 1 = 101 \ > 0 \quad (\text{energy increases; system more congested})
\end{equation}

At most one path can be served per cycle, so $\Delta$V \textgreater  0 is possible in this example. Long-term stability does not require $\Delta$V \textless  0 every cycle. What is required is that, in the long-run average, E[QᵀP] is sufficient to outweigh the injection Qᵀ$\lambda$; when Q is large, MWM continuously tilts service toward the longest queue, so the average energy removed exceeds what is injected.

\textbf{Injection--service trade-off decomposition:}
\begin{align}
&Q^\mathsf{T}\lambda = 100\times 1 + 50\times 1 = 150 \quad (\text{injection, positive}) \nonumber \\
&Q^\mathsf{T} P = 100\times 1 + 50\times 0 = 100 \quad (\text{removal, negative}) \nonumber \\
&Q^\mathsf{T}(\lambda - P) = 50 \quad (\text{injection wins this cycle})
\end{align}

\subsection{Traffic Matrix, Eigenvalues, and the Birkhoff--von Neumann Theorem}

\subsubsection{Two Different Matrices}

\bigskip\noindent
\par\leftskip=2em\relax\noindent \fittab{\begin{tabular}{p{\dimexpr 0.333\linewidth-2\tabcolsep}p{\dimexpr 0.333\linewidth-2\tabcolsep}p{\dimexpr 0.333\linewidth-2\tabcolsep}}
\toprule
Matrix & Meaning & Determined by \\
\midrule
$\Lambda$ (traffic matrix) & \textbf{Demand-side} mean arrival rate $\lambda$\_ij for each direction (i,j): packets that \textit{want} to go i\ensuremath{\rightarrow}j per cycle & External traffic; an objective fact \\
P(t) (scheduling matrix) & The matching actually executed in cycle t; row/column sums always $\leq$ 1 (crossbar forwards at most 1/cycle/port) & Computed in real time by the scheduler \\
\bottomrule
\end{tabular}}\par\leftskip=2em\relax{}

\bigskip

MWM optimizes ⟨Q(t), P(t)⟩, not ⟨$\Lambda$, P(t)⟩. The chip at runtime can only observe the current queue Q(t); it cannot know future traffic.

\subsubsection{Constraints on the Traffic Matrix}

$\Lambda$ is an N$\times$N non-negative matrix. Physically, $\Lambda$\_ij is a \textbf{demand-side} rate: the number of packets that \textit{want} to go from input i to output j per cycle (a count over a measurement window divided by the window length). Because an input port may simultaneously have traffic destined for several outputs, the row sum $\Sigma$\_j $\Lambda$\_ij can \textbf{exceed 1}, which means input i is offered more traffic than its line rate (1 packet/cycle) can carry, i.e. the port is overloaded. The \textbf{admissibility} requirement is precisely that this does not happen:

\begin{itemize}
\item Row sums $\leq$ 1 (no input port is offered more than its line rate)
\item Column sums $\leq$ 1 (no output port is offered more than its drain rate)
\end{itemize}

When both hold, $\Lambda$ is called sub-doubly-stochastic, a statistical description of traffic \textit{demand}. When a row or column sum exceeds 1, no scheduler can keep the queues bounded (overloaded 2$\times$2 example in §3.4.3). This is a statement about \textit{offered} traffic; the \textit{actually forwarded} P(t) and its time average P̄ always have row and column sums $\leq$ 1, enforced by the crossbar's one-transmit-per-cycle-per-port physical limit.

\subsubsection{Spectral Radius and the Perron--Frobenius Theorem}

The admissibility condition in §3.4.2 is stated as 2N port-load inequalities (every row sum and column sum \textless  1). A natural follow-up is whether these can be reduced to a single number indicating the system load. The spectral radius $\rho$($\Lambda$), the largest eigenvalue magnitude, is the natural candidate, and this section examines to what extent it suffices.

If every row sum and column sum of $\Lambda$ is strictly less than 1, then from the matrix-norm bound (the spectral radius never exceeds any consistent matrix norm):
\begin{equation}
\rho(\Lambda) \le \|\Lambda\|_\infty = \max_i \sum_j |\Lambda_{ij}| < 1
\end{equation}

So admissibility (§3.4.2) \ensuremath{\Rightarrow} $\rho$($\Lambda$) \textless  1. For non-negative matrices the Perron--Frobenius theorem sharpens this: $\rho$($\Lambda$) lies between the minimum and maximum row sums, consistent with the bound above.

The converse fails: $\rho$($\Lambda$) \textless  1 does \textbf{not} imply admissibility. The admissibility condition for an IQ switch is inherently per-port, requiring all row sums and column sums to be strictly below 1, and the spectral radius is only a necessary consequence, not a complete characterization of the capacity region. ($\rho$($\Lambda$) averages over the spatial distribution of load; two matrices with the same $\rho$ can differ in whether a single port is overloaded.)

\textbf{Worked intuition (2$\times$2 diagonal traffic):}
\begin{equation}
\Lambda = \begin{bmatrix} 0.6 & 0 \\ 0 & 0.6 \end{bmatrix}
\qquad \text{row and column sums both } 0.6 < 1
\end{equation}

Eigenvalues are 0.6, so $\rho$($\Lambda$)=0.6 \textless  1. Physical meaning: even if all ports are loaded, on average only 60\% of port capacity is used, so the system has spare capacity.

If we instead have:
\begin{equation}
\Lambda = \begin{bmatrix} 0.9 & 0.3 \\ 0.3 & 0.9 \end{bmatrix}
\qquad \text{row sum } = 1.2 \text{ (overloaded!)}
\end{equation}

Row sum exceeds 1 \ensuremath{\rightarrow} some input port exceeds its physical injection capacity; no scheduler can maintain long-term stability (queues must overflow).

\begin{quote}
Note: Throughout the rest of this paper, admissible traffic refers to input-port row sums and output-port column sums all strictly less than 1.
\end{quote}

\subsubsection{Birkhoff--von Neumann (BvN) Theorem [8],[9]}

What does ``admissible traffic'' (§3.4.2) mean concretely? BvN answers: the demand $\Lambda$ can be carried by time-multiplexing a set of permutation matchings P\_k with fractions $\alpha$\_k. That existence is the foundation of stability, because it provides a long-run-average schedule that serves every direction at least at its arrival rate; the scheduler's task is to track it.

Any doubly stochastic matrix B (non-negative elements, row sums = column sums = 1) can be decomposed as a convex combination of permutation matrices:
\begin{equation}
B = \alpha_1 P_1 + \alpha_2 P_2 + \dots + \alpha_m P_m \quad (\alpha_k > 0,\ \textstyle\sum_k \alpha_k = 1)
\end{equation}

For N$\times$N matrices, the number of permutation matrices needed satisfies m $\leq$ N\textsuperscript{2} - 2N + 2 (upper bound on the number of permutation matrices in a BvN decomposition; the number of vertices of the Birkhoff polytope is N!).

For a sub-doubly-stochastic traffic matrix $\Lambda$ (non-negative elements, row sums and column sums all $\leq$ 1), the BvN decomposition shows that there exists a set of permutation matrices \{P\_k\} with weights $\alpha$\_k ($\Sigma$ $\alpha$\_k $\leq$ 1) such that $\Lambda$ can be represented by, or dominated by, a convex combination of those matrices:
\begin{equation}
\Lambda \le \alpha_1 P_1 + \alpha_2 P_2 + \dots + \alpha_m P_m
\end{equation}

Denoting this convex combination as S, we have S\_ij $\geq$ $\Lambda$\_ij for all (i,j). S is a valid long-run-average scheduling combination; it only asserts that in the average sense the traffic capacity condition can be satisfied, and does not prescribe the actual scheduling matrix P(t) for each individual cycle.

\textbf{Traffic region (BvN existence conclusion):} The relation above can be expressed in existential language: when $\Lambda$ satisfies the admissible port-load constraints, there exists a sub-doubly-stochastic matrix S = $\Sigma$ $\alpha$\_k P\_k ($\Sigma$ $\alpha$\_k $\leq$ 1) such that
\begin{equation}
S \ge \Lambda
\end{equation}

The condition that the scheduler must satisfy is: when the queues are long enough, make the inner-product negative-drift condition hold:
\begin{equation}
E[Q^{\mathsf{T}}P] > Q^{\mathsf{T}}\lambda \iff E[Q^{\mathsf{T}}(\lambda - P)] < 0
\end{equation}

where $\lambda$ is the arrival vector defined by $\Lambda$ (notation consistent with §3.2), and the expectation operator E[·] is explained in §5.6.4. When this holds, the Lyapunov energy function V = ‖Q‖\textsuperscript{2} decreases steadily and the queue length remains bounded.

MWM, at each scheduling cycle, solves max ⟨Q,P⟩ for the current queue state Q(t). Because this policy always selects the currently optimal matching, its single-step inner product value, in the same state, is never lower than that of any fixed convex combination S. Therefore, from a long-run statistical perspective, MWM can satisfy the negative-drift inequality above and guarantee system stability.

\textbf{Intuition:} $\Lambda$ is the injection intensity; the admissible load condition (§3.4.2) caps it; and the scheduler makes grants favor the most congested queues in the ⟨Q,P⟩ dimension, thereby controlling queue depth.

\subsubsection{Geometric Picture}

In the N\textsuperscript{2}-dimensional queue space, V(Q) = ‖Q‖\textsuperscript{2} forms a high-dimensional paraboloid (a ``bowl''). As long as $\Lambda$ is strictly inside the admissible traffic region and the scheduler is strong enough (e.g., MWM), the negative feedback of MWM always pulls the queue vector back toward the bottom of the bowl, keeping the queue lengths bounded.

\subsubsection{Admissible Traffic $\neq$ Any Scheduler Is Stable}

A common misconception: since all input/output port loads of $\Lambda$ are below 1 and the total capacity is sufficient, any choice of P, even random matching each cycle, will eventually converge given enough time.

\textbf{Correction:}

\bigskip\noindent
\par\leftskip=2em\relax\noindent \fittab{\begin{tabular}{p{\dimexpr 0.500\linewidth-2\tabcolsep}p{\dimexpr 0.500\linewidth-2\tabcolsep}}
\toprule
Statement & Correct/Incorrect \\
\midrule
Row sums and column sums all \textless  1 \ensuremath{\Rightarrow} some scheduler exists that stabilizes the system & Correct \\
Row sums and column sums all \textless  1 \ensuremath{\Rightarrow} any P(t) is stable & Incorrect \\
``Given enough time'' \ensuremath{\Rightarrow} queues necessarily reach 0 & Incorrect (with $\lambda$ \textgreater  0 and continuous arrivals, queues never drain to 0) \\
``Given enough time'' \ensuremath{\Rightarrow} queues are bounded & Incorrect: holds only for schedulers with queue feedback \\
\bottomrule
\end{tabular}}\par\leftskip=2em\relax{}

\bigskip

\textbf{Admissible} means: there exists a convex combination of matching rates that can carry $\Lambda$ in the average sense. It does not mean ``any service allocation works.''

The Lyapunov stability condition (when Q is large) is:
\begin{equation}
E[Q^{\mathsf{T}}P] > Q^{\mathsf{T}}\lambda
\end{equation}

\textbf{Worked counterexample (2$\times$2): admissible, but weight-unaware random matching is unstable}

The scheduling policy is weight-unaware random matching: each cycle the two perfect matchings I and X in a 2$\times$2 system are chosen with equal probability 0.5, independent of the values of Q\_ij (no MWM-style weighting).

\textbf{Step 0. Traffic matrix $\Lambda$ (admissible, arrivals on all four directions)}
\begin{equation}
\Lambda = \begin{bmatrix} 0.85 & 0.08 \\ 0.07 & 0.08 \end{bmatrix}
\qquad \text{row sums } 0.93 / 0.15;\ \text{column sums } 0.92 / 0.16 \;\ensuremath{\Rightarrow}\; \rho(\Lambda) < 1
\end{equation}

Meaning: (1,1) is the dominant heavy flow (0.85); cross-paths (1,2), (2,1), and (2,2) also carry small flows, so all four VOQs will be non-empty after running for a while.

\textbf{Step 1. Current queue state Q (typical congested state)}
\begin{equation}
Q = \begin{bmatrix} 8000 & 60 \\ 50 & 40 \end{bmatrix}
\end{equation}

Q\_11 is very large; (1,2), (2,1), (2,2) all have backlog, so no Q\_ij = 0.

\textbf{Step 2. Two perfect matchings and the random rule}

In a 2$\times$2 system, only two perfect matchings exist (I and X are both complete schedules and cannot be used simultaneously):
\begin{equation}
I = \begin{bmatrix} 1 & 0 \\ 0 & 1 \end{bmatrix}
\quad (\text{serves } (1,1)+(2,2),\ \text{straight-through})
\qquad
X = \begin{bmatrix} 0 & 1 \\ 1 & 0 \end{bmatrix}
\quad (\text{serves } (1,2)+(2,1),\ \text{cross})
\end{equation}

Each cycle P(I) = P(X) = 0.5, independent of the values of Q.

\bigskip\noindent
\par\leftskip=2em\relax\noindent \fittab{\begin{tabular}{p{\dimexpr 0.167\linewidth-2\tabcolsep}p{\dimexpr 0.167\linewidth-2\tabcolsep}p{\dimexpr 0.167\linewidth-2\tabcolsep}p{\dimexpr 0.167\linewidth-2\tabcolsep}p{\dimexpr 0.167\linewidth-2\tabcolsep}p{\dimexpr 0.167\linewidth-2\tabcolsep}}
\toprule
Matching this cycle & P\_11 & P\_12 & P\_21 & P\_22 & VOQs actually served (under Q above) \\
\midrule
I & 1 & 0 & 0 & 1 & Q\_11=8000, Q\_22=40 \\
X & 0 & 1 & 1 & 0 & Q\_12=60, Q\_21=50 (both have backlog, no wasted service) \\
\bottomrule
\end{tabular}}\par\leftskip=2em\relax{}

\bigskip

Long-run average:
\begin{equation}
E[P_{11}] = E[P_{12}] = E[P_{21}] = E[P_{22}] = 0.5
\end{equation}

Key point: even with Q\_11 = 8000, E[P\_11] is fixed at 0.5 and does not increase with Q\_11. Weight-unaware random matching does not allocate more bandwidth to the direction with high backlog.

\textbf{Step 3. Inner-product criterion: E[QᵀP] \textless  Qᵀ$\lambda$ (service deficit \ensuremath{\rightarrow} unstable)}

Treating Q as approximately constant over this period:
\begin{align}
Q^{\mathsf{T}}\lambda &= 8000\times 0.85 + 60\times 0.08 + 50\times 0.07 + 40\times 0.08 \nonumber \\
&= 6800 + 4.8 + 3.5 + 3.2 = 6811.5 \quad (\text{injection}) \nonumber \\
E[Q^{\mathsf{T}}P] &= 8000\times 0.5 + 60\times 0.5 + 50\times 0.5 + 40\times 0.5 \nonumber \\
&= 4000 + 30 + 25 + 20 = 4075 \quad (\text{expected removal}) \nonumber \\
E[Q^{\mathsf{T}}P] - Q^{\mathsf{T}}\lambda &= 4075 - 6811.5 = -2736.5 < 0
\end{align}

``\textless  0'' here means (expected service inner product - injection inner product) \textless  0, i.e., the average per-step grant falls short of arrivals (service deficit); the queue length in direction (1,1) and others drifts upward, and the system is unstable.

Q\_11 dominates the injection term (6800 out of 6811.5), but E[P\_11] is fixed at 0.5 \ensuremath{\rightarrow} the high-backlog direction is chronically under-served \ensuremath{\rightarrow} Q\_11 drifts upward without bound.

\textbf{Step 4. Single-cycle vs. expected behavior}

\bigskip\noindent
\par\leftskip=2em\relax\noindent \fittab{\begin{tabular}{p{\dimexpr 0.333\linewidth-2\tabcolsep}p{\dimexpr 0.333\linewidth-2\tabcolsep}p{\dimexpr 0.333\linewidth-2\tabcolsep}}
\toprule
Choice this cycle & Immediate service to (1,1) & Other behavior \\
\midrule
I & Q\_11·P\_11 = 8000 (sufficient this cycle) & Also serves (2,2) \\
X & Q\_11·P\_11 = 0 (no service to (1,1) backlog) & Serves (1,2), (2,1): both have backlog, not wasted, but occupies a cycle that could have served (1,1) \\
\bottomrule
\end{tabular}}\par\leftskip=2em\relax{}

\bigskip

The instability comes from the expectation: E[P\_11] = 0.5 \textless  $\lambda$\_11 = 0.85, not from ``choosing X at every cycle.'' Choosing X in this example is legitimate (it serves non-empty VOQs), but the frequency is wrong given that $\Lambda$ is skewed toward (1,1).

\textbf{Step 5. Conclusion}
\begin{align}
\lambda_{11} = 0.85,\ E[P_{11}] = 0.5 &\ \to\ (1,1)\ \text{unstable,}\ Q_{11}\to\infty \nonumber \\
\lambda_{22} = 0.08,\ E[P_{22}] = 0.5 &\ \to\ (2,2)\ \text{over-served, stable} \nonumber \\
\lambda_{12} = 0.08,\ E[P_{12}] = 0.5 &\ \to\ \text{slightly over-served} \nonumber \\
\lambda_{21} = 0.07,\ E[P_{21}] = 0.5 &\ \to\ \text{slightly over-served}
\end{align}

Instability is direction-specific: even when the total capacity is sufficient, weight-unaware random matching cannot reallocate matching probabilities when the backlog at (1,1) grows. By contrast, MWM under the same Q would choose I in almost every cycle (Q\_11 has the largest weight), so it never falls into the fixed-50/50 allocation.

\textbf{Summary:} The default assumption in this paper is ``admissible traffic (row sums and column sums all \textless  1) + scheduler with queue feedback \ensuremath{\Rightarrow} stable''; not ``any scheduler within the physical capacity limit converges.''

\section{Computing MWM: Linear Algebra and Graph Theory}

This chapter translates the algebraic language of §3 into concrete computation: how to compute the matching P for the current cycle. The objective of MWM, $\max_{P\in M} \langle Q,P\rangle$ (§3.2), is mathematically clear and falls into two categories: (I) \textbf{Continuous relaxation then discretization.} First relax the discrete matching to a continuous matrix, solve it, then project back to 0/1; implemented as the scaling method (§4.1, Sinkhorn \ensuremath{\rightarrow} doubly stochastic \ensuremath{\rightarrow} BvN) and the spectral method (§4.2, power iteration / SVD leading component). (II) \textbf{Graph-theoretic exact matching (§4.3).} Without going through continuous relaxation, directly find the optimal combination over the matching set $\mathcal{M}$. The doubly stochastic matrices, Birkhoff polytope, and BvN theorem used by the relaxation methods were established in §3.4, which is the mathematical foundation for this chapter.

\begin{quote}
The main difficulty for both relaxation methods lies not in ``finding the continuous solution'' but in ``how to discretize the continuous matrix into a valid permutation matrix within a single cycle with a limited number of iterations.'' This is precisely where the industrial iSLIP of §5 applies its hardware effort.
\end{quote}

\subsection{Linear-Algebraic Approach I: Scaling and BvN Decomposition}

This section presents the first continuous-relaxation method: first scale the queue matrix Q into a doubly stochastic matrix B using Sinkhorn iteration, then use the BvN theorem (§3.4.4) to decompose B as a convex combination of permutation matrices, of the form B = $\Sigma$ $\alpha$\_k P\_k, and then interpret the coefficients $\alpha$\_k as the time-multiplexing fractions for each permutation P\_k.

\subsubsection{Sinkhorn--Knopp Iteration (Continuous Relaxation) [12]}

Alternately apply row normalization and column normalization to a positive matrix Q until convergence to a doubly stochastic matrix B. The reason for alternating: applying only row normalization makes each row sum equal to 1 but the column sums are generally $\neq$ 1 (multiple inputs still tend toward the same output, violating the at-most-one-match-per-column rule); applying only column normalization has the opposite problem. Sinkhorn's approach is to alternate row and column normalization repeatedly, driving both row sums and column sums toward 1 simultaneously, converging to the doubly stochastic matrix B.

\textbf{Worked example:}
\begin{equation}
Q = \begin{bmatrix} 8 & 2 \\ 3 & 5 \end{bmatrix}
\end{equation}

\textbf{Round 1, row normalization} (divide each row by its row sum):
\begin{equation}
B^{(1)} = \begin{bmatrix} 0.8 & 0.2 \\ 0.375 & 0.625 \end{bmatrix}
\end{equation}

\textbf{Round 1, column normalization} (divide each column by its column sum):
\begin{equation}
B^{(2)} = \begin{bmatrix} 0.681 & 0.242 \\ 0.319 & 0.758 \end{bmatrix}
\end{equation}

Continue iterating to convergence; the exact limit is approximately $\begin{bmatrix}0.72 & 0.28 \\ 0.28 & 0.72\end{bmatrix}$ ($B^{(2)}$ row sums are not yet 1, showing that two rounds are insufficient for convergence). We round to 0.7/0.3 to simplify the subsequent BvN peeling example:
\begin{equation}
B \approx \begin{bmatrix} 0.7 & 0.3 \\ 0.3 & 0.7 \end{bmatrix}
\qquad (\text{row sums and column sums all equal } 1)
\end{equation}

\subsubsection{BvN Peeling}

In the 2$\times$2 case, only two valid permutation matrices exist:
\begin{equation}
P_1 = \begin{bmatrix} 1 & 0 \\ 0 & 1 \end{bmatrix} \quad (\text{straight-through})
\qquad
P_2 = \begin{bmatrix} 0 & 1 \\ 1 & 0 \end{bmatrix} \quad (\text{cross})
\end{equation}

\textbf{Peel P\textsubscript{1}:} take the minimum of the two B entries on the support set ⟨0,0⟩, ⟨1,1⟩ of P\textsubscript{1}: \texttt{$\alpha$\textsubscript{1} = min(0.7, 0.7) = 0.7}. This ensures B - $\alpha$\textsubscript{1}P\textsubscript{1} remains non-negative on the support set.
\begin{equation}
B_{\text{remain}} = B - 0.7 \cdot P_1 = \begin{bmatrix} 0 & 0.3 \\ 0.3 & 0 \end{bmatrix}
\end{equation}

\textbf{Peel P\textsubscript{2}:} \texttt{$\alpha$\textsubscript{2} = min(0.3, 0.3) = 0.3}.
\begin{equation}
B_{\text{remain}}' = B_{\text{remain}} - 0.3 \cdot P_2 = \begin{bmatrix} 0 & 0 \\ 0 & 0 \end{bmatrix}
\end{equation}

\textbf{Final result:}
\begin{equation}
B = 0.7\,P_1 + 0.3\,P_2
\end{equation}

\textbf{Hardware meaning:} In 10 clock cycles, 7 cycles execute P\textsubscript{1} and 3 cycles execute P\textsubscript{2}; the long-run average service capacity is B.

\textbf{Conclusion:} Sinkhorn + BvN is a continuous-relaxation / frame-scheduling approximation of MWM (same class as the spectral scheduling of §4.2), not exact MWM; exact MWM still requires an optimal search over the matching set (§4.3, Appendix B.1--B.2).

\textbf{Complete flow:}

\par{\leftskip=0pt\relax\noindent\hspace*{2em}\begin{minipage}{\dimexpr\linewidth-2em\relax}
\begin{verbatim}
Q (queue lengths)
  → Sinkhorn iteration → B (continuous doubly stochastic)
  → BvN decomposition → { α_k, P_k}
  → time -multiplexed scheduling
\end{verbatim}
\end{minipage}\par}

\subsection{Linear-Algebraic Approach II: Spectral Scheduling and Power Iteration}

A linear-algebra textbook finds singular vectors by writing the characteristic equation and solving a high-degree polynomial. A switch chip cannot do that in one cycle, but it \textit{can} multiply a matrix by a vector many times. Power iteration exploits exactly this gap: repeatedly applying Q and Qᵀ to a vector makes it align with the dominant singular direction of Q, namely the single input\ensuremath{\leftrightarrow}output coupling that carries most of the current congestion, without ever forming the full SVD. The method therefore trades many hardware-friendly matrix-vector products for one infeasible eigensolve, and then discretizes the resulting direction into a 0/1 matching.

\begin{quote}
\textbf{Terminology note:} There is no standardized name ``Spectral Scheduling'' in the industry. This section summarizes a common approach that combines spectral methods with matching: use the leading singular vectors u\textsubscript{1}, v\textsubscript{1} to capture global features, combine them with the constraint boundaries of the queue matrix Q, and ultimately solve for the discrete matching matrix P.
\end{quote}

\subsubsection{Overall Flow of Spectral Scheduling}

The intermediate quantities from power iteration and SVD are real-valued matrices or vectors; the crossbar needs an integer 0/1 P(t) scheduling matrix each cycle (as discussed in §4). This section proceeds step by step: ``find spectrum \ensuremath{\rightarrow} (optional) construct W \ensuremath{\rightarrow} discretize into P.'' The overall flow is:

\bigskip\noindent
\par\leftskip=2em\relax\noindent \fittab{\begin{tabular}{p{\dimexpr 0.333\linewidth-2\tabcolsep}p{\dimexpr 0.333\linewidth-2\tabcolsep}p{\dimexpr 0.333\linewidth-2\tabcolsep}}
\toprule
Stage & Operation & Subsection \\
\midrule
Find spectrum & Power iteration on \texttt{Q(t)} \ensuremath{\rightarrow} u\textsubscript{1}, v\textsubscript{1} & §4.2.3 \\
Path A (explicit W, common in offline/simulation) & \texttt{W = Q ⊙ (u\textsubscript{1} v\textsubscript{1}ᵀ)} & §4.2.4 \\
 & \ensuremath{\rightarrow} Greedy / W\ensuremath{\rightarrow}iSLIP / Hungarian on W & §4.2.5 \\
Path B (no W constructed, common on-chip) & \texttt{Q·y \ensuremath{\rightarrow} Rounding \ensuremath{\rightarrow} Qᵀ·x \ensuremath{\rightarrow} \dots \ensuremath{\rightarrow} P(t)} & §4.2.3 \\
Output & \texttt{P(t)} (binary matching matrix) & --- \\
\bottomrule
\end{tabular}}\par\leftskip=2em\relax{}

\bigskip

\subsubsection{Meaning of u\textsubscript{1} and v\textsubscript{1}}

u\textsubscript{1} and v\textsubscript{1} are the pair of left and right singular vectors corresponding to the largest singular value of Q: u\textsubscript{1} is associated with the input-port side, v\textsubscript{1} with the output-port side. They are computed from a joint decomposition of the entire Q matrix. They reflect the global coupling direction between inputs and outputs under the current congestion, not the scalar queue lengths of each row or column independently.

\subsubsection{Spectral Scheduling Step 1: Power Iteration with Spectral Smoothing (Rounding)}

\paragraph{(1) Power Iteration: How u\textsubscript{1} and v\textsubscript{1} Are Computed from Matrix Multiplication}

Take a random initial vector y\textsubscript{0} (length = number of output ports N) and alternate multiplying by Q and Qᵀ:
\begin{align}
x_1 &= Q\, y_0 \nonumber \\
y_1 &= Q^{\mathsf{T}} x_1 \nonumber \\
x_2 &= Q\, y_1 \nonumber \\
y_2 &= Q^{\mathsf{T}} x_2 \nonumber \\
&\;\;\vdots
\end{align}

At each step, x and y can be normalized (divided by their norms) to avoid numerical overflow.

\textbf{Convergence (connection to Eckart--Young):} after repeated Q·y / Qᵀ·x, the direction of x converges to the leading left singular vector u\textsubscript{1} and the direction of y converges to the leading right singular vector v\textsubscript{1}, without explicitly computing U, $\Sigma$, V first. By the Eckart--Young theorem, the rank-1 matrix $\sigma$\textsubscript{1} u\textsubscript{1} v\textsubscript{1}ᵀ is the best rank-1 approximation of Q in the Frobenius-norm sense; power iteration is extracting exactly this leading component.

If no rounding is applied and many rounds are run, x \ensuremath{\rightarrow} u\textsubscript{1} and y \ensuremath{\rightarrow} v\textsubscript{1} are still continuous vectors; they still need to be discretized into a 0/1 P(t).

\paragraph{(2) Inserting Discrete Rounding Into the Iteration}

Spectral-scheduling chips break power iteration into steps with discrete rounding inserted in between. After initializing \texttt{y}, repeat the following four steps each round:

\bigskip\noindent
\par\leftskip=2em\relax\noindent \fittab{\begin{tabular}{p{\dimexpr 0.250\linewidth-2\tabcolsep}p{\dimexpr 0.250\linewidth-2\tabcolsep}p{\dimexpr 0.250\linewidth-2\tabcolsep}p{\dimexpr 0.250\linewidth-2\tabcolsep}}
\toprule
Step & Formula & Algebraic essence & Hardware meaning \\
\midrule
1 & \texttt{x = Q · y} & Half-step of power iteration; x converging toward u\textsubscript{1} & Each input accumulates output-side ``pressure'' y with weighting \ensuremath{\rightarrow} input-side congestion measure x \\
2 & \texttt{x \ensuremath{\leftarrow} Round(x)} & Spectral-smoothing projection & Compress continuous x to 0/1 or a sparse direction (take max, Grant, etc.) \\
3 & \texttt{y = Qᵀ · x} & Other half-step; y converging toward v\textsubscript{1} & Output side aggregates in reverse based on discretized x \ensuremath{\rightarrow} response for next round \\
4 & \texttt{y \ensuremath{\leftarrow} Round(y)} & Optional symmetric rounding & Project y onto the discrete feasible region as well \\
\bottomrule
\end{tabular}}\par\leftskip=2em\relax{}

\bigskip

\paragraph{(3) Geometric Picture of Rounding: a Trade-off Between the Continuous Optimum and Hardware Mutual-Exclusion Constraints}

\bigskip\noindent
\par\leftskip=2em\relax\noindent \fittab{\begin{tabular}{p{\dimexpr 0.500\linewidth-2\tabcolsep}p{\dimexpr 0.500\linewidth-2\tabcolsep}}
\toprule
Scenario & Result \\
\midrule
No rounding, run 10 rounds & x $\approx$ u\textsubscript{1}, y $\approx$ v\textsubscript{1} (purely mathematical; not a valid schedule) \\
Round in the middle of each iteration & x and y satisfy the discrete row/column exclusion form while converging toward u\textsubscript{1} and v\textsubscript{1} \\
When stopped & The result carries SVD leading-component information and satisfies 0/1 mutual exclusion \\
\bottomrule
\end{tabular}}\par\leftskip=2em\relax{}

\bigskip

\begin{itemize}
\item Spectral component: power iteration extracts the global leading component from Q (same as W = Q⊙(u\textsubscript{1}v\textsubscript{1}ᵀ));
\item Discrete feasibility constraint: each Rounding step enforces row/column mutual exclusion.
\end{itemize}

\subsubsection{Spectral Scheduling Step 2: Dual Masking (Constructing the Weight Matrix W)}

This section, after implicitly extracting the spectral direction within the §4.2.3 loop, presents \textbf{Path A}: explicitly constructing the weight matrix W.

This method does not select extreme values from u\textsubscript{1} and v\textsubscript{1} directly, but uses them to reweight the original matrix Q:

\par{\leftskip=0pt\relax\noindent\hspace*{2em}\begin{minipage}{\dimexpr\linewidth-2em\relax}
\begin{verbatim}
W_ij = Q_ij · u_{1,i} · v_{1,j}     (Hadamard product: W = Q ⊙ (u₁ v₁ᵀ))
\end{verbatim}
\end{minipage}\par}

\begin{itemize}
\item u\_\{1,i\}·v\_\{1,j\}: the global association score for (i,j) from the spectral structure;
\item Q\_ij: physical mask; if Q\_ij = 0, then W\_ij = 0, masking out empty queues.
\end{itemize}

\textbf{Example A:} when Q\_12=0, even a large spectral term has no effect. First obtain the spectral vectors from power iteration (u\textsubscript{1} on the input side, v\textsubscript{1} on the output side):
\begin{equation}
Q = \begin{bmatrix} 10 & 0 \\ 3 & 5 \end{bmatrix},\qquad
u_1 \approx [0.93,\ 0.36]^{\mathsf{T}},\quad
v_1 \approx [0.99,\ 0.17]^{\mathsf{T}}
\end{equation}

Then $u_{1,1}v_{1,2}\approx 0.93\times 0.17\approx 0.16$ (non-zero), but it is vetoed by $Q_{12}=0$:
\begin{align}
W_{12} &= Q_{12} \times (u_{1,1} v_{1,2}) = 0 \times 0.16 = 0 &&\ensuremath{\leftarrow} \text{vetoed} \nonumber \\
W_{11} &= 10 \times (u_{1,1} v_{1,1}) \approx 10\times 0.93\times 0.99 \approx 9.2 &&\ensuremath{\leftarrow} \text{backlog gives it a large weight}
\end{align}

\textbf{Example B:} \texttt{Q = [[8,2],[3,5]]}; from power iteration $u_1\approx[0.85,\ 0.53]^{\mathsf{T}}$, $v_1\approx[0.89,\ 0.46]^{\mathsf{T}}$; compute $W=Q\odot(u_1 v_1^{\mathsf{T}})$:
\begin{equation}
W \approx \begin{bmatrix} 6.05 & 0.78 \\ 1.40 & 1.21 \end{bmatrix}
\end{equation}

Constructing W requires O(N\textsuperscript{2}) multiplications (parallelizable in hardware).

\subsubsection{Spectral Scheduling Step 3: From Continuous Weights to a 0/1 Matching}

The previous step combined Q and the spectral leading component into a weighted matrix W; the final step selects a 0/1 P(t) within the matching set. This is the same as §3.2, except the weight matrix is W rather than Q:
\begin{equation}
P(t) = \operatorname*{argmax}_{P \in \text{matching set}}\ \langle W, P\rangle
\end{equation}

Continuing Example B, comparing the weight sums for the two valid matchings:
\begin{align}
(1,1)+(2,2)\ \text{weight sum} &\approx 6.05+1.21 = 7.26 \nonumber \\
(1,2)+(2,1)\ \text{weight sum} &\approx 0.78+1.40 = 2.18 \nonumber \\
&\ensuremath{\Rightarrow}\ \text{still selects straight-through } P^* = I\text{, consistent with MWM on } Q
\end{align}

The following two approaches are commonly used.

\begin{itemize}
\item \textbf{Row/column masking discretization}: project W (or Q) onto a valid matching; the spectral scheduling experiments in §7 use this as the final step. For large N, on-chip implementation requires global compare-and-feedback, making timing convergence difficult (Appendix B.4).
\item \textbf{Use W as initial pointer for iSLIP}: after W provides the row/column congestion information, proceed with the R--G--A and pointer rules of §5.2--§5.3; row/column mutual exclusion is ensured by the iSLIP process.
\end{itemize}

\clearpage
\subsection{Graph-Theoretic Exact Solution}

\subsubsection{Graph-Theoretic Formulation}

The MWM optimization problem $\max_{P\in\mathcal{M}}\langle Q,P\rangle$ from §3.2 is equivalent to the maximum weight bipartite matching problem: left nodes are input ports, right nodes are output ports, edge (i,j) has weight Q\_ij; the feasible matching set $\mathcal{M}$ requires at most one edge per row and per column, and the objective is to maximize the total matching weight. Software and the MWM simulation baseline in §7 typically use polynomial-time Hungarian or auction algorithms [10],[11] for the exact solution. Steps for both algorithms follow, with complexity in Appendix B.4.

The common example below uses \texttt{Q = [[8,7,2],[6,3,5],[2,6,4]]}; its optimal matching is \texttt{(1,1)+(2,3)+(3,2) = 8+5+6 = 19}.

\textbf{Hungarian Algorithm (Kuhn--Munkres)}

\begin{itemize}
\item Maintain a ``potential'' for each row and column: row potential $\alpha$\_i, column potential $\beta$\_j (the dual variables in linear programming, also called ``labels'' in matching theory). These assign a value to each row and column; the algorithm maintains $\alpha$\_i + $\beta$\_j $\geq$ Q\_ij (the sum of row and column labels covers the weight of every edge) and converges to the optimal matching by gradually adjusting these potentials.
\item Initialization rule: $\alpha_i=\max\limits_j Q_{ij}$ (max weight in that row), $\beta_j\equiv 0$.
\item Define the reduced weight: $c_{ij}=\alpha_i+\beta_j -Q_{ij}\ge 0$; look for a matching only on ``zero edges'' where $c_{ij}=0$.
\item If zero edges cannot form a perfect matching, update the potentials along an augmenting tree to introduce new zero edges: subtract $\delta$ from $\alpha$ of rows in the tree, add $\delta$ to $\beta$ of columns in the tree, where $\delta$ is the minimum reduced weight from tree-rows to non-tree-columns.
\item During augmentation, previously selected edges can be removed and better edges selected, so the algorithm can correct earlier suboptimal choices.
\end{itemize}

\textbf{Example:}

\begin{enumerate}
\item Initialize potentials and compute initial reduced weights. Row potentials take the row-maximum; column potentials are 0: $\alpha=[\max(8,7,2),\,\max(6,3,5),\,\max(2,6,4)]=[8,6,6]$, $\beta=[0,0,0]$. Compute $c_{ij}=\alpha_i+\beta_j -Q_{ij}$:
\begin{align}
&c_{11}=0,\quad c_{12}=1,\quad c_{13}=6 \nonumber \\
&c_{21}=0,\quad c_{22}=3,\quad c_{23}=1 \nonumber \\
&c_{31}=4,\quad c_{32}=0,\quad c_{33}=2
\end{align}
Initial zero edges: $(1,1)$, $(2,1)$, $(3,2)$.

\item Matching conflict detected. Among the zero edges, input 1 and input 2 both connect only to output 1 (column 1); they compete for the same column. Matching $\{(1,1),(3,2)\}$ leaves input 2 unmatched; no perfect matching exists, so the potentials must be updated to introduce new zero edges.
\item Find minimum slack $\delta$ and update potentials. Starting from the unmatched input 2, follow zero edge $(2,1)$ to the occupied output 1, then follow its owner to input 1; the ``explored'' rows are \{input 1, input 2\} and the ``explored'' columns are \{output 1\}; the unexplored columns are \{output 2, output 3\}. $\delta$ is the minimum reduced weight from explored rows (inputs 1, 2) to unexplored columns (outputs 2, 3): $\delta=\min\{c_{12},c_{13},c_{22},c_{23}\}=1$. Update rule: subtract $\delta$ from explored-row $\alpha$ values; add $\delta$ to explored-column $\beta$ values: $\alpha$\_1 and $\alpha$\_2 each decrease by 1, $\beta$\_1 increases by 1, giving $\alpha=[7,5,6]$, $\beta=[1,0,0]$.
\item Recompute reduced weights and generate new zero edges. Using the updated potentials:
\begin{align}
&c_{11}=0,\quad c_{12}=0,\quad c_{13}=5 \nonumber \\
&c_{21}=0,\quad c_{22}=2,\quad c_{23}=0 \nonumber \\
&c_{31}=5,\quad c_{32}=0,\quad c_{33}=2
\end{align}
New zero edges: $(1,2)$ and $(2,3)$.

\item Find the optimal matching via an augmenting path. Match the previously unmatched input 2 to output 3 (a new zero edge); the zero edges now have no row/column conflicts. Optimal assignment: $(1,1)+(2,3)+(3,2)=8+5+6=19$.
\end{enumerate}

\textbf{Auction Algorithm (Bertsekas)}

\begin{itemize}
\item Model: input i is a buyer, output j is a good, valuation $v_{ij}=Q_{ij}$; $\pi_j$ is the current price of output j, initialized to 0.
\item Unassigned buyers bid in turn on the good with the highest net benefit $v_{ij} -\pi_j$; the bid increment equals the best net benefit minus the second-best (plus a small $\epsilon$). The price increments computed by one buyer are immediately available to the next buyer.
\item The highest bidder wins the good; outbid buyers enter the next round to bid again. Prices increase monotonically until all buyers are assigned.
\end{itemize}

\textbf{Example} (taking $\epsilon$ \ensuremath{\rightarrow} 0, ignoring small offsets):

\begin{enumerate}
\item Initial prices $\pi=[0,0,0]$, all three inputs unmatched.
\item Buyer 1's net benefits are $[8,7,2]$; best is good 1, second-best is good 2, so price of good 1 increases by $8 -7=1$; all prices become $\pi=[1,0,0]$; buyer 1 \ensuremath{\rightarrow} good 1.
\item Buyer 2 recomputes net benefits using buyer 1's updated price increment: $[6 -1,\,3,\,5]=[5,3,5]$; goods 1 and 3 are tied for best, pick good 3 (tie means increment $\approx$ $\epsilon$); buyer 2 \ensuremath{\rightarrow} good 3.
\item Buyer 3 computes net benefits $[2 -1,\,6,\,4]=[1,6,4]$; best is good 2, second is good 3, so price of good 2 increases by $6 -4=2$; buyer 3 \ensuremath{\rightarrow} good 2.
\item All buyers assigned; final matching $(1,1)+(2,3)+(3,2)=19$, consistent with the Hungarian result.
\end{enumerate}

\textbf{Row/Column Masking Greedy}

\begin{itemize}
\item Each iteration finds the global maximum over the entire matrix, selects it, then masks the entire row and column of that entry to $-\infty$; repeat N times to obtain a matching. This is the same greedy framework as §4.2.5 Path A and the §7 spectral scheduling discretization.
\item The masking is irreversible: there is no Hungarian augmentation or auction re-bidding, so the result may be suboptimal for $N\ge 3$.
\end{itemize}

\textbf{Example:} Using the same Q as above, the greedy algorithm happens to achieve the optimal value of 19 in this instance. But if the weight matrix is instead $W'=\begin{bmatrix} 8 & 7 & 7 \\ 8 & 4 & 0 \\ 8 & 0 & 1 \end{bmatrix}$ (global optimum still 19), and the first greedy step selects the global maximum $(2,1)=8$, the remainder can only achieve:
\begin{equation}
P_g' = (2,1)+(1,2)+(3,3) = 8+7+1 = 16 < 19
\end{equation}

Comparison of the ability to correct earlier suboptimal choices:

\bigskip\noindent
\par\leftskip=2em\relax\noindent \fittab{\begin{tabular}{p{\dimexpr 0.500\linewidth-2\tabcolsep}p{\dimexpr 0.500\linewidth-2\tabcolsep}}
\toprule
After incorrectly locking (2,1)=8 in step 1 & Correcting an early mistake \\
\midrule
Row/column masking greedy & Cannot reassign, fixed at 16 \\
Auction & Can reassign \\
Hungarian & Augmenting path can remove existing matched edges \\
\bottomrule
\end{tabular}}\par\leftskip=2em\relax{}

\bigskip

\subsubsection{Connection to Continuous Relaxation}

§4.1--§4.2 first relax Q into a doubly stochastic matrix or a spectrally weighted matrix W, and then must project back to a 0/1 P(t). If this projection uses row/column masking discretization (§4.2.5 Path A, §7 spectral scheduling), it belongs to the same category of approximation as ``row/column masking greedy'' above: for N $\geq$ 3 the result may be suboptimal and the masking steps are not reversible. On the weight lower bound, two greedy variants must be distinguished: a global greedy directly on the Q weights (Appendix B.3; repeatedly take the largest Q\_ij, then mask its row and column) yields a maximal matching $\geq$ 1/2 of MWM(Q), the classic 1/2-approximation result [44] (§6.5 Table 3 ``$\geq$50\%''). Spectral scheduling, by contrast, runs greedy on W = Q⊙(u\textsubscript{1}v\textsubscript{1}ᵀ), so its 1/2 bound is relative to MWM(W), not MWM(Q), and therefore gives no guarantee on ⟨Q,P⟩. Neither greedy guarantees the optimal solution.

\subsubsection{Transition to Industrial Implementation}

At this point, MWM has played two roles: on one hand, it provides the exact reference in the ⟨Q,P⟩ sense; on the other hand, it also exposes the main hardware tension: every cycle a globally coupled matching must be computed over the full matrix of weights. Datacom switches and on-chip crossbar schedulers generally cannot complete the steps of ``read the full weight matrix + perform multiple rounds of global iteration'' within a single slot/cycle: such a complex operation takes many pipeline stages for timing closure, and when N is large it also requires wide weight buses and centralized state control, which is inconsistent with the premise of single-cycle grant implementation at the ns level.

Therefore, industrial implementations typically do not pursue single-cycle exact MWM, but instead reformulate the problem as a local arbitration process with easier timing closure: read only 1-bit queue-nonempty status, use parallel Request--Grant--Accept rounds to satisfy row/column mutual exclusion, and rely on cross-cycle pointer state to approximate the long-run average service rate. The iSLIP algorithm in §5 is the representative of this engineering approach.

\section{iSLIP: Industrial Hardware Approximation}

\subsection{Why Not MWM?}

\bigskip\noindent
\par\leftskip=2em\relax\noindent \fittab{\begin{tabular}{p{\dimexpr 0.333\linewidth-2\tabcolsep}p{\dimexpr 0.333\linewidth-2\tabcolsep}p{\dimexpr 0.333\linewidth-2\tabcolsep}}
\toprule
Constraint & Datacom switch chip & On-chip crossbar arbiter (isomorphic) \\
\midrule
Clock cycle & {\textasciitilde}2--4 ns & System clock on the same order \\
Scale & 64$\times$64 etc. (line-card fabric) & N$\approx$4--8 common; N=8 already close to limit for centralized crossbar \\
MWM & O(N\textsuperscript{3}) exact matching; cannot complete in a single cycle (§4.3, Appendix B.4) & Same: full MWM per cycle is impractical \\
Row/column masking greedy & N compare-feedback rounds; prone to timing violations & Long combinational logic chains similarly constrained \\
Iteration count & r rounds of iSLIP per slot (datacom, typically 3--4) & r rounds of R--G--A grant logic per cycle \\
\bottomrule
\end{tabular}}\par\leftskip=2em\relax{}

\bigskip

MWM is the mathematical optimum, but it is hard to meet single-cycle timing requirements. iSLIP [3] uses parallel 1-bit arbitration plus a small number of R--G--A rounds per cycle to approximate MWM in a statistical sense, and this is the mainstream RTL approach for crossbar implementation. McKeown and Anderson's comparative study [16] shows that iSLIP can achieve 100\% throughput under uniform (i.i.d.) traffic. Industrial implementations typically use only 3--4 rounds per cycle.

\subsection{Hardware Structure: iSLIP and Two-Level Round-Robin Arbitration}

iSLIP [3] uses parallel row/column Request--Grant--Accept: each row and each column runs a one-dimensional arbiter in parallel; Grant/Accept conflicts are resolved by Round-Robin pointers and priority encoders.

Each input port has one input arbiter and each output port has one output arbiter, each holding a priority pointer.

\textbf{Three-step protocol (Request \ensuremath{\rightarrow} Grant \ensuremath{\rightarrow} Accept):}

\par{\leftskip=0pt\relax\noindent\hspace*{2em}\begin{minipage}{\dimexpr\linewidth-2em\relax}
\begin{verbatim}
Step 1 Request: VOQs with packets send a 1 -bit request to the target output

Step 2 Grant:   The output arbiter scans clockwise from its pointer position
                and sends a Grant to the first input with a pending request

Step 3 Accept:  The input arbiter scans clockwise from its pointer position
                and accepts the first Grant it receives
\end{verbatim}
\end{minipage}\par}

\textbf{Round-Robin scan rule:} the pointer indicates the priority starting position, not a forced endpoint. The hardware uses a priority encoder to skip ports with no request until it finds the first port with a pending request.

\subsection{Pointer Update Rule (the Core of iSLIP)}

A given arbiter's pointer advances by one position only when its Grant is actually accepted in the Accept phase; otherwise the pointer does not move.

This rule produces:

\begin{itemize}
\item \textbf{When idle:} the pointer rotates quickly, approximating fair Round-Robin;
\item \textbf{When congested:} after a successful Accept, the pointers are staggered (desynchronized); in the next cycle, Grant/Accept scanning starts from the new pointer position, giving other input/output pairs priority. Pairs (i,j) that still have backlog continue to send Requests, so they will be matched in subsequent cycles rather than permanently monopolizing a port.
\end{itemize}

\subsection{Multiple Rounds of Iteration Within the Same Cycle}

The ``multiple rounds'' in the iSLIP paper refers to multiple Request--Grant--Accept rounds within the same clock cycle, with the pointers not advancing:

\bigskip\noindent
\par\leftskip=2em\relax\noindent \fittab{\begin{tabular}{p{\dimexpr 0.500\linewidth-2\tabcolsep}p{\dimexpr 0.500\linewidth-2\tabcolsep}}
\toprule
Round & Role \\
\midrule
Round 1 & High-backlog (i,j) pairs get priority; corresponding edges are locked \\
Round 2 & Remaining inputs/outputs continue matching \\
Round 3 & Continue filling until the number of matches reaches the physical limit \\
\bottomrule
\end{tabular}}\par\leftskip=2em\relax{}

\bigskip

After all matching is complete and packets are actually forwarded, the pointers are updated together.

\textbf{Pointer update refined rules:}

\begin{itemize}
\item Only output-side arbiters where the Round 1 Accept was successful update their pointer at the end of the cycle;
\item Only input-side arbiters where the Round 1 Accept was successful update their pointer at the end of the cycle;
\item Ports where ``leftover'' matching succeeded in rounds 2 or 3 do not move their pointer.
\end{itemize}

Therefore, multiple rounds of iteration perform repeated Request--Grant--Accept for unmatched ports under the same set of pointers, to increase the number of matched edges (trace) in the current cycle. A single round of Grant\ensuremath{\rightarrow}Accept often produces only a proper subset of the legal matching set M.

\subsection{Worked Example: Two-Round Matching in a Single Cycle and Cross-Cycle P̄}

The two examples below share a 2$\times$2 topology.

\textbf{Example A:}

Request mask R (a Request is sent if there is backlog; here all four VOQs are non-empty):
\begin{equation}
R = \begin{bmatrix} 1 & 1 \\ 1 & 1 \end{bmatrix}
\end{equation}

Initial pointer values: g\textsubscript{1} and g\textsubscript{2} both point to in1; a\textsubscript{1} points to out1.

Within a single cycle, Grant can temporarily violate the row constraint: each column operates independently, ensuring only that the column sum $\leq$ 1. Output 1 scans from in1 and finds a request \ensuremath{\rightarrow} Grants (1,1); output 2 scans from in1 and finds a request \ensuremath{\rightarrow} Grants (1,2). The Grant matrix G is:
\begin{equation}
G = \begin{bmatrix} 1 & 1 \\ 0 & 0 \end{bmatrix}
\qquad \ensuremath{\leftarrow} \text{row 1 sum} = 2;\ \text{not yet in the valid matching set } \mathcal{M}
\end{equation}

Accept (each row selects at most one from the Grants it received): input 1 receives Grants from both out1 and out2; starting from a\textsubscript{1} pointing to out1, it accepts out1 \ensuremath{\rightarrow} first match this cycle is (1,1), giving P\textsuperscript{(}\textsuperscript{1}\textsuperscript{)}:
\begin{equation}
P^{(1)} = \begin{bmatrix} 1 & 0 \\ 0 & 0 \end{bmatrix}
\end{equation}

\textbf{Round 2 (pointers do not move this cycle):} already-matched input 1 and output 1 are excluded from further iterations; remaining input 2 / output 2 go through Grant\ensuremath{\rightarrow}Accept again to obtain (2,2); merged result:
\begin{equation}
P(t) = \begin{bmatrix} 1 & 0 \\ 0 & 1 \end{bmatrix}
\qquad \mathrm{trace} = 2
\end{equation}

\textbf{Example B: Computing P̄ from P(t) over 10 Cycles}

Setup: Q\_11 is very long (backlog continuously non-zero), Q\_12=0, Q\_21 and Q\_22 are shorter. The table records only whether input 1 is connected (i.e., P\_11 and P\_12, from the multi-round merged P(t)):

\bigskip\noindent
\par\leftskip=2em\relax\noindent \fittab{\begin{tabular}{p{\dimexpr 0.250\linewidth-2\tabcolsep}p{\dimexpr 0.250\linewidth-2\tabcolsep}p{\dimexpr 0.250\linewidth-2\tabcolsep}p{\dimexpr 0.250\linewidth-2\tabcolsep}}
\toprule
t & Matched edges & P\_11(t) & P\_12(t) \\
\midrule
1 & (1,1) & 1 & 0 \\
2 & (1,1) & 1 & 0 \\
3 & (2,2) & 0 & 0 \\
4 & (1,1) & 1 & 0 \\
5 & (1,1) & 1 & 0 \\
6 & (1,1) & 1 & 0 \\
7 & (2,1) & 0 & 0 \\
8 & (1,1) & 1 & 0 \\
9 & (1,1) & 1 & 0 \\
10 & (1,1) & 1 & 0 \\
\bottomrule
\end{tabular}}\par\leftskip=2em\relax{}

\bigskip
\begin{align}
\sum_t P_{11}(t) = 8 &\ \to\ \bar P_{11} = E[P_{11}] = 8/10 = 0.8 \nonumber \\
\sum_t P_{12}(t) = 0 &\ \to\ \bar P_{12} = 0
\end{align}

(Even with staggered pointers, (1,1) is still frequently served, so the high-backlog path has a high connection rate.)

If Q\_11 $\approx$ 1000 and Q\_12 = 0 over this period, then ⟨Q,P̄⟩ $\approx$ 1000$\times$0.8 = 800. Taking $\lambda$\_11 = 0.5, then Qᵀ$\lambda$ = 500 \textless  800, which is the numerical version of E[QᵀP] \textgreater  Qᵀ$\lambda$ from §3.4.6.

\subsection{Mathematical Relationship Between iSLIP and Q(t)}

§5.2--§5.4 described iSLIP's hardware rules (Request--Grant--Accept with round-robin pointers). This section connects those rules to the MWM math of §3. The bridge is a \textbf{feedback loop that runs across cycles}, not a formula within one cycle. Read it as five steps; the subsections that follow give the formal math for steps 1, 3, and 5.

\begin{enumerate}
\item \textbf{Sense (1-bit).} Each cycle the queue state Q produces a request mask R\_ij = 1[Q\_ij \textgreater  0]. iSLIP sees only \textit{whether} a queue is non-empty, not \textit{how long} it is, so within one cycle there is no formula P(t) = f(Q(t)) of the MWM form argmax⟨Q, P⟩.
\item \textbf{Decide (decoupled).} The output arbiter issues a Grant from its pointer position (enforcing the column constraint); the input arbiter issues an Accept from its pointer position (enforcing the row constraint). MWM's coupled ILP, in which row and column constraints act jointly on the same P, is split into two inexpensive independent projections. This decoupling is exactly why iSLIP fits in one cycle.
\item \textbf{Update (the feedback).} A pointer advances \textbf{only when its Grant is accepted}. A pair (i,j) just served yields priority next cycle; a pair that is backlogged but lost this cycle keeps issuing a Request and, as the pointers rotate past it, eventually wins. This is the dynamic feedback that separates iSLIP from weight-unaware random matching.
\item \textbf{Average (emerges across cycles).} Because of step 3, a direction (i,j) with persistent backlog is connected for a larger \textit{fraction} of cycles over time. The mathematical name for that fraction is E[P\_ij] = P̄\_ij. Unlike the weight-unaware random matching of §3.4.6, where E[P\_11] is frozen at 0.5 regardless of Q\_11, here P̄\_ij rises with backlog.
\item \textbf{Close (stability).} Substituting this backlog-responsive P̄ into the drift condition of §3.4: when $\Lambda$ is admissible, ⟨Q, P̄⟩ \textgreater  Qᵀ $\lambda$ holds, so the queues stay bounded.
\end{enumerate}

\subsubsection{Step 1. Sense: No Single-Cycle Algebraic Mapping}
\begin{equation}
P_{\text{MWM}}(t) = \operatorname*{argmax}_{P}\ \langle Q(t), P\rangle
\end{equation}

iSLIP's P(t) depends only on:

\begin{itemize}
\item Arbiter pointer positions g(t), a(t)
\item Request mask is 1 only based on the condition [Q\_ij \textgreater  0] (1 if there is backlog, 0 if not)
\end{itemize}

iSLIP is not aware of the absolute value of queue lengths, only whether there are packets. Therefore, within a single clock cycle t, there is no algebraic formula of the form \texttt{P(t) = f(Q(t))}.

\subsubsection{Step 2. Decide: Decoupling the Coupled ILP (Grant\ensuremath{\rightarrow}Accept)}

MWM solves a coupled ILP over the valid matching domain (row and column constraints act simultaneously on the same P; one cannot argmax row by row and then concatenate):
\begin{align}
\max\ & \langle Q, P\rangle = \sum_{i,j} Q_{ij}\, P_{ij} \nonumber \\
\text{s.t.}\ & \sum_j P_{ij} \le 1, \quad \forall i \quad (\text{row constraint}) \nonumber \\
& \sum_i P_{ij} \le 1, \quad \forall j \quad (\text{column constraint}) \nonumber \\
& P_{ij} \in \{0, 1\}
\end{align}

The Hungarian and auction algorithms find the global optimum on this coupled ILP (O(N\textsuperscript{3})). Defining the column/row feasible sets $C_{\text{col}}$, $C_{\text{row}}$, and $\mathcal{M} = C_{\text{col}} \cap C_{\text{row}}$, and letting $\Pi_C(\cdot)$ denote a greedy projection with an RR pointer (not a Euclidean projection), iSLIP runs a single round on the request mask $R_{ij}=\mathbf{1}[Q_{ij}>0]$:
\begin{equation}
P^{(1)} = \Pi_{C_{\text{row}}}\!\big(\Pi_{C_{\text{col}}}(R)\big),
\end{equation}

where Grant corresponds to $\Pi_{C_{\text{col}}}(R)$ (column constraint, G row-sums may be \textgreater  1) and Accept corresponds to $\Pi_{C_{\text{row}}}(G)$ (row constraint, $P^{(1)}\in\mathcal{M}$).

\subsubsection{Step 3. Update: The Pointer Feedback}

Steps 1 and 2 happen inside one cycle and produce a single P(t). What turns iSLIP from a one-shot maximal matcher into something that tracks MWM in the long run is how its state changes \textbf{between} cycles.

Recall the pointer rule of §5.3: an arbiter's pointer advances by one position \textbf{only when its Grant is accepted}; on a cycle with no successful Accept the pointer does not move. Two consequences follow:

\begin{itemize}
\item \textbf{Yield after service.} The pair (i,j) just served has, on both its input and output arbiter, the pointer pointing one step past (i,j). Next cycle the scan starts from there, so (i,j) is no longer first in line and yields to other pending pairs. A port cannot monopolize the crossbar.
\item \textbf{Retry until served.} A pair (i,j) that is backlogged but lost this cycle (its Grant was not accepted, or it was masked out) keeps issuing a Request every cycle, because R\_ij = 1 as long as Q\_ij \textgreater  0. As the pointers of its input and output arbiters rotate, (i,j) eventually becomes the first pending pair in both scans and wins. The longer it stays backlogged, the more cycles it requests, and the more certain it is to be served.
\end{itemize}

This is the dynamic feedback: \textbf{the queue state Q drives the request mask R; R together with the pointer positions decides P(t); a successful Accept moves the pointers; the new pointer positions shape the next cycle's P(t+1).} Over many cycles, service is steered toward the directions that remain backlogged, without iSLIP ever reading the magnitude of Q\_ij. That is the qualitative reason the next step's P̄\_ij can grow with backlog, which is precisely what the weight-unaware random matching cannot do.

\subsubsection{Step 4. Average: Expected Value and Time Average}

The stability condition in §3.4.6 contains \texttt{E[QᵀP]}. Here we carefully align the concepts of ``expected value'' and ``time average.''

\textbf{Random variable perspective}

Each cycle t, the scheduling result \texttt{P\_ij(t)} is a 0/1 random variable:
\begin{equation}
P_{ij}(t) = \begin{cases}
1 & \text{if input } i \text{ is successfully matched with output } j \text{ in this cycle}\\
0 & \text{otherwise}
\end{cases}
\end{equation}

The expected value is defined as:
\begin{equation}
E[P_{ij}] = \lim_{T\to\infty} \frac{1}{T} \sum_{t=1}^{T} P_{ij}(t)
\end{equation}

i.e., the long-run fraction of time channel (i,j) is connected.

\textbf{Relationship with the time-average matrix P̄}

Define the time-average service matrix (over T cycles):
\begin{equation}
\bar P = \frac{1}{T} \sum_{t=1}^{T} P(t)
\end{equation}

From the statistical perspective, the two are equivalent: \texttt{E[P\_ij] = P̄\_ij}. Below, \texttt{E[P\_ij]} and \texttt{P̄\_ij} are used interchangeably with the same meaning.

\textbf{From E[P\_ij] to E[QᵀP]}

The inner product is a sum over all (i,j); expectations can be taken term by term:
\begin{align}
E[Q^{\mathsf{T}}P] &= E\!\left[ \sum_{i,j} Q_{ij}\, P_{ij} \right] \nonumber \\
&= \sum_{i,j} Q_{ij}\, E[P_{ij}] \quad (\text{$Q_{ij}$ treated as a known constant}) \nonumber \\
&= \sum_{i,j} Q_{ij}\, \bar P_{ij} \nonumber \\
&= \langle Q, \bar P\rangle
\end{align}

Physical meaning: the longer the queue in a given direction, the larger the energy-removal inner product if that direction also has a higher long-run average service rate P̄\_ij.

\textbf{Why iSLIP can satisfy the drift where random matching cannot}

In the weight-unaware random matching, the expected matching probability E[P\_11] is fixed at 0.5 and does not increase with the backlog Q\_11, hence E[QᵀP] \textless  Qᵀ$\lambda$. In contrast, the pointer feedback of Step 3 makes P̄\_ij responsive to backlog: pairs (i,j) with sustained backlog obtain a higher P̄\_ij over time. In the long run this has the potential to satisfy E[QᵀP] \textgreater  Qᵀ$\lambda$, exceeding the performance ceiling of a fixed 50/50 I/X uniform allocation.

\subsubsection{Step 5. Close: The Full Logic Chain}
\begin{align}
&\text{Single cycle:}\ P(t) \in M,\ P = \Pi_{C_{\text{row}}}\!\big(\Pi_{C_{\text{col}}}(\mathbf{1}[Q>0])\big);\text{ does not read } Q_{ij} \text{ values} \nonumber \\
&\text{Within cycle:}\ \text{multiple rounds of Grant–Accept; pointers do not move} \nonumber \\
&\text{Across cycles:}\ \text{pointer advances after successful Accept} \to E[P_{ij}]=\bar P_{ij};\text{ larger backlog} \ensuremath{\Rightarrow} \text{higher } \bar P_{ij} \nonumber \\
&\text{Stability:}\ \Lambda\ \text{admissible, and}\ \langle Q,\bar P\rangle > Q^{\mathsf{T}}\lambda \to \text{queues bounded}
\end{align}

\subsection{MWM vs. iSLIP Comparison}

This section gives a point-by-point comparison of MWM and iSLIP.

\bigskip\noindent
\par\leftskip=2em\relax\noindent \fittab{\begin{tabular}{p{\dimexpr 0.333\linewidth-2\tabcolsep}p{\dimexpr 0.333\linewidth-2\tabcolsep}p{\dimexpr 0.333\linewidth-2\tabcolsep}}
\toprule
Feature & MWM & iSLIP \\
\midrule
Per-cycle computation & Hungarian / auction, O(N\textsuperscript{3}) (exact combinatorial) & Distributed logic gates, O(1) hardware delay \\
Weight awareness & Reads Q\_ij values exactly & Only senses Q\_ij \textgreater  0 \\
Relationship with Q & Every step is strictly argmax⟨Q, P⟩ & No single-step mapping; long run E[P\_ij]=P̄\_ij, higher for high-backlog channels (§5.5) \\
ILP solving & Coupled ILP global optimum (Hungarian O(N\textsuperscript{3})) & Coordination decoupling: Grant\ensuremath{\rightarrow}Accept alternating projection \\
Queue stability / throughput & Classical theory gives 100\% throughput [4] & Multi-round iteration + pointer desynchronization statistically approximates 100\% throughput [3] \\
Stability criterion (drift) & \texttt{E[QᵀP] \textgreater  Qᵀ$\lambda$} (when ‖Q‖ is large) & Same requirement \texttt{⟨Q, P̄⟩ \textgreater  Qᵀ$\lambda$}, but P̄ is the time-average matrix \\
Industrial deployment & Almost never & Widely adopted \\
\bottomrule
\end{tabular}}\par\leftskip=2em\relax{}

\bigskip

\begin{quote}
\textbf{Note (exact vs. approximate):} Exact MWM per-cycle implementation is a combinatorial algorithm (Hungarian/auction). Sinkhorn+BvN, spectral scheduling (power iteration), etc. are continuous-relaxation/approximation methods for MWM (see §4.2, §4.1, §6.4 Table 2); they are not equivalent to exact MWM.
\end{quote}

\subsubsection{iSLIP Family Variants (Summary Only)}

After McKeown's iSLIP [3], DRRM (dual RR pointers [17], performance analysis in [18]), FIRM (FCFS fairness [19]), SRR / DRDSRR (static RR, low-delay MSM [20],[21]) and others continue to improve on pointers/fairness within the 1-bit R--G--A skeleton. See Table 1 and references [17]--[21].

\section{Algorithm Taxonomy}

This section places all the scheduling algorithms that have appeared in the paper on a single map: first classifying them by information granularity, matching objective, iteration count, and architectural premise; then providing a classification tree, conceptual diagram, and cross-comparison table. This allows the reader, before entering the experiments, to understand where MWM, iSLIP, spectral scheduling, BvN/OT, and other switch architecture extensions each fit.

\subsection{Classification Dimensions}

\bigskip\noindent
\par\leftskip=2em\relax\noindent \fittab{\begin{tabular}{p{\dimexpr 0.333\linewidth-2\tabcolsep}p{\dimexpr 0.333\linewidth-2\tabcolsep}p{\dimexpr 0.333\linewidth-2\tabcolsep}}
\toprule
Dimension & Meaning & Typical values \\
\midrule
Weight awareness & Whether Q\_ij values are read within a single cycle & Full-matrix comparison (MWM); 1-bit Request (iSLIP); spectrally weighted W = Q⊙(u\textsubscript{1}v\textsubscript{1}ᵀ) (§4.2) \\
Matching objective & What is optimized each cycle & Maximum weight matching (MWM); maximal matching (iSLIP); doubly stochastic convex combination (BvN time-division) \\
Iteration count & R--G--A or equivalent rounds per cycle & Typically 3--4 (iSLIP); O(log N) rounds (maximal matching, literature); O(N\textsuperscript{3}) offline (Hungarian) \\
Cross-cycle memory & Whether previous cycle state is retained & None (weight-unaware random matching); RR pointer (iSLIP); previous cycle matching M(t-1) (Tassiulas [22]); MERGE (APSARA [24]) \\
Arbitration topology & Centralized vs. distributed & Centralized matching (Hungarian, global Q); distributed R--G--A (iSLIP, per-port arbitration) \\
Stability definition & What does ``100\%'' mean & Queue bounded under admissible traffic (MWM [4]); iSLIP limited to uniform i.i.d. [3]; other easily confused metrics see §6.5 Table 3 \\
Architecture premise & Queue architecture & Pure IQ+VOQ; CIOQ + speedup [29]; CICQ crosspoint queues [37] \\
\bottomrule
\end{tabular}}\par\leftskip=2em\relax{}

\bigskip

\needspace{26\baselineskip}
\subsection{Classification Tree}

Input-queued / crossbar scheduling is divided into five categories by ``information granularity + matching objective'':

\bigskip\noindent
\par\leftskip=2em\relax\noindent \fittab{\begin{tabular}{p{\dimexpr 0.333\linewidth-2\tabcolsep}p{\dimexpr 0.333\linewidth-2\tabcolsep}p{\dimexpr 0.333\linewidth-2\tabcolsep}}
\toprule
Category & Representative algorithms & Key properties \\
\midrule
{[I]} Exact optimal · reads full weight Q\_ij & Graph-theoretic MWM: Hungarian [10], auction [11] & O(N\textsuperscript{3}), 100\% admissible, hard to implement per cycle \\
 & MWM weight family: LQF / OCF / LPF [4],[5] & 100\% admissible, different weight definitions \\
 & Distributed iterative MWM: $\epsilon$-Auction / $\epsilon$-min-sum [41] & Converges to MWM; O(n\textsuperscript{2}/$\epsilon$) rounds \\
{[II]} Maximal matching · 0/1 approximation (industrial mainstream) & RR + pointers: iSLIP [3], DRRM [17], RRM [3] §II & 100\% under uniform traffic; pointer rules vary \\
 & Fairness: FIRM [19] & FCFS approximation, MSM class \\
 & Static/deterministic RR: SRR [20], DRDSRR [21] & Low delay; SRR unstable under non-uniform traffic \\
{[III]} Randomized + memory · long-run average approximates MWM & Tassiulas [22] / compare-and-keep-better & O(N), 100\% admissible, high delay \\
 & APSARA / SERENA [23],[24],[25]; randomized evaluation [26] & Derandomized + parallel; fluid model [24] \\
{[IV]} Continuous relaxation \ensuremath{\rightarrow} valid discrete P & Sinkhorn [12] + BvN time-division [31],[32],[35] & Offline decomposition, online periodic permutation \\
 & Two-stage load-balanced BvN [33],[34] & No-center N\textsuperscript{2} matching, order-preserving [34] \\
 & Spectral scheduling: power iteration + (optional) W + discretization (§4.2) & Research / small N; §7 comparison with iSLIP \\
{[V]} Architecture extensions (problem formulation changes) & CIOQ + speedup 2/4 [27],[28],[29],[30] & Emulates OQ / 100\% maximal matching \\
 & CICQ / LIPS [36],[37],[38],[39] & Independent input+output scheduling; speedup=2 \\
 & PIFO [40] (programmable dataplane) & Connects to CIOQ theory \\
\bottomrule
\end{tabular}}\par\leftskip=2em\relax{}

\bigskip

\needspace{33\baselineskip}
\subsection{Diagram (Conceptual)}

\penalty10000\relax
\par\medskip{\leftskip=0pt\relax\noindent\hspace*{2em}%
\begin{minipage}{\dimexpr\linewidth-2em\relax}
\centering\fitwidth{%
\begin{tikzpicture}[font=\footnotesize,>=Stealth,
  voq/.style={draw,rounded corners,inner sep=3pt,minimum width=52mm,
              minimum height=6mm,anchor=west}]
  \node[voq] (vq0) at (0,0) {[VOQ$\to$0]\,[VOQ$\to$1]\,$\cdots$\,[VOQ$\to$N$-$1]};
  \node[voq] (vq1) at (0,-10mm) {[VOQ$\to$0]\,[VOQ$\to$1]\,$\cdots$\,[VOQ$\to$N$-$1]};
  \node (vd) at (26mm,-16mm) {$\vdots$};
  \node[voq] (vqn) at (0,-22mm) {[VOQ$\to$0]\,$\cdots$\,[VOQ$\to$N$-$1]};
  \node[anchor=east] (i0) at (-8mm,0) {Input 0};
  \node[anchor=east] (i1) at (-8mm,-10mm) {Input 1};
  \node[anchor=east] (iN) at (-8mm,-22mm) {Input N$-$1};
  \draw[->] (i0.east)--(vq0.west);
  \draw[->] (i1.east)--(vq1.west);
  \draw[->] (iN.east)--(vqn.west);
  \node[draw,fill=black!5,align=center,minimum width=16mm,minimum height=22mm]
        (cb) at (80mm,-11mm) {N$\times$N\\Crossbar};
  \coordinate (cbw0) at ([yshift=4mm]cb.west);
  \coordinate (cbw1) at (cb.west);
  \coordinate (cbwN) at ([yshift=-4mm]cb.west);
  \draw[->] (vq0.east)--(cbw0);
  \draw[->] (vq1.east)--(cbw1);
  \draw[->] (vqn.east)--(cbwN);
  \node[right=10mm of cb] (out) {Output $0\cdots$N$-$1};
  \draw[->] (cb.east)--(out.west);
  \node[draw,dashed,align=center,below=10mm of cb] (sch) {Scheduler selects matching $P(t)$ each cycle\\at most one 1 per row and column};
  \draw[->] (sch.north)--(cb.south);
\end{tikzpicture}}

\end{minipage}\par}\medskip

{\small\textbf{Figure 1.} N$\times$N input-queued + VOQ + crossbar.{\footnotesize  Each input port maintains N VOQs [4], eliminating HOL [1]; the scheduler outputs the valid matching matrix $P(t)\in\{0,1\}^{N\times N}$.}}

\penalty10000\relax\vspace{3\baselineskip}
\par\medskip{\leftskip=0pt\relax\noindent\hspace*{2em}%
\begin{minipage}{\dimexpr\linewidth-2em\relax}
\centering\fitwidth{%
\begin{tikzpicture}[font=\footnotesize,>=Stealth,
  box/.style={draw,rounded corners,align=center,inner sep=4pt}]
  \node[box,fill=black!5] (top) {Queue matrix $Q(t)$\quad·\quad Service matrix $P(t)$\quad·\quad Lyapunov $V=\lVert Q\rVert^2$};
  \node[box,below=14mm of top] (m2) {[Thread II]\\Graph-theoretic exact MWM\\§4.3\\Hungarian [10]};
  \node[box,left=10mm of m2] (m1) {[Thread I]\\Linear algebra / stability\\§2–§3, §4.1–§4.2\\MWM derivation [4],[7]};
  \node[box,right=10mm of m2] (m3) {[Thread III]\\iSLIP 0/1 iteration\\§5\\$\bar P$ with backlog\\(long-run avg) [3]};
  \draw[->] (top.south west) -- (m1.north);
  \draw[->] (top.south) -- (m2.north);
  \draw[->] (top.south east) -- (m3.north);
  \node[box,below=10mm of m2] (join) {§4 convergence · Sinkhorn / BvN / Spectral};
  \draw[->] (m1.south) -- (join.north west);
  \draw[->] (m2.south) -- (join.north);
  \draw[->] (m3.south) -- (join.north east);
\end{tikzpicture}}

\end{minipage}\par}\medskip

{\small\textbf{Figure 2.} Relationship among the three main threads of this paper.}

\needspace{24\baselineskip}
\subsection{Cross-Algorithm Comparison Table}

The table below compares the main-thread algorithms, experimental subjects, and several related scheduling algorithms, focusing on queue-read granularity, iteration budget, throughput/stability context, and hardware scale.

\bigskip\noindent
\par\leftskip=2em\relax\noindent \fittab{\begin{tabular}{p{\dimexpr 0.167\linewidth-2\tabcolsep}p{\dimexpr 0.167\linewidth-2\tabcolsep}p{\dimexpr 0.167\linewidth-2\tabcolsep}p{\dimexpr 0.167\linewidth-2\tabcolsep}p{\dimexpr 0.167\linewidth-2\tabcolsep}p{\dimexpr 0.167\linewidth-2\tabcolsep}}
\toprule
Algorithm & Reads weights & Per-cycle iterations & Throughput/stability & Hardware scale & Role \\
\midrule
MWM (LQF) & Full Q\_ij & 1 (if solvable) & 100\% admissible [4] & O(N\textsuperscript{3}) & Theoretical upper bound \\
Hungarian/auction & Full Q\_ij & Multiple until convergence & Same as MWM & O(N\textsuperscript{3}) & Simulation/offline [10],[11] \\
iSLIP & 1-bit & Configurable (typically 3) & High throughput under uniform; §7 diverges under non-uniform & O(log N) rounds & Industrial baseline [3] \\
Spectral scheduling & Q + u\textsubscript{1},v\textsubscript{1} & r power-iteration steps & §7: close to MWM & O(r·N\textsuperscript{2}) & This paper §7 \\
Entropy-regularized OT & Full Q\_ij & r\_sink Sinkhorn rounds & §7: closest to MWM delay & O(r\_sink·N\textsuperscript{2})+exp & This paper §7 \\
Sinkhorn+BvN & Continuous B & Offline + periodic permutation & Frame-level time-division [31],[32] & O(N\textsuperscript{2})/iteration & Relaxation method [12] \\
Tassiulas / APSARA & Full Q\_ij & 1 (APSARA parallelizable) & 100\% admissible [22],[24] & O(N)--parallel & Global Q, high delay \\
\bottomrule
\end{tabular}}\par\leftskip=2em\relax{}

\bigskip

{\small\textbf{Table 2.} Cross-algorithm comparison for main-thread algorithms}

\clearpage
\subsection{Complexity and Throughput Metrics Quick Reference}

The three percentages (58.6\%, 50\%, 100\%) in the table have different meanings (architecture HOL upper bound, maximal-matching weight-ratio lower bound, 100\% in the admissible stability sense) and cannot be directly compared to the experimentally measured throughput--load curves.

\bigskip\noindent
\par\leftskip=2em\relax\noindent \fittab{\begin{tabular}{p{\dimexpr 0.200\linewidth-2\tabcolsep}p{\dimexpr 0.200\linewidth-2\tabcolsep}p{\dimexpr 0.200\linewidth-2\tabcolsep}p{\dimexpr 0.200\linewidth-2\tabcolsep}p{\dimexpr 0.200\linewidth-2\tabcolsep}}
\toprule
Metric & Value & Applies to & Meaning & Reference \\
\midrule
HOL saturation throughput & $\approx$58.6\% (=2-√2) & Input FIFO, no VOQ & Architectural upper bound & [1] \\
Maximal matching weight ratio & $\geq$50\% of MWM & Single weighted maximal matching & Lower bound on approximation ratio vs. MWM & §4.3 \\
MWM stability & 100\% & Any admissible $\Lambda$ (row and column sums \textless  1) & Queues bounded; theorem & [4],[7] \\
iSLIP stability & 100\% & Uniform i.i.d. only; no general guarantee for non-uniform (§7) & Queues bounded & [3] \\
Exact MWM (Hungarian/auction) & O(N\textsuperscript{3}) & Exact MWM & Acceptable in software; infeasible per cycle & §4.3, Appendix B.4 \\
iSLIP iteration & O(log N) rounds & Maximal matching & O(1) distributed per round & [3] \\
Sinkhorn & O(N\textsuperscript{2}) / iteration & Doubly stochastic & Continuous relaxation & §4.1.1 \\
BvN decomposition & $O(N^{4.5})$ etc. & Offline & Number of permutation matrices $\leq$ N\textsuperscript{2}-2N+2 & [31] \\
\bottomrule
\end{tabular}}\par\leftskip=2em\relax{}

\bigskip

{\small\textbf{Table 3.} Definitions of easily confused numerical values in this paper}

\section{Comparative Experiments: Spectral Scheduling, iSLIP, and Entropy-Regularized OT (Adaptive Temperature)}

The experiments in this chapter are based on the standard IQ+VOQ crossbar model. Maximum weight matching (MWM) solved by the Hungarian algorithm serves as the performance reference, and the paper compares iSLIP, spectral scheduling, and entropy-regularized optimal transport (OT) in throughput and delay. The per-cycle iteration count for iSLIP and spectral scheduling is uniformly capped at 3 (r=3); the Sinkhorn row/column normalization for entropy-regularized OT is set to r\_sink=10.

Entropy-regularized scheduling refers to processing the queue matrix via the transformation $K\propto\exp(Q/\varepsilon_{\mathrm{eff}})$ (the Softassign idea [43]; entropy-regularized OT / Sinkhorn distance [42]), then combining a finite number of Sinkhorn iterations to obtain an approximate doubly stochastic weight matrix, and finally computing the discrete matching result by greedy rounding. Entropy-regularized scheduling is chosen over the continuous-relaxation method of §4.1 because, unlike that approach (which directly scales Q and performs BvN frame decomposition), this method is closer to the ⟨Q,P⟩ optimization objective of MWM.

\subsection{Research Questions and Design Principles}

The experiments are organized around three research questions:

\textbf{Q1:} Under non-uniform admissible traffic, do approximate scheduling algorithms that read the full queue information matrix outperform iSLIP (which collects only 1-bit queue status), and do they track the MWM reference baseline more closely?

\textbf{Q2:} Under the iteration budget constraints above, what are the characteristics of each algorithm's throughput--$\rho$\_load and delay--$\rho$\_load curves?

\textbf{Q3:} Can increasing the iteration count bring meaningful performance gains?

In the experiments, only the core scheduling mapping schedule(Q)\ensuremath{\rightarrow}P is replaced; all other simulation configurations, traffic parameters, and statistics remain consistent. The IQ+VOQ model from §2.2 is used, with no queue depth limit and no hardware backpressure, studying only the performance differences among the scheduling algorithms themselves. MWM, spectral scheduling, and entropy-regularized OT read the full 32-bit queue values; iSLIP collects only the 1-bit queue-nonempty status.

\bigskip\noindent
\par\leftskip=2em\relax\noindent \fittab{\begin{tabular}{p{\dimexpr 0.15\linewidth-2\tabcolsep}p{\dimexpr 0.35\linewidth-2\tabcolsep}p{\dimexpr 0.14\linewidth-2\tabcolsep}p{\dimexpr 0.20\linewidth-2\tabcolsep}p{\dimexpr 0.16\linewidth-2\tabcolsep}}
\toprule
Algorithm & Abstract form schedule(Q)\ensuremath{\rightarrow}P & Q read resolution & Per-cycle operation scale & Experimental role \\
\midrule
iSLIP (r=3) & P $\approx$ $\Pi$\_row($\Pi$\_col(1\{Q\textgreater 0\})), R--G--A + RR pointer & 1-bit nonempty (Q\_ij\textgreater 0) & O(r·N) 1-bit RR arbitration & Mature industrial baseline \\
Spectral scheduling (r=3) & P = Greedy(Q ⊙ u\textsubscript{1}v\textsubscript{1}ᵀ), power iter + greedy & Full matrix + rank-1 spectral feature & O(r·N\textsuperscript{2}) MAC + O(N\textsuperscript{2}) greedy & Linear-algebraic approximation (§4.2.4) \\
Entropy-OT (r\_sink=10, $\epsilon$=1) & P = Greedy(Sinkhorn(exp(Q/$\epsilon$))) & Full matrix + entropic weighting & O(r\_sink·N\textsuperscript{2}) exp/norm + O(N\textsuperscript{2}) greedy & Entropy-OT scheduler (§7.5) \\
MWM (Hungarian) & P = Hungarian(Q), max-weight bipartite matching & Full matrix, exact values & O(N\textsuperscript{3}) optimal matching & Simulation reference baseline \\
\bottomrule
\end{tabular}}\par\leftskip=2em\relax{}

\bigskip

{\small\textbf{Table 4.} Four schedulers: abstract form, information granularity, and per-cycle operation scale (same schedule(Q)\ensuremath{\rightarrow}P interface)}

\needspace{26\baselineskip}
\subsection{System Model and Simulator}

The simulation uses a discrete-event platform built in-house; the unified simulation parameters are:

\bigskip\noindent
\par\leftskip=2em\relax\noindent \fittab{\begin{tabular}{p{\dimexpr 0.500\linewidth-2\tabcolsep}p{\dimexpr 0.500\linewidth-2\tabcolsep}}
\toprule
Item & Specification \\
\midrule
Topology & N$\times$N IQ + VOQ, single crossbar, speedup=1, globally synchronous clock \\
Port count N & 8 (core experiment scale; scale rationale below) \\
Per-cycle timing flow & (1) Traffic arrives and enters queue \ensuremath{\rightarrow} (2) Scheduler reads queue matrix and outputs matching P \ensuremath{\rightarrow} (3) Dequeue according to matching result \\
VOQ configuration & No depth limit; dynamically expanded during simulation; hardware backpressure and packet dropping are not modeled \\
Delay statistics definition & Each VOQ maintains a FIFO timestamp; single-cell delay = current clock cycle - head-of-queue arrival cycle \\
Throughput statistics definition & Normalized throughput = output cells in statistics window / (total cycles $\times$ port count N), range [0,1] \\
Warm-up cycles & 10\textsuperscript{4} clock cycles (warm-up data discarded; ensures system is in steady state) \\
Effective statistics window & 10\textsuperscript{5} clock cycles \\
Random seed mechanism & 20 independent random seeds per ($\rho$\_load, traffic model, algorithm) configuration \\
Result presentation & Mean + 95\% confidence interval (CI = 1.96·SE) across all experiments \\
\bottomrule
\end{tabular}}\par\leftskip=2em\relax{}

\bigskip

\textbf{Scale rationale (N=8):} This paper targets SoC on-chip interconnect, not large-scale data-center switches. The mainstream scale for on-chip single-stage crossbars is N=4 to 8; N=8 is already close to the physical limit for on-chip VOQ arrays and single-cycle arbitration. In practical hardware design, larger port counts typically use multi-stage or other topologies, which are outside the scope of this paper. N=8 is sufficient to compare the performance difference between weight-unaware scheduling and full-information scheduling.

\subsection{The Four Schedulers Compared}

The information granularity, computation scale, and experimental role of the four typical scheduling algorithms are shown in Table 4. Core operational specifications for each algorithm:

\begin{itemize}
\item \textbf{iSLIP:} Follows the standard R-G-A three-round iteration mechanism; pointers update only after the first-round grant is accepted; relies on pointer desynchronization to avoid pointer synchronization; recognizes only queue-nonempty status, not queue length.
\item \textbf{Spectral scheduling:} Extracts the rank-1 spectral feature of the queue matrix via 3 rounds of power iteration with L2 normalization; constructs the weight matrix W=Q⊙(u\textsubscript{1}v\textsubscript{1}ᵀ) and then discretizes using row/column masking per §4.2.5 Path A to obtain a valid matching.
\item \textbf{Entropy-regularized OT:} Constructs an exponential kernel from the queue matrix (Softassign [43]); generates an approximate doubly stochastic matrix via r\_sink rounds of Sinkhorn iteration [42]; obtains the matching by greedy rounding. An adaptive temperature mechanism is introduced to mitigate numerical degradation and overflow under fixed parameters and to make the weights more aligned with the ⟨Q,P⟩ optimization objective.
\item \textbf{MWM:} Solves the global maximum weight matching each cycle; provides the reference result as the simulation baseline.
\end{itemize}

\subsection{Traffic Models}

Four typical traffic models are used. Unbalanced traffic is the core scenario for distinguishing performance across algorithms; the hotspot overload scenario is used to observe capacity boundaries.

\bigskip\noindent
\par\leftskip=2em\relax\noindent \fittab{\begin{tabular}{p{\dimexpr 0.250\linewidth-2\tabcolsep}p{\dimexpr 0.250\linewidth-2\tabcolsep}p{\dimexpr 0.250\linewidth-2\tabcolsep}p{\dimexpr 0.250\linewidth-2\tabcolsep}}
\toprule
Traffic model & Traffic definition & Contention characteristics & Admissible range (N=8) \\
\midrule
Uniform & $\lambda$\_ij = $\rho$\_load/N & No structured contention; experimental baseline & Any $\rho$\_load $\leq$ 1 \\
Diagonal ($\alpha$=0.7) & $\lambda$\_ii = $\alpha$·$\rho$\_load; $\lambda$\_ij = (1-$\alpha$)$\rho$\_load/(N(N-1)), i$\neq$j & Near-permutation structure; low port contention & Any $\rho$\_load $\leq$ 1 \\
Hotspot (h=0.2, j*=0) & $\lambda$\_\{i,j*\} = h·$\rho$\_load; $\lambda$\_\{i,j\} = (1-h)$\rho$\_load/(N-1), j$\neq$j* & Structural bottleneck at a single output port; global traffic aggregation & $\rho$\_load $\leq$ 1/(Nh) = 0.625; beyond this j* column overloads (still used for stress testing to observe capacity boundary) \\
Unbalanced (w=0.5) & $\lambda$\_ii = $\rho$\_load(w + (1-w)/N); $\lambda$\_ij = $\rho$\_load(1-w)/N, i$\neq$j & Row/column sums constant at $\rho$\_load (doubly stochastic row/column sums); diagonal-dominant load plus cross-port contention & Any $\rho$\_load $\leq$ 1 \\
\bottomrule
\end{tabular}}\par\leftskip=2em\relax{}

\bigskip

\subsection{Implementation Notes}

For reproducibility, the algorithm implementation and data collection conventions are as follows:

\begin{itemize}
\item \textbf{Delay statistics:} Single-cell delay is computed based on per-VOQ independent FIFO timestamps following FIFO order, to maintain consistent delay statistics under congestion scenarios.
\item \textbf{Standard iSLIP mechanism:} Implemented according to the original pointer update rule to avoid degenerating into RRM, keeping the implementation consistent with the industrial standard.
\item \textbf{Entropy-regularized OT numerical stability:} exp(·) subtracts the global maximum before exponentiation to prevent overflow; relative temperature $\epsilon$\_eff = $\epsilon$·max(1, Qmax/W\textsubscript{0}) (W\textsubscript{0}=10) is used so that the magnitude of the exponential term is independent of queue scale, resolving the iterative degradation problem with fixed $\epsilon$.
\item \textbf{Unified throughput definition:} Throughput is uniformly normalized to [0,1] for cross-scenario comparison.
\item \textbf{Experimental state isolation:} Each ($\rho$\_load, traffic, seed) experiment reinitializes the scheduler; no internal algorithm state carries over between groups.
\end{itemize}

\subsection{Result 1: Throughput vs. Load}

\par\medskip{\leftskip=0pt\relax\noindent\hspace*{2em}%
\begin{minipage}{\dimexpr\linewidth-2em\relax}
\includegraphics[width=\linewidth]{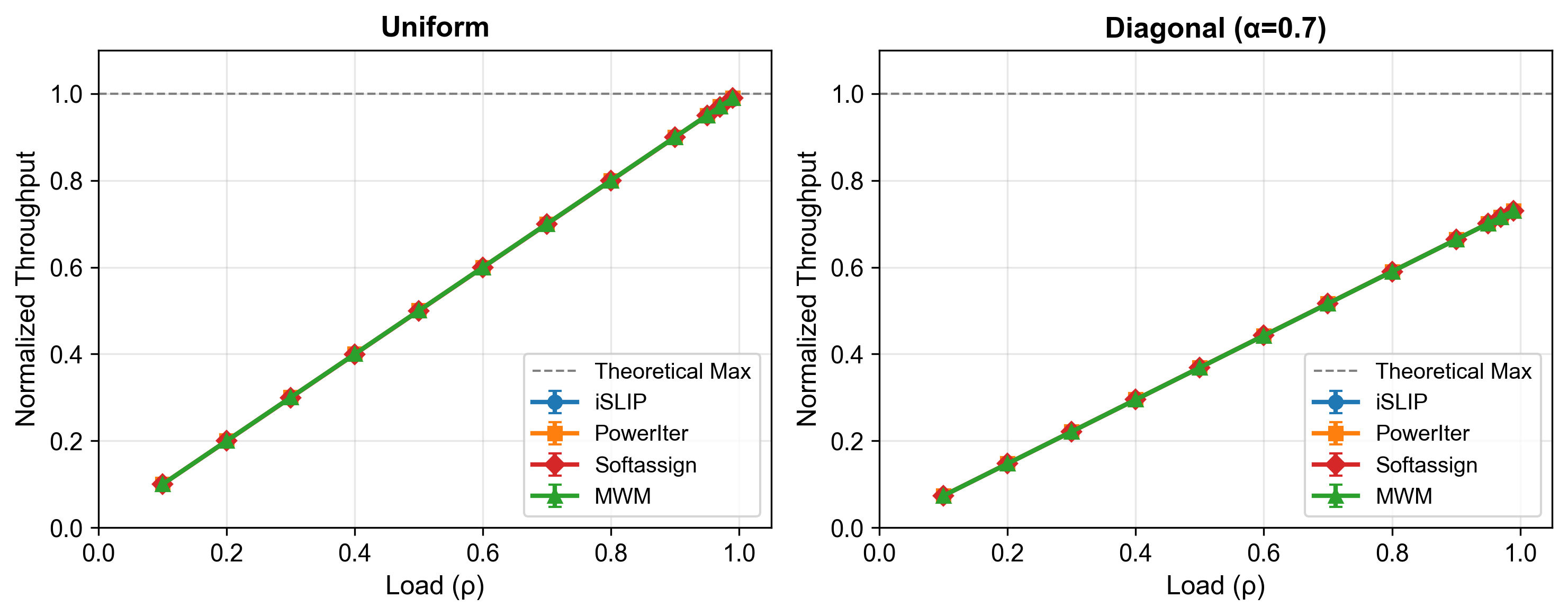}\par\medskip
\includegraphics[width=\linewidth]{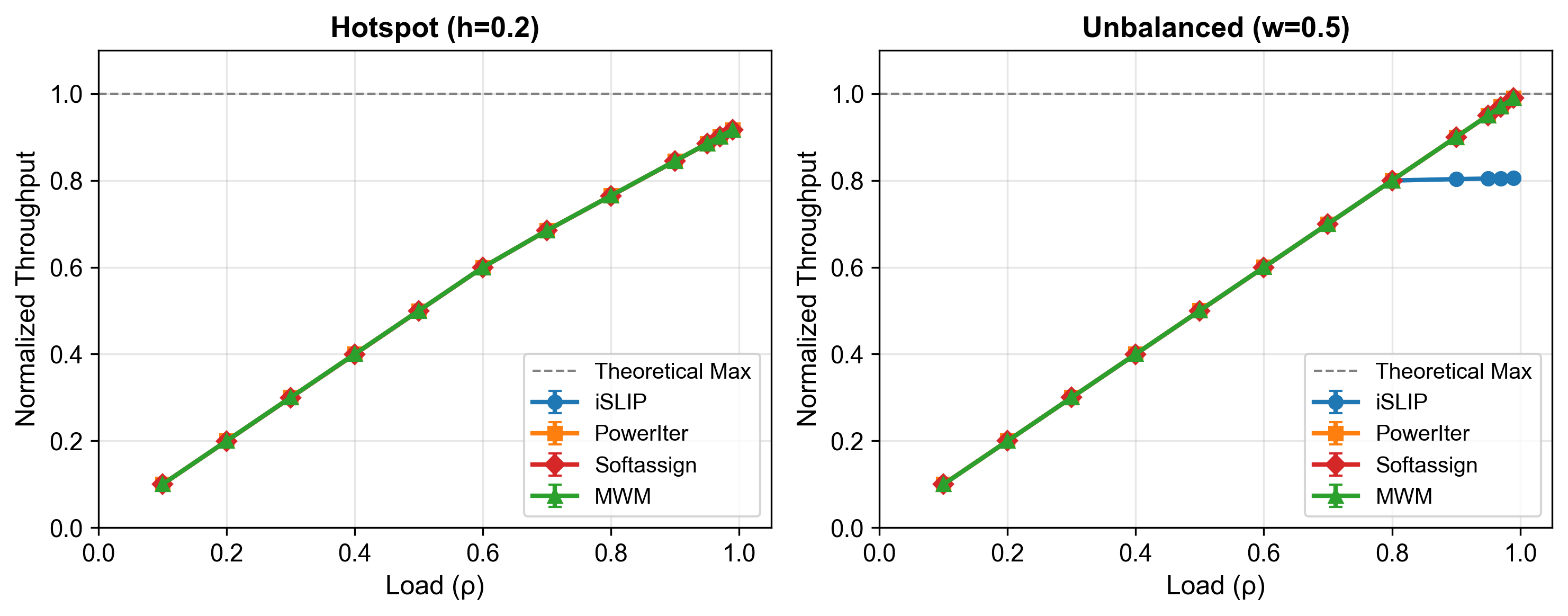}
\par\medskip
{\small\textbf{Figure 3.} Normalized throughput vs.{\footnotesize  $\rho$\_load (four traffic patterns $\times$ four algorithms; error bars are 95\% CI). Top: Uniform, Diagonal ($\alpha$=0.7); Bottom: Hotspot (h=0.2), Unbalanced (w=0.5).}}
\par\medskip
\end{minipage}\par\medskip\par}

Two observations from Figure 3. (1) Except under unbalanced w=0.5, the normalized throughput of MWM, spectral scheduling, and entropy-regularized OT tracks the load closely as $\rho$\_load increases; the flat tops in the diagonal and hotspot curves come from the traffic definition or output bottlenecks, not from algorithmic differences. (2) Under w=0.5, iSLIP hits a {\textasciitilde}80\% throughput ceiling when $\rho$\_load $\geq$ 0.9, consistent with the literature benchmarks [16],[30]; this pattern is used to amplify the difference between weight-unaware MSM and full-information algorithms. Under other non-uniform traffic patterns, iSLIP can still approach full throughput. Representative values for w=0.5, $\rho$\_load=0.99 are in Table 7.

\subsection{Result 2: Average Delay}

\bigskip\noindent
\par\leftskip=2em\relax\noindent \fittab{\begin{tabular}{p{\dimexpr 0.167\linewidth-2\tabcolsep}p{\dimexpr 0.167\linewidth-2\tabcolsep}p{\dimexpr 0.167\linewidth-2\tabcolsep}p{\dimexpr 0.167\linewidth-2\tabcolsep}p{\dimexpr 0.167\linewidth-2\tabcolsep}p{\dimexpr 0.167\linewidth-2\tabcolsep}}
\toprule
Traffic type & $\rho$\_load & MWM & Entropy-reg. OT & Spectral & iSLIP \\
\midrule
Uniform & 0.50 & 0.8 & 0.9 & 1.1 & 1.0 \\
Uniform & 0.70 & 1.8 & 2.2 & 2.6 & 2.6 \\
Uniform & 0.80 & 3.1 & 3.8 & 4.7 & 4.9 \\
Uniform & 0.90 & 7.0 & 8.2 & 12.0 & 13.1 \\
Uniform & 0.95 & 14.8 & 16.1 & 27.7 & 40.3 \\
Uniform & 0.99 & 78.3 & 75.7 & 159.0 & 291.2 \\
Unbalanced & 0.70 & 1.5 & 1.7 & 1.9 & 2.8 \\
Unbalanced & 0.80 & 2.5 & 3.1 & 3.4 & 69.2 \\
Unbalanced & 0.90 & 5.4 & 6.8 & 8.6 & 5846 \\
Unbalanced & 0.95 & 11.3 & 13.2 & 19.9 & 7901 \\
Unbalanced & 0.99 & 59.8 & 59.3 & 119.0 & 9171 \\
\bottomrule
\end{tabular}}\par\leftskip=2em\relax{}

\bigskip

{\small\textbf{Table 5.} Average cell delay (cycles), N=8, mean over 20 seeds (Uniform/Unbalanced traffic)}

\par\medskip{\leftskip=0pt\relax\noindent\hspace*{2em}%
\begin{minipage}{\dimexpr\linewidth-2em\relax}
\includegraphics[width=\linewidth]{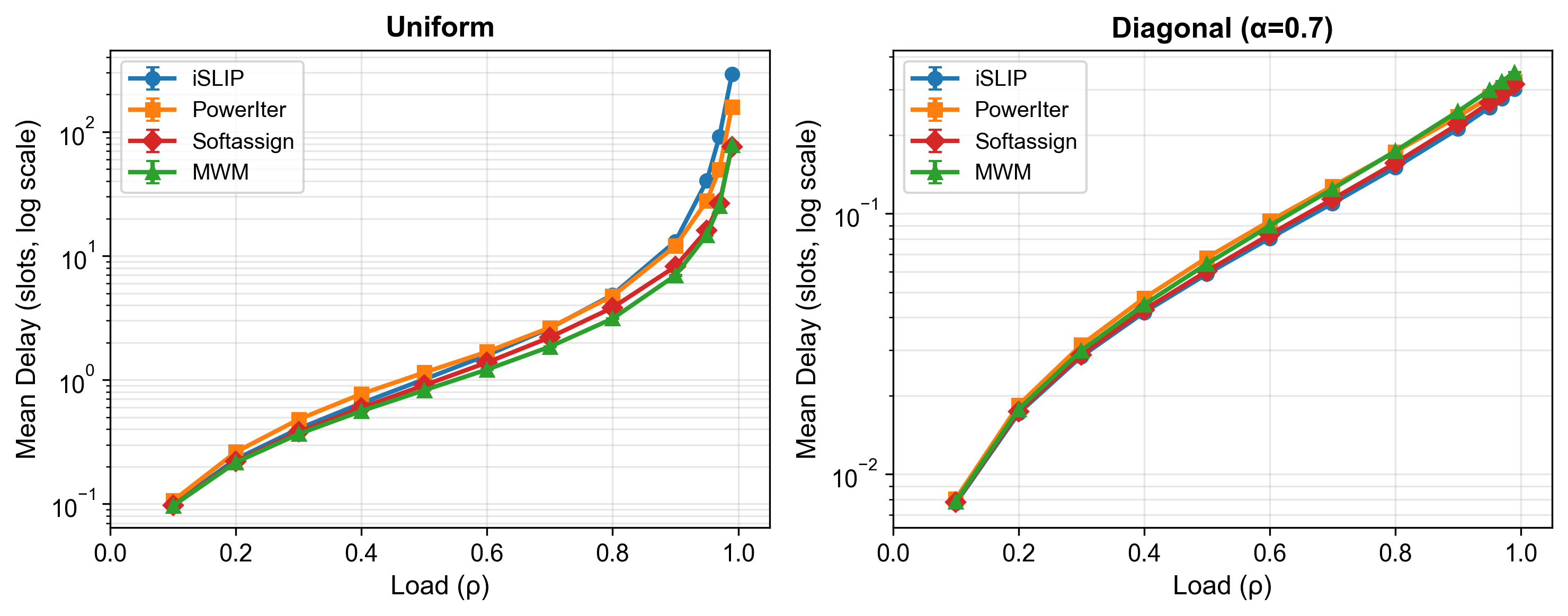}\par\medskip
\includegraphics[width=\linewidth]{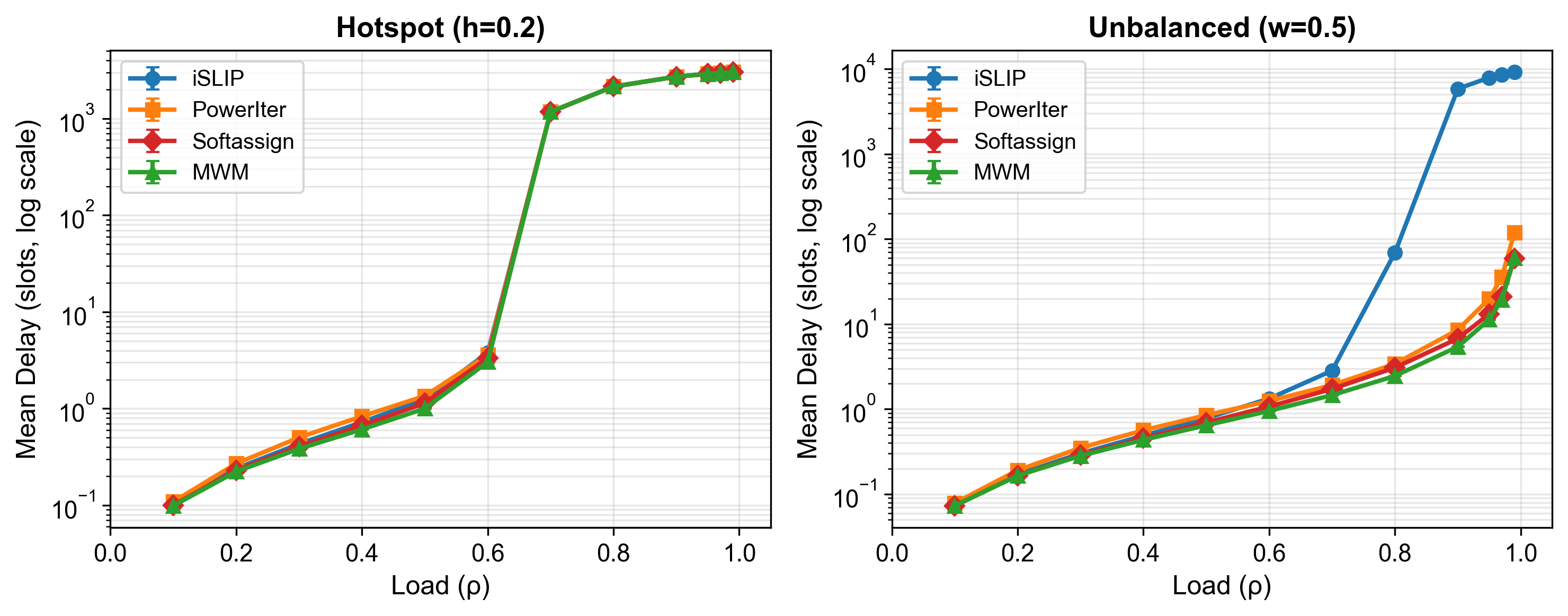}
\par\medskip
{\small\textbf{Figure 4.} Average delay vs.{\footnotesize  $\rho$\_load (log-scale y-axis, four traffic patterns $\times$ four algorithms; error bars are 95\% CI). Top: Uniform, Diagonal ($\alpha$=0.7); Bottom: Hotspot (h=0.2), Unbalanced (w=0.5).}}
\par\medskip
\end{minipage}\par\medskip\par}

Across most loads, the delay ordering is MWM ≲ entropy-regularized OT \textless  spectral scheduling \textless  iSLIP; the gap widens as load and traffic non-uniformity increase. Under unbalanced w=0.5 at high $\rho$, iSLIP can exhibit delays on the order of 10\textsuperscript{3}--10\textsuperscript{4} cycles, which represents a VOQ backlog imbalance condition and should not be directly compared to the single-digit values of MWM/OT; a more appropriate comparison is at moderate load, e.g., $\rho$\_load=0.8. Under uniform $\rho$\_load=0.99, the OT result is slightly below MWM, within the 95\% CI, because MWM optimizes ⟨Q,P⟩ rather than average delay. Under hotspot overload, the delay differences among algorithms are small.

\subsection{Result 3: Sensitivity to Iteration Count}

To study the effect of the iteration budget on performance, a sweep experiment is conducted at fixed $\rho$\_load=0.8 with uniform traffic.

\bigskip\noindent
\par\leftskip=2em\relax\noindent \fittab{\begin{tabular}{p{\dimexpr 0.200\linewidth-2\tabcolsep}p{\dimexpr 0.200\linewidth-2\tabcolsep}p{\dimexpr 0.200\linewidth-2\tabcolsep}p{\dimexpr 0.200\linewidth-2\tabcolsep}p{\dimexpr 0.200\linewidth-2\tabcolsep}}
\toprule
Algorithm & Budget=1 & 3 & 8 & 16 \\
\midrule
iSLIP (r rounds R--G--A) & 22.3 & 4.86 & 4.85 & --- \\
Spectral scheduling (r power-iteration rounds) & 4.96 & 4.74 & 4.73 & 4.73 \\
Entropy-regularized OT (r\_sink Sinkhorn rounds) & 3.84 & 3.83 & 3.83 & 3.83 \\
MWM (reference baseline) & 3.12 & --- & --- & --- \\
\bottomrule
\end{tabular}}\par\leftskip=2em\relax{}

\bigskip

{\small\textbf{Table 6.} Iteration sweep results (uniform traffic): normalized throughput is 0.80 for all; values are average delay (cycles)}

\par\medskip{\leftskip=0pt\relax\noindent\hspace*{2em}%
\begin{minipage}{\dimexpr\linewidth-2em\relax}
\includegraphics[width=\linewidth]{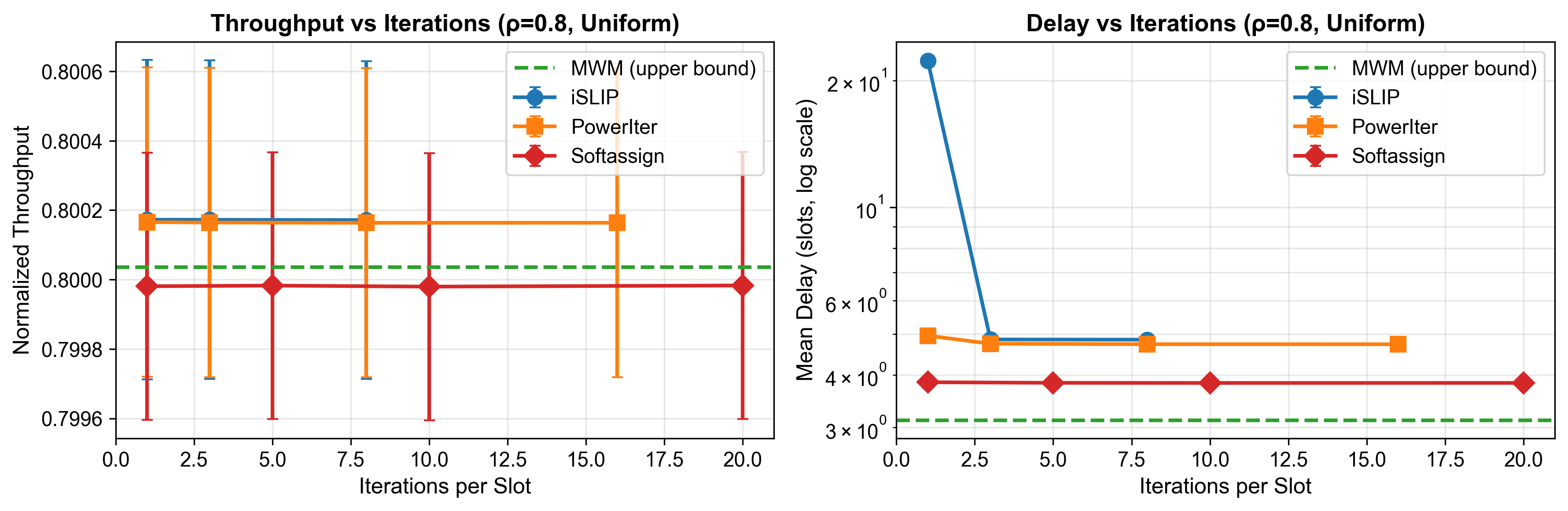}
\par\medskip
{\small\textbf{Figure 5.} Performance vs.{\footnotesize  iteration count (left: throughput; right: delay on log axis; MWM is a horizontal baseline).}}
\par\medskip
\end{minipage}\par\medskip\par}

Table 6 and Figure 5 show that iSLIP needs approximately 3 R--G--A rounds before delay settles to a stable value; spectral scheduling and entropy-regularized OT approach their minimum values within 1--3 rounds. This justifies the choice of r=3 and r\_sink=10 in the main experiments.

\subsection{Experiment Summary}

The table below summarizes representative throughput and delay under the same configuration (N=8, unbalanced w=0.5, $\rho$\_load=0.99, r=3).

\bigskip\noindent
\par\leftskip=2em\relax\noindent \fittab{\begin{tabular}{p{\dimexpr 0.200\linewidth-2\tabcolsep}p{\dimexpr 0.200\linewidth-2\tabcolsep}p{\dimexpr 0.200\linewidth-2\tabcolsep}p{\dimexpr 0.200\linewidth-2\tabcolsep}p{\dimexpr 0.200\linewidth-2\tabcolsep}}
\toprule
Algorithm & Per-cycle abstract operations & Normalized throughput & Average delay (cycles) & Role \\
\midrule
MWM & O(N\textsuperscript{3}) & $\approx$0.99 & 59.8 & Simulation reference \\
Entropy-reg. OT (adaptive temp.) & O(r\_sink·N\textsuperscript{2})+exp & $\approx$0.99 & 59.3 & Closest approximation to MWM \\
Spectral scheduling & O(r·N\textsuperscript{2})+greedy & $\approx$0.99 & 119.0 & Linear-algebraic approximate scheduler \\
iSLIP & O(r·N) 1-bit ops & $\approx$0.806 & 9171 & Industrial weight-unaware iterative baseline \\
\bottomrule
\end{tabular}}\par\leftskip=2em\relax{}

\bigskip

{\small\textbf{Table 7.} Comparative experiment summary (N=8, unbalanced traffic, r=3)}

\subsection{Analysis-Level Hardware Pareto (No RTL Synthesis)}

§7.3 provides algorithm-level O(·) counts; the discrete-event experiments do not model ASIC area. To address the ``is on-chip implementation worthwhile'' trade-off question raised in the abstract, this section assigns the following relative weights to abstract arithmetic operation counts and queue-read bit widths without conducting RTL pre-synthesis (no RTL calibration; for order-of-magnitude ranking only): 1-bit RR arbitration=1, MAC=5, exp+Sinkhorn normalization=20, O(N\textsuperscript{2}) greedy compare=2.

Selection rationale: using 1-bit round-robin arbitration as baseline 1, MAC, exp normalization, and compare-tree are scaled up by approximately 5$\times$, 20$\times$, and 2$\times$ respectively; the relative ordering of C\_rel within the same algorithm family is stable as N varies. In practice, depending on the process node, pipeline depth, or area-vs-speed strategy, the absolute factors can differ by several times; Table 8 is qualitative Pareto only, not an absolute area prediction.

Per-cycle arbiter-core abstract hardware cost formulas (excluding VOQ SRAM depth overhead):
\begin{align}
&C_{\text{islip}}(N) = r\cdot N \nonumber \\
&C_{\text{spec}}(N) = 2\,r\,N^2\,w_{\text{mac}} + 2\,N^2\,w_{\text{greedy}} = 34\,N^2 \nonumber \\
&C_{\text{ot}}(N) = r_{\text{sink}}\,N^2\,w_{\text{exp}} + 2\,N^2\,w_{\text{greedy}} = 204\,N^2
\end{align}

\bigskip\noindent
\par\leftskip=2em\relax\noindent \fittab{\begin{tabular}{p{\dimexpr 0.143\linewidth-2\tabcolsep}p{\dimexpr 0.143\linewidth-2\tabcolsep}p{\dimexpr 0.143\linewidth-2\tabcolsep}p{\dimexpr 0.143\linewidth-2\tabcolsep}p{\dimexpr 0.143\linewidth-2\tabcolsep}p{\dimexpr 0.143\linewidth-2\tabcolsep}p{\dimexpr 0.143\linewidth-2\tabcolsep}}
\toprule
Algorithm & C\_rel (arithmetic) & Q read storage (bit) & Storage vs. iSLIP & Normalized throughput† & Average delay† (slots) & Delay / MWM† \\
\midrule
iSLIP & 1.0 & 64 & 1$\times$ & $\approx$0.806 & 9171 & $\approx$153$\times$ \\
Spectral scheduling & 91 & 2048 & 32$\times$ & $\approx$0.990 & 119.0 & $\approx$2.0$\times$ \\
Entropy-reg. OT & 544 & 2048 & 32$\times$ & $\approx$0.989 & 59.3 & $\approx$1.0$\times$ \\
MWM (simulation upper bound) & O(N\textsuperscript{3}), infeasible per cycle & 2048 & 32$\times$ & $\approx$0.990 & 59.8 & 1.0$\times$ \\
\bottomrule
\end{tabular}}\par\leftskip=2em\relax{}

\bigskip

{\small\textbf{Table 8.} Analysis-level Pareto (N=8 measured; cost scales with N via formulas)}

Note: columns marked with † (normalized throughput, average delay, delay/MWM) are simulation measurements from §7 at N=8; all other cost columns scale with N via the complexity formulas.

\textbf{Result interpretation (N=8 measurements + cost formula scaling with N)}

\begin{itemize}
\item \textbf{Hardware cost:} At the same port scale, spectral scheduling's relative arithmetic cost is 45--91$\times$ that of iSLIP (N=4--8); entropy-regularized OT is as high as 272--544$\times$ (N=4--8). Both full-information algorithms also read Q at 32$\times$ the bit width of iSLIP. As N increases, the core hardware costs of MAC, exp, etc. scale at the O(N\textsuperscript{2}) level; iSLIP's cost ($\propto$N) means the ratio grows monotonically with N.
\item \textbf{Timing path depth (qualitative):} iSLIP uses r serial R-G-A iterations each with O(log N) round-robin arbitration; lowest timing path depth. Spectral scheduling uses r serial N$\times$N matrix MAC operations plus a greedy compare tree; medium depth. Entropy-regularized OT includes 10 serial exp operations and normalization iterations; deepest timing path, and if the greedy feedback iteration mechanism of §4.2.5 is added on top, the timing cost increases further (not counted in the experiments).
\end{itemize}

\textbf{Engineering trade-off summary:} Spectral scheduling and entropy-regularized OT have C\_rel values 45--544$\times$ that of iSLIP (scaling with N per the formula above), and a 32$\times$ wider queue read bus (Table 8). Per-item throughput, delay, and iteration sensitivity results are in §7.6--§7.8; single-point values for w=0.5, $\rho$\_load=0.99 are in Table 7. Under uniform load with area and timing constraints, iSLIP typically already meets performance requirements. Under non-uniform high-congestion scenarios where the hardware can accommodate the MAC/exp overhead, spectral scheduling and OT are worth further evaluation.

\section{Conclusion}

The first theoretical thread of this paper uses the Lyapunov energy-decrease mechanism to explain the mathematical basis for the MWM scheduling policy. The admissible traffic condition defines the boundary within which input and output ports can operate without overload; BvN theory further shows that a long-run average scheduling combination exists that can carry the corresponding traffic, namely a convex combination of valid matching matrices. These results characterize the core prerequisites and capability conditions for the system's queues to remain stable in the long run, but they do not directly yield real-time grant decisions suitable for single-cycle hardware operation.

\addvspace{12pt plus 3pt minus 1pt}\noindent For continuous-relaxation scheduling algorithms, this paper discusses the key constraints on hardware implementation. Sinkhorn iteration, spectral scheduling, and similar algorithms can only produce doubly stochastic matrices or weighted association matrices, which are continuous floating-point matrices that cannot be directly converted into the 0/1 discrete matching results that hardware needs. To satisfy the per-cycle hardware rule of row/column mutual exclusion with exactly one valid match per row/column, an additional 0/1 projection step is required: the continuous weight matrix is mapped to a valid matching matrix P(t) (at most one 1 per row and per column) through methods such as weighted greedy selection or valid-match extraction.

\addvspace{12pt plus 3pt minus 1pt}\noindent The second theoretical thread observes that graph-theoretic exact MWM can serve as a simulation upper bound, but this multi-round, full-matrix-weight-reading solving process cannot meet the timing requirements of per-cycle on-chip grant. Polynomial-time solvability is not equivalent to single-cycle implementability.

The third engineering thread focuses on the industrial mainstream iSLIP iterative scheduling mechanism. iSLIP collects only the queue-nonempty status signal each cycle, relying on multi-round Request--Grant--Accept iteration and the desynchronization mechanism of round-robin pointers to perform iterative matching. Under long-run uniform traffic, pairs (i,j) with persistent backlog can obtain a higher average matching probability P̄, roughly satisfying the drift condition $E[\langle Q,P\rangle] > Q^{\mathsf{T}}\lambda$ (equivalent to $\langle Q,\lambda\rangle$, with $\lambda$ flattened from $\Lambda$). However, limited by its 1-bit queue awareness, which cannot access full queue weight information, this steady-state approximate relationship no longer holds in certain non-uniform admissible traffic scenarios.

Four directions for future work are identified:

(1) The stability and instability boundary of iterative matching scheduling under non-uniform admissible traffic.

(2) Supplementing the experiments with simulations of finite queue depth, backpressure mechanisms, and bursty/converging traffic; conducting a fair performance comparison with the APSARA [24] algorithm under identical conditions of global queue awareness and a fixed iteration count.

(3) Extending the arbitration algorithm comparison to new switch architectures such as CIOQ/CICQ [27],[28],[29],[39] and PIFO programmable queues [40].

(4) Incorporating distributed iterative MWM [41] and optimal load-balanced switching [2] into the ⟨Q,P⟩ notation and comparing them with the main threads of this paper.

\section*{Appendix A. Terminology Reference}

The terms below recur throughout the paper; they are collected here in one place for convenience.

\bigskip\noindent
\par\leftskip=2em\relax\noindent \fittab{\begin{tabular}{p{\dimexpr 0.500\linewidth-2\tabcolsep}p{\dimexpr 0.500\linewidth-2\tabcolsep}}
\toprule
Term & Meaning \\
\midrule
crossbar arbitration / grant & per-cycle selection of which input--output pairs may forward \\
flit (burst = 1) & the basic data unit forwarded in one cycle; equals one cell when burst = 1 \\
input-queued (IQ) switch & switch architecture with queues at the input side \\
virtual output queue (VOQ) & a separate queue per (input, output) pair, eliminating HOL blocking \\
backlog & queued packet/cell awaiting a crossbar grant \\
maximum weight matching (MWM) & the matching P that maximizes ⟨Q, P⟩ \\
iSLIP & iterative round-robin matching with R--G--A rounds [3] \\
maximal matching (MSM-class) & a matching to which no edge can be added without violating the row/column constraint \\
admissible traffic & traffic matrix $\Lambda$ with every row sum and column sum $\leq$ 1 \\
spectral radius $\rho$($\Lambda$) & modulus of the largest eigenvalue of $\Lambda$ \\
normalized load factor $\rho$\_load & per-input-port offered load, swept in the experiments \\
Request--Grant--Accept (R--G--A) & the three-phase iSLIP handshake \\
per-cycle / per-slot iteration count & number of R--G--A rounds completed within one clock cycle \\
crossbar arbiter; per-cycle grant & the hardware block that emits the grant matrix P(t) each cycle \\
programmable dataplane (e.g. PIFO [40]) & a queue/scheduling abstraction reconfigurable at line rate \\
per clock cycle (per-cycle) & one scheduling slot; the unit over which a matching is computed \\
weight-unaware & the scheduler reads only 1-bit queue-nonempty status, not queue length \\
bit-plane matching & per-weight-bit iterative arbitration \\
row/column masking & after selecting (i, j), mask its row and column so neither is selected again \\
weight-unaware random matching & a scheduler that picks a matching ignoring Q weights \\
\bottomrule
\end{tabular}}\par\leftskip=2em\relax{}

\bigskip

\section*{Appendix B. Graph-Theoretic Exact Matching: Ideas and On-Chip Barriers}

\subsection*{B.1 Hungarian Algorithm (Kuhn--Munkres) [10]}

Idea: maintain row potentials \texttt{$\alpha$\_i} and column potentials \texttt{$\beta$\_j}; define the reduced weight \texttt{c\_ij = $\alpha$\_i + $\beta$\_j - Q\_ij $\geq$ 0}; look for a perfect matching only on ``zero edges'' where \texttt{c\_ij = 0}. If no perfect matching is found, update the potentials by the minimum slack \texttt{$\delta$} to expand the zero-edge set, then flip the matching along an alternating path (augmenting path). During augmentation, previously selected edges can be removed and other edges selected, i.e., early local choices can be corrected by backtracking. Full procedure and worked example in §4.3 graph-theoretic formulation; complexity in B.4.

\subsection*{B.2 Auction Algorithm (Bertsekas) [11]}

Idea: buyer = input i, good = output j, valuation \texttt{v\_ij = Q\_ij}, price \texttt{$\pi$\_j}. An unassigned buyer bids on the good with the highest net benefit \texttt{v\_ij - $\pi$\_j}; the highest bidder wins the good, and the unsuccessful bidder re-bids in the next round; temporary assignments can be revoked. \texttt{$\epsilon$} controls convergence precision and speed. Full procedure and worked example in §4.3 graph-theoretic formulation; complexity in B.4.

\subsection*{B.3 Global Greedy Row/Column Masking}

Idea: each iteration finds the global max over the entire matrix, selects it, then masks the row and column of the selected entry (set to \texttt{-$\infty$}); repeat N times. This is the same as §4.2.5 Path A and §7 spectral scheduling discretization. The masking is irreversible (there is no Hungarian augmentation or auction cancellation), so the result may be suboptimal for $N \ge 3$. Full procedure and worked example in §4.3 graph-theoretic formulation; complexity in B.4.

\subsection*{B.4 Why Exact Matching is Hard to Implement On-Chip}

Commercial switch chips almost never use Hungarian/auction/serial masking greedy for per-slot/per-cycle fabric scheduling; the reasons are summarized below.

\textbf{Algorithm form: iterative, not single-cycle closed-form}

\bigskip\noindent
\par\leftskip=2em\relax\noindent \fittab{\begin{tabular}{p{\dimexpr 0.333\linewidth-2\tabcolsep}p{\dimexpr 0.333\linewidth-2\tabcolsep}p{\dimexpr 0.333\linewidth-2\tabcolsep}}
\toprule
Approach & What it computes & Why it is unsuitable for a 2--4 ns single cycle \\
\midrule
Hungarian (B.1) & Row/column potentials + augmentation & O(N\textsuperscript{3}) serial; each round depends on the full global state from the previous round \\
Auction (B.2) & Bidding + price $\pi$\_j & Indeterminate iteration count; $\epsilon$ and price feedback \\
Masking greedy (B.3) & max \ensuremath{\rightarrow} mask \ensuremath{\rightarrow} max again & N compare-feedback rounds \ensuremath{\rightarrow} timing violation \\
\bottomrule
\end{tabular}}\par\leftskip=2em\relax{}

\bigskip

\textbf{Data requirements:} must read the full matrix weights (8--10 bits, etc.), not 0/1 Requests; at $N=64$, approximately $64^2 W$ bits/cycle.

\textbf{Hardware structure:} centralized maintenance of potentials/prices/augmentation state; an update to any (i,j) affects the global state.

\textbf{Operation type:} residual min/max, price accumulation, etc.; inconsistent with on-chip preference for ``compare + shift.''

\textbf{Complexity (rough estimate for $N=64$):} $O(N^3) \approx 2.6\times 10^5$ basic operations (with serial dependencies).

\section*{Appendix C. Non-Square Crossbar (M$\times$N, M \textgreater  N)}

The main text throughout assumes an N$\times$N IQ+VOQ. When M$\neq$N, two common approaches are used: (i) \textbf{Pad to a square matrix:} add dummy output columns to extend Q to M$\times$M; run Sinkhorn+BvN on the extended matrix and ignore dummy columns during scheduling; (ii) \textbf{Generalized doubly stochastic:} directly define $B_{M\times N}$ (column sums = 1, row sums $\leq$ 1) and decompose by the Von Neumann theorem into a convex combination of partial permutation matrices. When M\textgreater N, at least M-N inputs are idle per cycle, which can cause HOL-like problems.

\section*{Generative AI Use Statement}

During the preparation of this work, the authors used Cursor (a commercial IDE with integrated large-language-model assistants). The tools assisted with: drafting and revising tutorial exposition (§2--§8, appendices, and revision response sorting); bilingual Chinese--English editing; manuscript structure (section renumbering, cross-references, terminology consistency, and formatting); literature and notation cross-checks (suggested wording only; each citation was verified by the authors against primary sources); and auxiliary support for the public experiment repository (C++/Python scripts and plotting workflows in Code and Data Availability). Research questions, algorithm settings, simulation parameters, and numerical results presented in §7 were independently defined, executed, and interpreted solely by the authors; LLM-generated text or code was not incorporated without human review. Generative AI tools are not listed as authors. After using these tools, the authors reviewed and edited the manuscript and code as needed and take full responsibility for the content of this publication.

\mwmsection*{Code and Data Availability}

The C++ model, test files, and Python plotting scripts for reproducing all experiments in §7 of this paper are publicly available under the MIT License:

\begin{itemize}
\item Code repository (read-only): \href{https://github.com/xiaotongyuan/mwm-islip-experiment}{https://github.com/xiaotongyuan/mwm-islip-experiment} (Release v1.0.0)
\item Permanent archive: \href{https://zenodo.org/records/20472143}{https://zenodo.org/records/20472143} (DOI: \texttt{10.5281/zenodo.20472143})
\end{itemize}

The raw result files (\texttt{main\_results.csv}, \texttt{iteration\_sweep.csv}) and Figures 3--5 are not uploaded to the repository, but can be reproduced from the fixed 20 random seeds via
\texttt{make run \&\& make run-softassign \&\& python py/plot\_results.py} (see repository README for details). Experiment configuration in §7.2; implementation details in §7.5.

\clearpage
\thispagestyle{empty}
\vspace*{2cm}
\begin{center}
{\LARGE\bfseries Chinese Version / 中文版}\\[1.5em]
{\large Original manuscript language: Chinese}\\[0.5em]
{\normalsize The English translation appears in Part I above.}
\end{center}
\vfill
\clearpage
\renewcommand{\theHsection}{cn\arabic{section}}\renewcommand{\theHsubsection}{cn\arabic{section}.\arabic{subsection}}\renewcommand{\theHsubsubsection}{cn\arabic{section}.\arabic{subsection}.\arabic{subsubsection}}\setcounter{section}{0}\section{介绍}

\subsection{背景与动机}

在片上互连架构中，N$\times$N 集中式 crossbar 需要在单个时钟周期内完成高速仲裁与通路匹配，是片上数据传输的核心组件。仲裁的核心任务为筛选合法的输入输出匹配对，每个周期内每组端口仅允许单个 flit 的数据传输。硬件的整体队列状态可以用队列矩阵 Q(t) 表示，每周期仲裁选择可以表示为 0/1 二元匹配矩阵 P(t)。受制于较小迭代次数的硬件限制，现有工业迭代算法均为最优匹配的近似求解方案，无法部署匈牙利算法等复杂度极高的最优匹配求解策略。因此，本文核心研究动机为：在固定迭代次数的公平对比条件下，量化分析各类近似调度算法对 MWM 最优性能的逼近能力，以及各自面对的局限性。

交换机 IQ 调度理论的标准化模型已经得到长期研究，时隙内的调度本质为求取二分图合法匹配的最优解。传统的单输入先入先出队列存在严重队头阻塞（HOL）问题 [1],[13]，而虚拟输出队列（VOQ）架构通过为每一组 (i,j) 端口对配置独立队列，从架构上消除 HOL [4],[14]。经典理论表明，基于 VOQ 的 MWM 调度可实现 100\% 吞吐性能 [4],[7]，但算法复杂度极高，无法适配实时仲裁场景。

为平衡硬件开销与调度性能，iSLIP [3] 等低复杂度的迭代算法成为工业主流方案。此类算法仅依靠查询队列非空状态完成轮询迭代，硬件成本低、时序收敛性好，在均匀可容许流量下可保障队列稳定；在特定非均匀场景下性能会退化。与其它近似调度方案比较时，需要在受控实验中量化边界。

\subsection{三类研究脉络}

当前 crossbar 仲裁与 IQ 调度在文献里常割裂为三类独立的研究脉络，缺少跨领域可对照写法。

理论研究脉络：现有基于排队论与 Lyapunov 稳定性的研究，侧重于稳态与吞吐分析，系统刻画 MWM 策略的全局最优性与吞吐上界 [7],[8],[9]。但是纯理论分析普遍忽略硬件周期迭代的实际约束。

工程实现脉络：iSLIP 等迭代算法聚焦 RTL 可实现性，是成熟的硬件架构 [3]，但是通常依赖仿真评估性能，缺少用数学语言对其近似机理与性能边界进行的解释。

片上网络研究脉络：SoC/NoC 领域聚焦 flit 传输、crossbar 架构与缓存深度设计，术语与数据通信领域不一致，同时同样缺少 MWM 最优理论、迭代近似算法的数学解释。

总结，纯数学教程通常不强调 crossbar 的单时钟周期内少量迭代次数的硬件约束，纯 SoC 文档则少有 MWM 稳定性的数学解释。要把 iSLIP 或谱分析引入 arbiter 设计，需要在上述脉络之间建立一套可对照的数学表述。

\subsection{主要贡献}

\begin{itemize}
\item \textbf{统一的数学语言}：以队列矩阵 Q、匹配矩阵 P、内积 ⟨Q,P⟩ 与 Lyapunov 能量 V=‖Q‖\textsuperscript{2} 为语言，把 MWM 的调度目标与稳定性、连续松弛（Sinkhorn 迭代/BvN 分解、谱调度）与 iSLIP 的 0/1 迭代纳入同一套代数框架下比较，缝合上一节中提到的三条割裂的研究脉络。
\item \textbf{可手算复现的推导链路}：从 MWM 调度目标和可容许流量下的稳定性条件出发，推进到 iSLIP 的 Grant--Accept 机制与跨周期平均服务矩阵 P̄，整个过程配以可手算复现的示例。
\item \textbf{易混概念的统一对照}：将全文出现的多种「吞吐」百分比（HOL 架构上限、极大匹配近似比、admissible 稳定意义下的 100\% 等）集中给出对照定义，以作区分。
\item \textbf{公平预算下的定量对照与代价分析}：在 r=3、r\_sink=10 的迭代次数下，以 MWM 为性能参考基准对比 iSLIP、谱调度与熵正则 OT 的吞吐与时延，并完成算术运算开销与队列读取位宽的分析级 Pareto 比较。
\end{itemize}

\subsection{文章结构}

全文按线性代数建模、图论原理、迭代算法的工程实现三大主线展开，具体结构如下：

\bigskip\noindent
\par\leftskip=2em\relax\noindent \fittab{\begin{tabular}{p{\dimexpr 0.333\linewidth-2\tabcolsep}p{\dimexpr 0.333\linewidth-2\tabcolsep}p{\dimexpr 0.333\linewidth-2\tabcolsep}}
\toprule
主线 & 核心问题 & 对应章节 \\
\midrule
I. 线性代数 & 系统建模、MWM 理论来源与稳定性，连续松弛（缩放与谱）向离散匹配的落地逻辑 & §2，§3，§4.1、§4.2 \\
II. 图论 & 精确 MWM 的图论表述、与工业实现的衔接 & §4.3，附录 B \\
III. iSLIP & 0/1 迭代近似的原理及稳定性 & §5 \\
IV. 分类、实验与结论 & 算法分类、指标对比、对比实验、结论与未来方向 & §6--§8（非方阵扩展见附录 C） \\
\bottomrule
\end{tabular}}\par\leftskip=2em\relax{}

\bigskip

\subsection{预备知识与符号约定}

读者需具备线性代数入门（向量、矩阵乘法、内积、特征值）及基本概率论的知识背景。

核心符号：

\bigskip\noindent
\par\leftskip=2em\relax\noindent \fittab{\begin{tabular}{p{\dimexpr 0.500\linewidth-2\tabcolsep}p{\dimexpr 0.500\linewidth-2\tabcolsep}}
\toprule
符号 & 含义 \\
\midrule
Q(t) & N$\times$N 矩阵；Q\_ij(t) = (i,j) 在时刻 t 的 backlog（积压的待发送 cell，见下行）；片上即指 VOQ / egress queue 深度 \\
backlog & (i,j) 方向上积压的、待 crossbar 匹配转发的 packet/cell；以 Q\_ij 计数。Q\_ij\textgreater 0 即有 backlog、可发 Request \\
P(t) & 本 cycle 0/1 grant 矩阵；每行、每列至多一个 1 \\
P̄ & 时间平均服务矩阵；P̄\_ij = E[P\_ij] \\
$\Lambda$ & 流量/到达率矩阵；$\lambda$\_ij 为 (i,j) 方向平均到达率 \\
$\lambda$ & 到达向量（$\Lambda$ 展平） \\
$\rho$($\Lambda$) & 流量矩阵 $\Lambda$ 的谱半径 \\
$\rho$\_load & 归一化负载因子（每个输入端口的利用率，实验中的扫描量）；与谱半径 $\rho$($\Lambda$) 区分 \\
MWM & maximum weight matching；每步 max ⟨Q,P⟩ \\
MSM & 极大匹配类调度（maximal matching schedulers）：iSLIP 等单周期只求极大匹配、而非全局 MWM 最优 \\
IQ / VOQ & input-queued switch；virtual output queue \\
iSLIP & McKeown 的迭代式 round-robin 匹配算法 [3] \\
HOL & head-of-line，头阻塞 \\
R--G--A & Request--Grant--Accept \\
admissible & 可容许流量 \\
\bottomrule
\end{tabular}}\par\leftskip=2em\relax{}

\bigskip

\needspace{20\baselineskip}
\subsection{术语对照}

下表对齐片上网络与数据通信领域的概念：

\bigskip\noindent
\par\leftskip=2em\relax\noindent \fittab{\begin{tabular}{p{\dimexpr 0.333\linewidth-2\tabcolsep}p{\dimexpr 0.333\linewidth-2\tabcolsep}p{\dimexpr 0.333\linewidth-2\tabcolsep}}
\toprule
片上 / NoC & Datacom / IQ switch（正文与文献） & 备注 \\
\midrule
flit（burst=1） & cell / packet & 每 cycle 至多传 1 个基本单位的数据 \\
master / slave IP；initiator / target & input / output port & N$\times$N crossbar 假设 \\
crossbar matrix；集中式 crossbar & switch fabric；IQ crossbar & 非 mesh NoC 路由 \\
egress queue per (i,j)；Q & VOQ 长度 Q & N/A \\
grant / arbitration / scheduler & switch scheduling；matching & iSLIP 仍称为 scheduling \\
链路带宽；cycle 内可完成仲裁次数 & 端口额定带宽；每时隙迭代次数 & 实验：对齐 r 次 \\
饱和吞吐、反压 & throughput；admissible $\Lambda$；可容许 & N/A \\
\bottomrule
\end{tabular}}\par\leftskip=2em\relax{}

\bigskip

下文中与硬件实现相关的用语解释如：每周期、无权重感知、bit-plane matching、行/列掩码等，请见附录 A。

\subsection{相关工作}

现有 IQ 交换机调度研究可按队列读取精度与单周期迭代次数分类，详见 §6.2 与 §6.4 表 2。

\bigskip\noindent
\par\leftskip=2em\relax\noindent \fittab{\begin{tabular}{p{\dimexpr 0.250\linewidth-2\tabcolsep}p{\dimexpr 0.250\linewidth-2\tabcolsep}p{\dimexpr 0.250\linewidth-2\tabcolsep}p{\dimexpr 0.250\linewidth-2\tabcolsep}}
\toprule
主题 & 代表工作 & 核心思想 & 在本文中的位置 \\
\midrule
Max-weight / Lyapunov 稳定性 & Tassiulas--Ephremides [7]；McKeown 等 100\% 吞吐 [4] & 基于队列内积的最大权重优化与 Lyapunov 稳定性；VOQ 架构可在容许流量下实现满吞吐 & §3.2--§3.4 重述，并衔接 BvN 松弛理论 [8],[9] \\
迭代匹配调度（工业主流） & iSLIP [3]；McKeown 博士论文 [6]；文献对比研究 [16] & 基于 R-G-A 迭代与轮询指针的离散匹配机制，以低开销换取可实现性 & §5 核心理论；§7 仿真对比基准 \\
iSLIP 同族变体 & DRRM [17]；FIRM [19]；SRR/DRDSRR [20],[21]；无饿死 MSM [15] & 通过双指针、公平约束等策略优化时延与公平性 & §5.7.1 算法变体拓展与归纳 \\
全局队列随机调度 & Tassiulas [22]、APSARA/SERENA [23],[24],[25]；随机化评估 [26] & 利用全局队列信息与随机策略实现低复杂度权重匹配 & §6.2、§8 \\
分布式渐进最优匹配 & $\epsilon$-Auction / min-sum [41] & 多轮迭代逐步收敛至精确最大权重匹配 & §6.2、§8 \\
连续松弛离散匹配 & Sinkhorn [12]、BvN 调度 [31]--[35] & 矩阵双随机松弛与 BvN 置换分解 & §4.1 理论；§7 另设熵正则 OT exp(Q/$\epsilon$) 后 r\_sink 轮归一化实验基线 \\
交换架构扩展 & CIOQ [27],[28],[29]、CICQ/LIPS [36],[38],[39] & 通过内部队列架构改造缓解全局匹配压力 & §6.2、§8 \\
可编程队列调度 & PIFO [40] & 可编程数据平面队列调度机制 & §8 未来研究方向拓展 \\
\bottomrule
\end{tabular}}\par\leftskip=2em\relax{}

\bigskip

{\small\textbf{表 1。}相关工作对照（节选）}

\section{问题背景与系统模型}

\subsection{HOL 阻塞、58.6\% 上限与 VOQ 动机}

经典空分交换中，若每个输入端口仅维护单一 FIFO（无 per-output 队列结构），当队首报文指向已被占用的输出时，会阻塞同端口上所有后续报文，即 head-of-line（HOL）阻塞 [1],[13]。Karol 等分析得到：在大 N 极限下，这种输入 FIFO 架构的饱和吞吐约为 2-√2 $\approx$ 58.6\%（架构上限，非调度器算法本身）。输出队列带宽利用率可接近 100\%，但最坏情况下 N 个输入同时指向同一个 output port，每个输出缓冲需要 N $\times$ 端口速率的写带宽，实现代价过高。

VOQ [4],[14] 为每个 (i,j) 维护 Q\_ij，从架构上消除 HOL，调度问题回归到为每周期在匹配集内选 P(t)。在可容许流量下，以 Q\_ij 为权重的 MWM 算法可以达到 100\% 吞吐 [4],[7]。本文默认 N$\times$N IQ + VOQ + 单 crossbar 结构，与 SoC 的 (i,j) egress queue 的设定同构。

\subsection{系统模型}

考虑 N$\times$N IQ switch：每个 input port i 到 output port j 维护独立 VOQ [4]，Q\_ij(t) 为 (i,j) 的 backlog。调度器每 clock cycle t 选一组合法匹配 P(t)（0/1 grant 矩阵，每行、每列至多一个 1），决定本 cycle 转发的 cell（burst=1 时为 1 flit）。

下文沿三条主线展开：（I）线性代数建模、MWM 导出与稳定性；（II）图论精确算法及硬件约束；（III）iSLIP 工业近似及其长期平均意义下的统计性质。

\section{MWM 的代数原理与稳定性}

这章建立全文的代数与稳定性基础，回答两个问题：MWM 为什么可作为数学上的调度目标，以及在什么流量条件下系统能保持稳定。§3.1 给出排队系统的线性代数模型（队列矩阵 Q、流量 $\Lambda$、匹配 P 与演化方程）；§3.2 定义 MWM 及其权重变体；§3.3 用 Lyapunov 能量函数解释 MWM 与系统能量下降之间的关系，是稳定性分析的代数来源；§3.4 借助特征值与 Birkhoff--von Neumann 定理刻画可容许流量的范围，并说明"可容许范围内 $\neq$ 任意调度都稳定"。其中双随机矩阵、Birkhoff 多面体与 BvN 定理（§3.4）也是 §4 连续松弛算法的数学基础。

\subsection{排队系统的线性代数建模}

\subsubsection{状态向量}

把 N$\times$N 个队列长度展平（或保持矩阵形式），得到状态：
\begin{equation}
Q(t)\in\mathbb{R}^{N^2}\ (\text{展平为向量})\quad\text{或}\quad Q(t)\in\mathbb{R}^{N\times N}\ (\text{保持矩阵}).
\end{equation}

Q(t) 随时间变化：新报文到达时增大，被转发时减小。

\begin{quote}
\textbf{内积与能量记号约定}：下文为书写简洁，对矩阵 Q、P 沿用向量记法 $Q^{\mathsf{T}}P$ 表示二者的标量内积，它指把 Q、P 展平为 $\mathbb{R}^{N^2}$ 向量后的内积，等价于 Frobenius 内积 $\langle Q,P\rangle=\operatorname{tr}(Q^{\mathsf{T}}P)=\sum_{i,j}Q_{ij}P_{ij}$。相应地，能量函数 $V=Q^{\mathsf{T}}Q$ 指 $\langle Q,Q\rangle=\operatorname{tr}(Q^{\mathsf{T}}Q)=\lVert Q\rVert_F^2$（即全文中的 $\lVert Q\rVert^2$）。当 Q、P 视作矩阵时，$Q^{\mathsf{T}}P$、$Q^{\mathsf{T}}Q$ 均按此约定理解为标量，而非矩阵乘积。
\end{quote}

\subsubsection{到达率与服务}

\begin{itemize}
\item 到达率 $\lambda$（向量）或 $\Lambda$（矩阵）：每个 (i,j) 方向上单位时间平均到达的报文数。
\item 服务矩阵 P(t)：本周期实际转发的匹配矩阵。硬件中每个被选中的 (i,j) 恰好转发 1 个 cell/flit，因此 \texttt{P\_ij(t) ∈ \{0,1\}}，且每行、每列至多一个 1。
\end{itemize}

\subsubsection{状态演化方程}

离散时间下，一个时钟周期内的队列更新为 [4],[7]（McKeown 的 IQ/VOQ cycle 模型；Tassiulas--Ephremides 的 max-weight 理论）：
\begin{equation}
Q(t+1) = Q(t) + \lambda - P(t)
\end{equation}

严格随机模型下应为 $Q(t+1)=\max\bigl(Q(t)+A(t) -P(t),\,0\bigr)$，其中 $A(t)$ 为随机到达量、$\mathbb{E}[A(t)]=\lambda$；下文为便于手算与推导，用确定性的 $\lambda$ 代替 $\mathbb{E}[A(t)]$，并暂不考虑队列下溢（长度变负）的情况。

为什么是"减法"而不是 M(t)·Q(t)？

\begin{itemize}
\item 矩阵乘法 M·Q 表示按比例抽走队列内容（如每次处理 20\%）。
\item 交换机硬件是离散的：每个周期每个端口最多转发 1 个固定大小的 cell，这个行为与队列绝对长度无关（下文的匹配选择与列队长度相关，两者是不同的阶段）。
\item 因此用向量减法 \texttt{Q - P} 更准确地对应"每周期减去固定个数"的物理行为。
\end{itemize}

\subsection{最大权重匹配算法}

\subsubsection{权重与优化目标}

调度算法每周期只做"是否传输"的 0/1 决策，但在选择匹配方案时可以引入权重偏好。MWM（Maximum Weight Matching） [4] 把队列长度本身作为权重：
\begin{equation}
\max_{P\in M}\ \sum_{i,j} Q_{ij}(t)\, P_{ij}(t) \;=\; \max_{P\in M}\ \langle Q(t),\, P(t)\rangle
\end{equation}

其中 M 是所有合法匹配矩阵（置换矩阵或其子集）的集合。\texttt{⟨Q, P⟩} 表示 Q 与 P 的内积（对应元素相乘再求和）。

物理直觉：队列越长的 (i,j) 方向，优先获得本周期的转发机会越大。

\subsubsection{入门算例（2 队列 / 单服务器标量示意）}

\begin{quote}
\textbf{记号说明}：本例仅以简化的"2 队列争用 1 个服务器"标量示意（每周期只发 1 个报文），非 §3.1.1 定义的 N$\times$N 队列矩阵与置换矩阵。
\end{quote}

当前队列向量（输入 1 极度拥堵，输入 2 空闲）：
\begin{equation}
Q = \begin{bmatrix} 10 \\ 2 \end{bmatrix}
\end{equation}

每周期只能转发 1 个报文，两种方案：

\bigskip\noindent
\par\leftskip=2em\relax\noindent \fittab{\begin{tabular}{p{\dimexpr 0.250\linewidth-2\tabcolsep}p{\dimexpr 0.250\linewidth-2\tabcolsep}p{\dimexpr 0.250\linewidth-2\tabcolsep}p{\dimexpr 0.250\linewidth-2\tabcolsep}}
\toprule
方案 & 服务向量 P & 下一时刻 Q & 能量 V = ‖Q‖\textsuperscript{2} \\
\midrule
A：优先队列 1 & \texttt{[1, 0]ᵀ} & \texttt{[9, 2]ᵀ} & 85 \\
B：优先队列 2 & \texttt{[0, 1]ᵀ} & \texttt{[10, 1]ᵀ} & 101 \\
\bottomrule
\end{tabular}}\par\leftskip=2em\relax{}

\bigskip

初始能量：
\begin{equation}
V(0) = 10^2 + 2^2 = 104
\end{equation}

MWM 选方案 A（内积 \texttt{10$\times$1 + 2$\times$0 = 10 \textgreater  2}），能量下降 \texttt{104 - 85 = 19}；选 B 只下降 3。

\textbf{结论}：MWM 在物理含义上就是每一步选择都沿着系统"总能量"下降最快的方向。

\subsubsection{MWM 权重变体：LQF、OCF 与 LPF}

McKeown 等 [4] 在 VOQ 语境下比较了多种 max-weight 权重定义，均属 MWM 族（每步 max ⟨Q, P⟩，权重函数不同）：

\bigskip\noindent
\par\leftskip=2em\relax\noindent \fittab{\begin{tabular}{p{\dimexpr 0.250\linewidth-2\tabcolsep}p{\dimexpr 0.250\linewidth-2\tabcolsep}p{\dimexpr 0.250\linewidth-2\tabcolsep}p{\dimexpr 0.250\linewidth-2\tabcolsep}}
\toprule
权重 & 定义要点 & 100\% 吞吐 & 公平性 / 硬件 \\
\midrule
LQF & Q\_ij 本身（最长队列优先） & ✓ [4] & 可能饿死短队列 \\
OCF & 最旧 cell 优先（Oldest Cell First） & ✓ [4] & 公平性更好 \\
LPF & 在最大匹配集合内选最大端口权 [5] & ✓ [5] & 比 LQF 更易硬件实现 \\
\bottomrule
\end{tabular}}\par\leftskip=2em\relax{}

\bigskip

\subsection{李雅普诺夫稳定性与 MWM 的代数起源}

我们要让队列保持有界，即 ‖Q(t)‖ 不无限膨胀。标准做法是定义一个度量总积压的"能量"函数，研究它何时能下降；这样就把"系统稳不稳"转化成对每周期变化量 $\Delta$V 的一个代数条件。本节先在无到达（§3.3.2）、再在有到达（§3.3.3）的情形下推导 $\Delta$V，并解读该条件蕴含的调度规则。

\subsubsection{能量函数（Lyapunov 函数）}

定义系统总能量为积压的平方：
\begin{equation}
V(t) = Q(t)^{\mathsf{T}} Q(t) = \lVert Q(t)\rVert^2
\end{equation}

这是一个标准的二次型。``能量下降"字面含义就是"队列在排空''。

\subsubsection{无到达流量时的推导（$\lambda$ = 0）}

状态方程简化为 \texttt{Q(t+1) = Q(t) - P(t)}。

展开 V(t+1)：
\begin{align}
V(t+1) &= (Q - P)^{\mathsf{T}} (Q - P) \nonumber \\
&= Q^{\mathsf{T}}Q - 2\,Q^{\mathsf{T}}P + P^{\mathsf{T}}P
\end{align}
\vspace{1pt plus 1pt minus 1pt}
\noindent 能量变化量：
\begin{equation}
\Delta V = V(t+1) - V(t) = -2\,Q^{\mathsf{T}}P + \lVert P\rVert^2
\end{equation}

\needspace{12\baselineskip}
\noindent 分析两项：

\bigskip\noindent
\par\leftskip=2em\relax\noindent \fittab{\begin{tabular}{p{\dimexpr 0.500\linewidth-2\tabcolsep}p{\dimexpr 0.500\linewidth-2\tabcolsep}}
\toprule
项 & 性质 \\
\midrule
‖P‖\textsuperscript{2} & 假设每周期完成完美匹配（N 条边），此项为常数 N \\
-2·QᵀP & 唯一可控项；要令 $\Delta$V 尽可能负，需要最大化 QᵀP \\
\bottomrule
\end{tabular}}\par\leftskip=2em\relax{}

\bigskip

\begin{quote}
\textbf{注}：\texttt{‖P‖\textsuperscript{2} = N} 的前提是每周期均实现完美匹配。若某些端口空闲，因为 \texttt{‖P‖\textsuperscript{2} \textless  N}，$\Delta$V 中的常数项更小，稳定结论不受影响（‖P‖\textsuperscript{2} 更小，对 $\Delta$V$\leq$0 更有利），但更完整的 Lyapunov 推导需对 \texttt{‖P‖\textsuperscript{2}} 用一个固定上界 N 处理。相关处理见 §3.3--§3.4；iSLIP 在均匀 i.i.d. 流量下的稳定性见 McKeown 等 [3]。
\end{quote}

因此：
\begin{equation}
\max_{P\in M} Q^{\mathsf{T}}P \iff \text{MWM}
\end{equation}

\textbf{结论}：MWM 不是凭空设计的，而是从二次型能量下降最快的代数条件中推导出来的 [4],[7]。

\subsubsection{有到达流量时的完整公式}

完整状态方程 \texttt{Q(t+1) = Q(t) + $\lambda$ - P(t)} 代入展开：
\begin{equation}
\Delta V = 2\,Q^{\mathsf{T}}(\lambda - P) + \lVert \lambda - P\rVert^2
\end{equation}

\begin{itemize}
\item Qᵀ$\lambda$ \textgreater  0（当有 backlog 的方向上有到达时）：新流量注入能量，使队列膨胀。
\item -QᵀP \textless  0：调度算法抽走能量，使队列缩短。
\end{itemize}

系统稳定的基本条件是：调度抽走的能量（长期平均）需要压过注入的能量。

\subsubsection{手算：有到达流量时的 $\Delta$V 与注入--服务权衡}

\begin{quote}
\textbf{记号说明}：本例沿用 §3.2.2 的简化：2 路队列争用 1 个服务器 的标量示意（Q、$\lambda$、P 为 2 维向量），非 §3.1.1 定义的 N$\times$N 队列矩阵与置换矩阵；完整 2$\times$2 交换机算例见 §3.4.6。
\end{quote}

设 N=2，某周期队列与流量（展平为向量）：
\begin{equation}
Q = \begin{bmatrix} 100 \\ 50 \end{bmatrix}, \quad
\lambda = \begin{bmatrix} 1 \\ 1 \end{bmatrix}, \quad
P = \begin{bmatrix} 1 \\ 0 \end{bmatrix}
\end{equation}

含义：两路队列都很长；本周期两路各到达 1 个报文；调度只服务了输入 1（MWM 会如此选择）。

\addvspace{22pt plus 4pt minus 2pt}\noindent Step 1. 代入 $\Delta$V 公式
\begin{equation}
\Delta V = 2\,Q^{\mathsf{T}}(\lambda - P) + \lVert \lambda - P\rVert^2
\end{equation}

\addvspace{22pt plus 4pt minus 2pt}\noindent Step 2. 算内积项 Qᵀ($\lambda$ - P)
\begin{align}
&\lambda - P = [0,\ 1]^{\mathsf{T}} \nonumber \\
&Q^{\mathsf{T}}(\lambda - P) = 100\times 0 + 50\times 1 = 50 \ > 0
\end{align}

\ensuremath{\rightarrow} 本周期净效果是注入能量（队列 2 来了包却没被服务）。

\addvspace{22pt plus 4pt minus 2pt}\noindent Step 3. 算常数项
\begin{equation}
\lambda - P = [0,\ 1]^{\mathsf{T}} \ \to\ \lVert \lambda - P\rVert^2 = 0^2 + 1^2 = 1
\end{equation}

\needspace{18\baselineskip}

\addvspace{22pt plus 4pt minus 2pt}\noindent Step 4. 合并
\begin{equation}
\Delta V = 2\times 50 + 1 = 101 \ > 0 \quad (\text{能量上升，系统更拥堵})
\end{equation}

单 cycle 至多服务一路，故本例可出现 $\Delta$V\textgreater 0。长期稳定并不要求每个 cycle 都有 $\Delta$V\textless 0。要求的是长期平均上 E[QᵀP] 足以抵消 Qᵀ$\lambda$ 的注入；当 Q 很大时，MWM 会把服务持续倾斜到最长队列，使平均抽走能量超过注入。

注入--服务权衡分解：
\begin{align}
&Q^\mathsf{T}\lambda = 100\times 1 + 50\times 1 = 150 \quad (\text{注入，正贡献}) \nonumber \\
&Q^\mathsf{T} P = 100\times 1 + 50\times 0 = 100 \quad (\text{抽走，负贡献}) \nonumber \\
&Q^\mathsf{T}(\lambda - P) = 50 \quad (\text{本周期注入胜出})
\end{align}

\subsection{流量矩阵、特征值与 Birkhoff--von Neumann 定理}

\subsubsection{两个不同的矩阵}

\bigskip\noindent
\par\leftskip=2em\relax\noindent \fittab{\begin{tabular}{p{\dimexpr 0.333\linewidth-2\tabcolsep}p{\dimexpr 0.333\linewidth-2\tabcolsep}p{\dimexpr 0.333\linewidth-2\tabcolsep}}
\toprule
矩阵 & 含义 & 由谁决定 \\
\midrule
$\Lambda$（流量矩阵） & \textbf{需求侧}各 (i,j) 方向的平均到达率 $\lambda$\_ij：平均每周期\textit{想}从 i 去 j 的包数（总包数/总时间窗口） & 外部流量，客观存在 \\
P(t)（调度矩阵） & 第 t 周期实际执行的匹配；行/列和始终 $\leq$ 1（crossbar 每周期每端口最多转发 1 个） & 调度算法实时计算 \\
\bottomrule
\end{tabular}}\par\leftskip=2em\relax{}

\bigskip

MWM 优化的是 ⟨Q(t), P(t)⟩，不是 ⟨$\Lambda$, P(t)⟩。 芯片在运行时只能看到当前队列 Q(t)，无法预知未来流量。

\subsubsection{流量矩阵的约束}

$\Lambda$ 是 N$\times$N 非负矩阵。物理上，$\Lambda$\_ij 是\textbf{需求侧}速率：每周期\textit{想}从输入 i 去 output j 的包数（某测量窗口内的计数除以窗口长度）。由于一个输入口可能同时有去往多个输出端的流量，行和 $\Sigma$\_j $\Lambda$\_ij \textbf{可以 \textgreater  1}。这表示了输入 i 被注入的流量超过其线速（每周期 1 包）能承载的量，即该端口过载。\textbf{可容许}（admissible）要求的恰恰是这件事不发生：

\begin{itemize}
\item 每行和 $\leq$ 1（每个输入端口被注入的流量不超过线速）
\item 每列和 $\leq$ 1（每个输出端口被注入的流量不超过其排出速率）
\end{itemize}

两者同时成立时，$\Lambda$ 称为亚双随机矩阵（sub-doubly stochastic），是流量\textit{需求}的统计描述。当某行或某列和 \textgreater  1 时，任何调度都无法使队列有界（见 §3.4.3 的 2$\times$2 过载算例）。注意这是对\textit{被注入/需求}流量的陈述；\textit{实际转发}的 P(t) 及其时间平均 P̄ 的行/列和始终 $\leq$ 1，由 crossbar"每周期每端口最多一个 trans"的物理极限强制保证。

\subsubsection{谱半径与 Perron--Frobenius 定理}

§3.4.2 的可容许条件是 2N 条端口负载不等式（每行和、每列和都 \textless  1）。追问：能不能把这堆不等式压缩成一个数来回答"系统有多堵"？谱半径 $\rho$($\Lambda$)，即最大特征值的模是自然候选。

若 $\Lambda$ 的所有行和、列和严格小于 1，则由矩阵范数界（谱半径不超过任意相容范数）直接得到：
\begin{equation}
\rho(\Lambda) \le \|\Lambda\|_\infty = \max_i \sum_j |\Lambda_{ij}| < 1
\end{equation}

即 可容许（§3.4.2）\ensuremath{\Rightarrow} $\rho$($\Lambda$) \textless  1。对非负矩阵，Perron--Frobenius 定理进一步给出 $\rho(\Lambda)$ 落在最小/最大行和之间，与上界一致。

但反过来不成立：$\rho$($\Lambda$) \textless  1 \textbf{并不}蕴含可容许。IQ switch 的可容许条件本质上是逐端口的，即每行和、每列和都严格小于 1；谱半径只是必要推论，不是容量域的完整刻画。（$\rho$($\Lambda$) 把负载"压在哪"给平均掉了；两个 $\rho$ 相同的矩阵，可能一个某端口过载、一个不过载。）

手算直觉（2$\times$2 对角流量）：
\begin{equation}
\Lambda = \begin{bmatrix} 0.6 & 0 \\ 0 & 0.6 \end{bmatrix}
\qquad \text{行和、列和均为 } 0.6 < 1
\end{equation}

特征值为 0.6，故 $\rho$($\Lambda$)=0.6 \textless  1。物理含义：即使所有端口满载，平均也只用到 60\% 端口容量，系统有"呼吸空间"。

若改成：
\begin{equation}
\Lambda = \begin{bmatrix} 0.9 & 0.3 \\ 0.3 & 0.9 \end{bmatrix}
\qquad \text{行和 } = 1.2 \text{（已过载！）}
\end{equation}

行和超过 1 \ensuremath{\rightarrow} 某个输入端口超出物理注入能力，任何调度都无法长期稳定（队列必溢出）。

\begin{quote}
注意：本文后续的 admissible / 可容许流量默认指输入端口行和、输出端口列和均小于 1。
\end{quote}

\subsubsection{Birkhoff--von Neumann（BvN）定理 [8],[9]}

「可容许流量」（§3.4.2）具体是什么意思？BvN 给出标准答案：需求 $\Lambda$ 可以通过若干置换匹配 P\_k 按比例 $\alpha$\_k 分时复用来承载。这个存在性是稳定性的脊柱：存在一个长期平均调度，使每个方向至少按其到达率被服务；调度器的任务就是去逼近它。

任意双随机矩阵 B（元素非负，行和 = 列和 = 1）均可分解为若干置换矩阵的凸组合：
\begin{equation}
B = \alpha_1 P_1 + \alpha_2 P_2 + \dots + \alpha_m P_m \quad (\alpha_k > 0,\ \textstyle\sum_k \alpha_k = 1)
\end{equation}

针对 N 阶方阵，分解所需置换矩阵的数量满足 m $\leq$ N\textsuperscript{2} - 2N + 2（BvN 分解所需置换矩阵数量的上界；Birkhoff 多面体顶点数为 N!）。

对亚双随机流量矩阵 $\Lambda$（元素非负，行和、列和均不超过 1），BvN 分解说明存在一组置换矩阵 \{P\_k\} 与对应权重 $\alpha$\_k（$\Sigma$ $\alpha$\_k $\leq$ 1），使得 $\Lambda$ 可由上述矩阵的凸组合表示或支配（dominated）：
\begin{equation}
\Lambda \le \alpha_1 P_1 + \alpha_2 P_2 + \dots + \alpha_m P_m
\end{equation}

记 S 为该凸组合，则对所有 (i,j) 都有 S\_ij $\geq$ $\Lambda$\_ij。S 是合法的平均调度仲裁组合；它只说明平均意义下流量容量条件可以满足，不规定每个 cycle 实际的调度矩阵 P(t)。

流量范围（BvN 存在性结论）：上述关系可以用如下存在性语言表达，当 $\Lambda$ 满足可容许端口负载约束时，存在亚双随机矩阵 S = $\Sigma$ $\alpha$\_k P\_k（$\Sigma$ $\alpha$\_k $\leq$ 1）使
\begin{equation}
S \ge \Lambda
\end{equation}

而调度器需要满足的条件是： 当队列足够长时，让内积形式的负漂移条件成立：
\begin{equation}
E[Q^{\mathsf{T}}P] > Q^{\mathsf{T}}\lambda \iff E[Q^{\mathsf{T}}(\lambda - P)] < 0
\end{equation}

其中 $\lambda$ 为由 $\Lambda$ 定义的到达向量（记号与 §3.2 保持一致），期望算子 E[·] 的说明参见 §5.6.4。此时 Lyapunov 能量函数 V = ‖Q‖\textsuperscript{2} 持续递减，队列长度能够保持有界。

MWM 在每个调度周期内，针对当前队列状态 Q(t) 求解 max ⟨Q,P⟩。由于该策略在每一状态下都选取当前最优匹配，同状态下其单步内积取值，不会低于任意固定凸组合 S。因此从长期统计角度，MWM 能够满足前述负漂移不等式，保障系统稳定。

直觉上：$\Lambda$ 是注入强度，可容许负载条件（§3.4.2）给其封顶；调度器让 grant 在 ⟨Q,P⟩ 维度上优先服务最拥塞的队列，从而控制队列水位。

\subsubsection{几何图像}

在 N\textsuperscript{2} 维队列空间中，V(Q) = ‖Q‖\textsuperscript{2} 构成高维抛物面（``碗''）。只要 $\Lambda$ 位于可容许流量区域内部，且调度器足够强（如 MWM），MWM 的负反馈总能把队列向量拉回碗底附近，队列长度有界。

\subsubsection{可容许流量 $\neq$ 任意调度都稳定}

常见误解：既然 $\Lambda$ 的输入/输出端口负载都小于 1，总容量够，那任意选取 P、甚至每周期随机匹配，只要时间足够长总会收敛。

纠正：

\bigskip\noindent
\par\leftskip=2em\relax\noindent \fittab{\begin{tabular}{p{\dimexpr 0.500\linewidth-2\tabcolsep}p{\dimexpr 0.500\linewidth-2\tabcolsep}}
\toprule
说法 & 对错 \\
\midrule
行和、列和均 \textless  1 \ensuremath{\Rightarrow} 存在某种调度使系统稳定 & 正确 \\
行和、列和均 \textless  1 \ensuremath{\Rightarrow} 任意 P(t) 都稳定 & 错误 \\
``时间够长''\ensuremath{\Rightarrow} 队列一定变成 0 & 错误（$\lambda$\textgreater 0 有持续到达时，队列不会排空至 0） \\
``时间够长''\ensuremath{\Rightarrow} 队列有界 & 错误；仅对具备队列反馈的调度策略成立 \\
\bottomrule
\end{tabular}}\par\leftskip=2em\relax{}

\bigskip

可容许（admissible） 的含义是：存在一组匹配速率的凸组合能在平均意义下承载 $\Lambda$，不是``随便怎么分配服务都行''。

稳定性的 Lyapunov 条件（Q 很大时）是：
\begin{equation}
E[Q^{\mathsf{T}}P] > Q^{\mathsf{T}}\lambda
\end{equation}

手算反例（2$\times$2）：可容许，但无权重随机匹配时不稳定

调度策略是无权重的随机匹配：每周期在 2$\times$2 的两个完美匹配 I、X 中均匀抽取，概率不随 Q\_ij 数值变化（不实现 MWM 式加权）。

\addvspace{22pt plus 4pt minus 2pt}\noindent Step 0. 流量矩阵 $\Lambda$（可容许，四向均有到达）
\begin{equation}
\Lambda = \begin{bmatrix} 0.85 & 0.08 \\ 0.07 & 0.08 \end{bmatrix}
\qquad \text{行和 } 0.93 / 0.15;\ \text{列和 } 0.92 / 0.16 \;\ensuremath{\Rightarrow}\; \rho(\Lambda) < 1
\end{equation}

含义：(1,1) 为主热流（0.85）；交叉 (1,2)、(2,1) 与 (2,2) 也有小流量，长期运行后四个 VOQ 均可非空。

\addvspace{22pt plus 4pt minus 2pt}\noindent Step 1. 当前队列状态 Q（典型拥堵态）
\begin{equation}
Q = \begin{bmatrix} 8000 & 60 \\ 50 & 40 \end{bmatrix}
\end{equation}

Q\_11 极长；(1,2)、(2,1)、(2,2) 都有 backlog，不存在 Q\_ij = 0 的"空 VOQ"。

\addvspace{22pt plus 4pt minus 2pt}\noindent Step 2. 两个完美匹配与随机规则

2$\times$2 每周期只能在两个完美匹配中选择一个（I 与 X 均为完整调度，不能同时使用）：
\begin{equation}
I = \begin{bmatrix} 1 & 0 \\ 0 & 1 \end{bmatrix}
\quad (\text{服务 } (1,1)+(2,2),\ \text{直通})
\qquad
X = \begin{bmatrix} 0 & 1 \\ 1 & 0 \end{bmatrix}
\quad (\text{服务 } (1,2)+(2,1),\ \text{交叉})
\end{equation}

每周期 P(I) = P(X) = 0.5，与 Q 的数值无关。

\bigskip\noindent
\par\leftskip=2em\relax\noindent \fittab{\begin{tabular}{p{\dimexpr 0.167\linewidth-2\tabcolsep}p{\dimexpr 0.167\linewidth-2\tabcolsep}p{\dimexpr 0.167\linewidth-2\tabcolsep}p{\dimexpr 0.167\linewidth-2\tabcolsep}p{\dimexpr 0.167\linewidth-2\tabcolsep}p{\dimexpr 0.167\linewidth-2\tabcolsep}}
\toprule
本周期匹配 & P\_11 & P\_12 & P\_21 & P\_22 & 本例 Q 下实际服务的 VOQ \\
\midrule
I & 1 & 0 & 0 & 1 & Q\_11=8000，Q\_22=40 \\
X & 0 & 1 & 1 & 0 & Q\_12=60，Q\_21=50（都有 backlog，非空转） \\
\bottomrule
\end{tabular}}\par\leftskip=2em\relax{}

\bigskip

长期平均：
\begin{equation}
E[P_{11}] = E[P_{12}] = E[P_{21}] = E[P_{22}] = 0.5
\end{equation}

关键：即使 Q\_11 = 8000，E[P\_11] 仍固定 0.5，不会随 Q\_11 增大。无权重随机匹配不会在高 backlog 的 (1,1) 上多分配带宽。

\addvspace{22pt plus 4pt minus 2pt}\noindent Step 3. 内积判据：E[QᵀP] \textless  Qᵀ$\lambda$（服务赤字 \ensuremath{\rightarrow} 不稳定）

将 Q 视为本段时间内近似常数：
\begin{align}
Q^{\mathsf{T}}\lambda &= 8000\times 0.85 + 60\times 0.08 + 50\times 0.07 + 40\times 0.08 \nonumber \\
&= 6800 + 4.8 + 3.5 + 3.2 = 6811.5 \quad (\text{注入}) \nonumber \\
E[Q^{\mathsf{T}}P] &= 8000\times 0.5 + 60\times 0.5 + 50\times 0.5 + 40\times 0.5 \nonumber \\
&= 4000 + 30 + 25 + 20 = 4075 \quad (\text{期望抽走}) \nonumber \\
E[Q^{\mathsf{T}}P] - Q^{\mathsf{T}}\lambda &= 4075 - 6811.5 = -2736.5 < 0
\end{align}

上式「\textless  0」指（期望服务内积 - 注入内积）\textless  0，即单步平均 grant 弱于到达（服务赤字），(1,1) 等方向上队列长度漂移向上、系统不稳定。

Q\_11 主注入项（6800 / 6811.5），但 E[P\_11] 被固定为 0.5 \ensuremath{\rightarrow} 高 backlog 方向长期服务不足 \ensuremath{\rightarrow} Q\_11 漂移向上，无限膨胀。

\addvspace{22pt plus 4pt minus 2pt}\noindent Step 4. 单周期 vs 期望

\bigskip\noindent
\par\leftskip=2em\relax\noindent \fittab{\begin{tabular}{p{\dimexpr 0.333\linewidth-2\tabcolsep}p{\dimexpr 0.333\linewidth-2\tabcolsep}p{\dimexpr 0.333\linewidth-2\tabcolsep}}
\toprule
本周期选择 & 对 (1,1) 的即时服务 & 系统其它行为 \\
\midrule
I & Q\_11·P\_11 = 8000（本周期足够） & 同时服务 (2,2) \\
X & Q\_11·P\_11 = 0（本周期不服务 (1,1) 的 backlog） & 服务 (1,2)、(2,1)：有 backlog、非浪费，但抢占了本来可用于 (1,1) 的 cycle \\
\bottomrule
\end{tabular}}\par\leftskip=2em\relax{}

\bigskip

不稳定来自期望：E[P\_11] = 0.5 \textless  $\lambda$\_11 = 0.85，而非``每一拍都选 X''。选 X 在本例中是合理的服务非空 VOQ，但在 $\Lambda$ 偏 (1,1) 时比例错误。

\addvspace{22pt plus 4pt minus 2pt}\noindent Step 5. 结论
\begin{align}
\lambda_{11} = 0.85,\ E[P_{11}] = 0.5 &\ \to\ (1,1)\ \text{不稳定，}Q_{11}\to\infty \nonumber \\
\lambda_{22} = 0.08,\ E[P_{22}] = 0.5 &\ \to\ (2,2)\ \text{服务过剩，稳定} \nonumber \\
\lambda_{12} = 0.08,\ E[P_{12}] = 0.5 &\ \to\ \text{略过剩} \nonumber \\
\lambda_{21} = 0.07,\ E[P_{21}] = 0.5 &\ \to\ \text{略过剩}
\end{align}

不稳定是按方向的：即使总容量够，但是无权重随机匹配不能在 (1,1) 的 backlog 变长时重新分配匹配概率。对照之下，MWM 在相同 Q 下会几乎每周期都选 I（Q\_11 权重最大），因此不会陷入 E[P\_11]=0.5 的固定分配。

小结：本文默认的是 ``可容许流量（行和、列和均 \textless  1）+ 具备队列反馈的调度策略 \ensuremath{\Rightarrow} 稳定''；不是 ``物理极限以内任意调度都收敛''。

\section{MWM 的计算：线性代数与图论}

本章把 §3 建立的代数语言落到"如何真正算出本周期的匹配 P"。MWM 的目标 max\_\{P∈M\} ⟨Q,P⟩（§3.2）在数学上明确，可归为两类方法：（I）连续松弛再离散化。先把离散匹配放宽为连续矩阵、求解后再投影回 0/1；实现方法为缩放方法（§4.1，Sinkhorn\ensuremath{\rightarrow}双随机\ensuremath{\rightarrow}BvN）与谱方法（§4.2，幂迭代 / SVD 主成分）。（II）图论精确匹配（§4.3）。不经过连续松弛，直接在匹配集 $\mathcal{M}$ 上求最优组合。其中松弛方法所用的双随机矩阵、Birkhoff 多面体与 BvN 定理已在 §3.4 建立，是本章的数学基础。

\begin{quote}
两种松弛方法的共同难点不在"求连续解"，而在"如何在单周期、有限迭代内把连续矩阵离散化为合法置换矩阵"。这一步正是 §5 工业 iSLIP 的硬件着力点。
\end{quote}

\subsection{线性代数解法一：缩放与 BvN 分解}

本节给出第一种连续松弛方法：先把队列矩阵 Q 用 Sinkhorn 迭代缩放成双随机矩阵 B，再用 BvN 定理（§3.4.4）把 B 分解为置换矩阵的凸组合，形如 B = $\Sigma$ $\alpha$\_k P\_k 的分解，再将系数 $\alpha$\_k 理解为各置换 P\_k 的时分复用比例。

\subsubsection{Sinkhorn--Knopp 迭代（连续松弛）[12]}

将正矩阵 Q 交替做行归一化、列归一化，收敛到双随机矩阵 B。之所以要交替：只做行归一化能让每行和为 1，但列和通常 $\neq$ 1（多个输入仍会倾向同一输出，违反每列至多一个匹配）；只做列归一化则反之。Sinkhorn 的做法是反复交替行、列归一化，使行和与列和同时逼近 1，收敛到双随机矩阵 B。

算例：
\begin{equation}
Q = \begin{bmatrix} 8 & 2 \\ 3 & 5 \end{bmatrix}
\end{equation}

第 1 轮  行归一化（每行除以行和）：
\begin{equation}
B^{(1)} = \begin{bmatrix} 0.8 & 0.2 \\ 0.375 & 0.625 \end{bmatrix}
\end{equation}

第 1 轮  列归一化（每列除以列和）：
\begin{equation}
B^{(2)} = \begin{bmatrix} 0.681 & 0.242 \\ 0.319 & 0.758 \end{bmatrix}
\end{equation}

继续迭代至收敛；精确极限约为 $\begin{bmatrix}0.72 & 0.28 \\ 0.28 & 0.72\end{bmatrix}$（$B^{(2)}$ 行和尚未为 1，说明仅迭代两轮尚未收敛）。以下取整为 0.7/0.3 便于后续 BvN 剥离手算：
\begin{equation}
B \approx \begin{bmatrix} 0.7 & 0.3 \\ 0.3 & 0.7 \end{bmatrix}
\qquad (\text{行和、列和均为 } 1)
\end{equation}

\subsubsection{BvN 剥离}

2$\times$2 情况下，合法置换矩阵只有两个：
\begin{equation}
P_1 = \begin{bmatrix} 1 & 0 \\ 0 & 1 \end{bmatrix} \quad (\text{直通})
\qquad
P_2 = \begin{bmatrix} 0 & 1 \\ 1 & 0 \end{bmatrix} \quad (\text{交叉})
\end{equation}

剥离 P\textsubscript{1}：取 P\textsubscript{1} 支撑集 ⟨0,0⟩、⟨1,1⟩ 上两个 B 元素的最小值 \texttt{$\alpha$\textsubscript{1} = min(0.7, 0.7) = 0.7}。这样 B - $\alpha$\textsubscript{1}P\textsubscript{1} 在支撑集上仍非负数。
\begin{equation}
B_{\text{remain}} = B - 0.7 \cdot P_1 = \begin{bmatrix} 0 & 0.3 \\ 0.3 & 0 \end{bmatrix}
\end{equation}

剥离 P\textsubscript{2}：\texttt{$\alpha$\textsubscript{2} = min(0.3, 0.3) = 0.3}。
\begin{equation}
B_{\text{remain}}' = B_{\text{remain}} - 0.3 \cdot P_2 = \begin{bmatrix} 0 & 0 \\ 0 & 0 \end{bmatrix}
\end{equation}

最终结果：
\begin{equation}
B = 0.7\,P_1 + 0.3\,P_2
\end{equation}

硬件含义：在 10 个时钟周期中，7 个周期执行 P\textsubscript{1}，3 个周期执行 P\textsubscript{2}，长期平均服务能力即为 B。

\textbf{结论}：Sinkhorn+BvN 是 MWM 的连续松弛 / 帧调度近似方法（与 §4.2 谱调度同类），不是精确 MWM；精确 MWM 仍须在匹配集上求最优（§4.3、附录 B.1--B.2、§5.6）。

完整流程：

\par{\leftskip=0pt\relax\noindent\hspace*{2em}\begin{minipage}{\dimexpr\linewidth-2em\relax}
\begin{verbatim}
Q（队列长度）
  → Sinkhorn 迭代 → B（连续双随机矩阵）
  → BvN 分解 → { α_k, P_k}
  → 时分复用调度
\end{verbatim}
\end{minipage}\par}

\subsection{线性代数解法二：谱调度与幂迭代}

线代教材求奇异向量的做法是列特征方程、解高次多项式。交换机芯片单周期内无法完成，但是可以完成"矩阵乘向量"计算，而且可以反复多次计算。幂迭代方法因此就可以被运用在这里：把 Q 与 Qᵀ 反复作用在一个向量上，向量会逐渐对齐到 Q 的主奇异方向，也就是当前拥塞最集中的那一条输入\ensuremath{\leftrightarrow}输出耦合，全程不必显式做 SVD。整个过程就变成了用"很多次硬件友好的矩阵--向量乘法"代替"一次无法实现的特征值求解"，再把得到的方向离散成 0/1 匹配。

\begin{quote}
术语说明：业界没有统一的"Spectral Scheduling"标准名称；本节归纳的是谱方法 + 匹配的常见做法：用主奇异向量 u\textsubscript{1}、v\textsubscript{1} 提供全局特征，结合队列矩阵 Q 的约束边界，最终求解离散匹配矩阵 P。
\end{quote}

\subsubsection{谱调度总流程}

幂迭代与 SVD 的中间量都是实数矩阵或向量；crossbar 每 cycle 需要的是整数 0/1 的 P(t) 调度矩阵（§4 已述）。本节按「求谱 \ensuremath{\rightarrow}（可选）造 W \ensuremath{\rightarrow} 离散成 P」的分步说明，总流程如下：

\bigskip\noindent
\par\leftskip=2em\relax\noindent \fittab{\begin{tabular}{p{\dimexpr 0.333\linewidth-2\tabcolsep}p{\dimexpr 0.333\linewidth-2\tabcolsep}p{\dimexpr 0.333\linewidth-2\tabcolsep}}
\toprule
阶段 & 操作 & 小节 \\
\midrule
求谱 & \texttt{Q(t)} 幂迭代 \ensuremath{\rightarrow} u\textsubscript{1}, v\textsubscript{1} & §4.2.3 \\
路径 A（显式 W，离线/仿真计算常见） & \texttt{W = Q ⊙ (u\textsubscript{1} v\textsubscript{1}ᵀ)} & §4.2.4 \\
 & \ensuremath{\rightarrow} 贪心 / W\ensuremath{\rightarrow}iSLIP / 匈牙利 on W & §4.2.5 \\
路径 B（不构造 W，片上常用） & \texttt{Q·y \ensuremath{\rightarrow} Rounding \ensuremath{\rightarrow} Qᵀ·x \ensuremath{\rightarrow} \dots \ensuremath{\rightarrow} P(t)} & §4.2.3 \\
输出 & \texttt{P(t)}（0/1 匹配矩阵） & --- \\
\bottomrule
\end{tabular}}\par\leftskip=2em\relax{}

\bigskip

\subsubsection{u\textsubscript{1}、v\textsubscript{1} 的含义}

u\textsubscript{1}、v\textsubscript{1} 是 Q 的最大奇异值对应的一对左右奇异向量：u\textsubscript{1} 挂在输入端口侧，v\textsubscript{1} 挂在输出端口侧。它们由整个 Q 矩阵的联合分解计算出。反映的是当前拥塞下输入--输出之间的全局耦合方向，而不是 Q 矩阵各行、各列各自独立的标量队列长度。

\subsubsection{谱调度第一步：幂迭代与谱平滑截断（Power Iteration with Rounding）}

\paragraph{（1）幂迭代：u\textsubscript{1}、v\textsubscript{1} 如何从矩阵乘法里计算出来}

取随机初始向量 y\textsubscript{0}（长度 = 输出端口数 N），交替乘 Q 与 Qᵀ：
\begin{align}
x_1 &= Q\, y_0 \nonumber \\
y_1 &= Q^{\mathsf{T}} x_1 \nonumber \\
x_2 &= Q\, y_1 \nonumber \\
y_2 &= Q^{\mathsf{T}} x_2 \nonumber \\
&\;\;\vdots
\end{align}

每步可对 x、y 做归一化（除以范数），避免数值爆炸。

\textbf{收敛性（与 Eckart--Young 的关系）：} 反复 Q·y / Qᵀ·x 后，x 方向收敛到最大左奇异向量 u\textsubscript{1}，y 方向收敛到最大右奇异向量 v\textsubscript{1}，不必先显式算出 U、$\Sigma$、V。由 Eckart--Young 定理，秩-1 矩阵 $\sigma$\textsubscript{1} u\textsubscript{1} v\textsubscript{1}ᵀ 是 Q 在 Frobenius 范数意义下的最佳秩-1 近似：幂迭代正是在提取这一主成分。

若不加截断、连续跑很多轮，x \ensuremath{\rightarrow} u\textsubscript{1}、y \ensuremath{\rightarrow} v\textsubscript{1} 就是连续向量，仍然需要再离散成 0/1 的 P(t)。

\paragraph{（2）在迭代中间插入离散截断}

谱调度芯片把幂迭代拆成数步，并在中间插入离散截断。初始化 \texttt{y} 后，每轮重复执行下列四步：

\bigskip\noindent
\par\leftskip=2em\relax\noindent \fittab{\begin{tabular}{p{\dimexpr 0.250\linewidth-2\tabcolsep}p{\dimexpr 0.250\linewidth-2\tabcolsep}p{\dimexpr 0.250\linewidth-2\tabcolsep}p{\dimexpr 0.250\linewidth-2\tabcolsep}}
\toprule
步骤 & 公式 & 代数本质 & 硬件含义 \\
\midrule
1 & \texttt{x = Q · y} & 幂迭代半步；x 正在凝聚 u\textsubscript{1} & 各输入把输出侧"压力"y 加权累加 \ensuremath{\rightarrow} 输入端拥堵程度 x \\
2 & \texttt{x \ensuremath{\leftarrow} 截断(x)} & 谱平滑投影 & 将连续值 x 压成 0/1 或稀疏方向（取最大、Grant 等） \\
3 & \texttt{y = Qᵀ · x} & 幂迭代另半步；y 正在凝聚 v\textsubscript{1} & 输出侧根据已离散化的 x 反向聚合 \ensuremath{\rightarrow} 下一轮响应 \\
4 & \texttt{y \ensuremath{\leftarrow} Rounding(y)} & 可选：对称截断 & 同样把 y 投影到离散可行域 \\
\bottomrule
\end{tabular}}\par\leftskip=2em\relax{}

\bigskip

\paragraph{（3）截断的代数图像：连续最优与硬件互斥约束之间的折中}

\bigskip\noindent
\par\leftskip=2em\relax\noindent \fittab{\begin{tabular}{p{\dimexpr 0.500\linewidth-2\tabcolsep}p{\dimexpr 0.500\linewidth-2\tabcolsep}}
\toprule
情形 & 结果 \\
\midrule
不加截断、跑 10 轮 & x $\approx$ u\textsubscript{1}，y $\approx$ v\textsubscript{1}（纯数学，非法调度） \\
每轮中间截断 & x、y 在向 u\textsubscript{1}、v\textsubscript{1} 收敛的过程中满足行/列互斥的离散形态 \\
停下来时 & 既带 SVD 主成分信息，又满足 0/1 互斥 \\
\bottomrule
\end{tabular}}\par\leftskip=2em\relax{}

\bigskip

\begin{itemize}
\item 谱信息分量：幂迭代从 Q 中提取全局主成分信息（同 W = Q⊙(u\textsubscript{1}v\textsubscript{1}ᵀ)）；
\item 离散可行性约束：每步 Rounding 用于满足行/列互斥。
\end{itemize}

\subsubsection{谱调度第二步：对偶掩码（Dual Masking），构造权重矩阵 W}

本节在 §4.2.3 循环内隐式提取谱方向后，介绍 \textbf{路径 A}：显式构造权重矩阵 W。

本方法并非在 u\textsubscript{1}、v\textsubscript{1} 里选取极值，而是利用它们对原矩阵 Q 做加权处理：

\par{\leftskip=0pt\relax\noindent\hspace*{2em}\begin{minipage}{\dimexpr\linewidth-2em\relax}
\begin{verbatim}
W_ij = Q_ij · u_{1,i} · v_{1,j}     （Hadamard 积：W = Q ⊙ (u₁ v₁ᵀ)）
\end{verbatim}
\end{minipage}\par}

\begin{itemize}
\item u\_\{1,i\}·v\_\{1,j\}：谱结构给的 (i,j) 全局关联度得分；
\item Q\_ij：物理掩码 Q\_ij=0 则 W\_ij=0，一票否决空队列。
\end{itemize}

算例 A：Q\_12=0 时谱项再大也无效。先由幂迭代得谱向量（u\textsubscript{1} 在输入侧、v\textsubscript{1} 在输出侧）：
\begin{equation}
Q = \begin{bmatrix} 10 & 0 \\ 3 & 5 \end{bmatrix},\qquad
u_1 \approx [0.93,\ 0.36]^{\mathsf{T}},\quad
v_1 \approx [0.99,\ 0.17]^{\mathsf{T}}
\end{equation}

于是 $u_{1,1}v_{1,2}\approx 0.93\times 0.17\approx 0.16$（非零），但仍被 $Q_{12}=0$ 一票否决：
\begin{align}
W_{12} &= Q_{12} \times (u_{1,1} v_{1,2}) = 0 \times 0.16 = 0 &&\ensuremath{\leftarrow} \text{被否决} \nonumber \\
W_{11} &= 10 \times (u_{1,1} v_{1,1}) \approx 10\times 0.93\times 0.99 \approx 9.2 &&\ensuremath{\leftarrow} \text{有 backlog 才脱颖而出}
\end{align}

算例 B：\texttt{Q = [[8,2],[3,5]]}，先由幂迭代得 $u_1\approx[0.85,\ 0.53]^{\mathsf{T}}$、$v_1\approx[0.89,\ 0.46]^{\mathsf{T}}$，据此算 $W=Q\odot(u_1 v_1^{\mathsf{T}})$：
\begin{equation}
W \approx \begin{bmatrix} 6.05 & 0.78 \\ 1.40 & 1.21 \end{bmatrix}
\end{equation}

构造 W 是 O(N\textsuperscript{2}) 次乘法（硬件可并行）。

\subsubsection{谱调度第三步：从连续权重到 0/1 匹配}

上一步 W 把 Q 与谱主成分合并为有权矩阵；最后一步是在匹配集内选出 0/1 的 P(t)。与 §3.2 相同，只是把权重矩阵从 Q 换为 W：
\begin{equation}
P(t) = \operatorname*{argmax}_{P \in \text{匹配集}}\ \langle W, P\rangle
\end{equation}

继续算例 B，比较两种合法匹配的权重和：
\begin{align}
(1,1)+(2,2)\ \text{权重和} &\approx 6.05+1.21 = 7.26 \nonumber \\
(1,2)+(2,1)\ \text{权重和} &\approx 0.78+1.40 = 2.18 \nonumber \\
&\ensuremath{\Rightarrow}\ \text{仍选直通 } P^* = I\text{，与用 } Q \text{ 做 MWM 计算结果一致}
\end{align}

以下为两种常用方法。

\begin{itemize}
\item 行/列掩码离散化：在 W（或 Q）上投影为合法匹配；§7 谱调度实验采用此类收尾。大 N 时因为片上实现需要全局比较--反馈，时序较难收敛（附录 B.4）。
\item 将 W 作为 iSLIP 初始指针：用 W 给出各行/列拥堵信息后，仍走 §5.2--§5.3 的 R--G--A 与指针规则；行列互斥由 iSLIP 过程保证。
\end{itemize}

\clearpage
\subsection{图论精确解}

\subsubsection{图论表述}

MWM 优化问题与 §3.2 的 $\max_{P\in\mathcal{M}}\langle Q,P\rangle$ 等价于图论中的二分图最大权匹配问题：左节点为输入端口、右节点为输出端口，边 $(i,j)$ 权值为 $Q_{ij}$；可行匹配集合 $\mathcal{M}$ 要求每行、每列至多选取一条边，优化目标为匹配总权重的最大化。软件与 §7 的 MWM 仿真基准通常用多项式时间的匈牙利或拍卖算法 [10],[11] 求精确解。两类算法步骤如下，复杂度见附录 B.4。

\addvspace{8pt plus 2pt minus 1pt}下面以 \texttt{Q = [[8,7,2],[6,3,5],[2,6,4]]} 为公共示例，其最优匹配为 \texttt{(1,1)+(2,3)+(3,2) = 8+5+6 = 19}。

\needspace{4\baselineskip}\addvspace{12pt plus 3pt minus 1pt}\nobreak\textbf{匈牙利算法（Kuhn--Munkres）流程}

\begin{itemize}
\item 为每行、每列各设一个"势"：行势 $\alpha_i$、列势 $\beta_j$（即线性规划里的对偶变量，匹配理论中也称"顶标"）。它给每行、每列预设一个数值，算法始终维持 $\alpha_i+\beta_j\ge Q_{ij}$（行列顶标之和盖住每条边的权重），并通过逐步调整这组顶标来逼近最优匹配。
\item 初始化规则：$\alpha_i=\max\limits_j Q_{ij}$（该行最大权值）、$\beta_j\equiv 0$。
\item 定义约化权： $c_{ij}=\alpha_i+\beta_j -Q_{ij}\ge 0$，只在 $c_{ij}=0$ 的"零边"上寻找匹配。
\item 若零边无法形成完美匹配，沿增广树更新势以引入新零边：树内行的 $\alpha$ 减 $\delta$、树内列的 $\beta$ 加 $\delta$，其中 $\delta$ 取"由树内行连向树外列"的最小约化权。
\item 增广时可撤销先前已选的边、改选更优边，因此能纠正早期的局部误选。
\end{itemize}

示例：

\begin{enumerate}
\item 初始化势、计算初始约化权。行势取每行最大权值，列势取 0：$\alpha=[\max(8,7,2),\,\max(6,3,5),\,\max(2,6,4)]=[8,6,6]$、$\beta=[0,0,0]$。逐项求解 $c_{ij}=\alpha_i+\beta_j -Q_{ij}$：
\begin{align}
&c_{11}=0,\quad c_{12}=1,\quad c_{13}=6 \nonumber \\
&c_{21}=0,\quad c_{22}=3,\quad c_{23}=1 \nonumber \\
&c_{31}=4,\quad c_{32}=0,\quad c_{33}=2
\end{align}
得初始零边 $(1,1)$, $(2,1)$, $(3,2)$。

\item 发现匹配冲突。零边里输入 1、输入 2 都只能连输出 1（列 1），二者争抢同一列，匹配 $\{(1,1),(3,2)\}$ 会把输入 2 落单，无法形成完美匹配，需要更新势以引入新零边。
\item 求最小松弛 $\delta$ 并更新势。从落单的输入 2 出发，沿零边 $(2,1)$ 连到被占的输出 1，再顺着它的占用者连到输入 1；于是"已探索到"的是行 $\{$输入 1、输入 2$\}$ 与列 $\{$输出 1$\}$，尚未探索的是列 $\{$输出 2、输出 3$\}$。$\delta$ 取"从已探索到的行（输入 1、2）通向尚未探索的列（输出 2、3）``那些元素的最小约化权为：$\delta=\min\{c_{12},c_{13},c_{22},c_{23}\}=1$。更新规则为"已探索的行 $\alpha$ 减 $\delta$、已探索的列 $\beta$ 加 $\delta$''，即 $\alpha_1,\alpha_2$ 各减 1、$\beta_1$ 加 1，得到 $\alpha=[7,5,6]$、$\beta=[1,0,0]$。
\item 重新计算约化权、生成新的零边。用更新后的势重算 $c_{ij}$：
\begin{align}
&c_{11}=0,\quad c_{12}=0,\quad c_{13}=5 \nonumber \\
&c_{21}=0,\quad c_{22}=2,\quad c_{23}=0 \nonumber \\
&c_{31}=5,\quad c_{32}=0,\quad c_{33}=2
\end{align}
新增零边 $(1,2)$ 与 $(2,3)$。

\item 沿增广路得到最优匹配。把落单的输入 2 匹配到新零边对应的输出 3，此时零边无行列冲突，得到最优指派 $(1,1)+(2,3)+(3,2)=8+5+6=19$。
\end{enumerate}

\needspace{4\baselineskip}\addvspace{12pt plus 3pt minus 1pt}\nobreak\textbf{拍卖算法（Bertsekas）流程}

\begin{itemize}
\item 建模：输入 $i$ 为买家、输出 $j$ 为商品，估价 $v_{ij}=Q_{ij}$；$\pi_j$ 为输出 $j$ 的当前标价，初始全为 0。
\item 未分配买家轮流按照净收益 $v_{ij} -\pi_j$ 选最优商品出价，加价幅度 = 最优净收益 - 次优净收益（再加一个极小量 $\varepsilon$）。前一位买家对所有商品计算出的加价幅度会直接给后一位买家使用。
\item 同一商品价高者中标，流拍的买家进入下一轮重新出价；标价单调抬升，直到所有买家完成指派。
\end{itemize}

示例（取 $\varepsilon\to 0$，忽略微量偏移）：

\begin{enumerate}
\item 初始标价 $\pi=[0,0,0]$，三个输入均未匹配。
\item 买家 1 的净收益为 $[8,7,2]$，最优为商品 1、次优为商品 2，故 $\pi_1$ 加价为 $8 -7=1$，所有商品加价为 $\pi=[1,0,0]$，买家 1 \ensuremath{\rightarrow} 商品 1。
\item 买家 2 根据买家1的实时加价幅度重新计算净收益 $[6 -1,\,3,\,5]=[5,3,5]$，商品 1 与商品 3 并列最优，取商品 3（并列故加价约为 $\varepsilon$），买家 2 \ensuremath{\rightarrow} 商品 3。
\item 买家 3 计算净收益 $[2 -1,\,6,\,4]=[1,6,4]$，最优为商品 2、次优为商品 3，故 $\pi_2$ 抬升 $6 -4=2$，买家 3 \ensuremath{\rightarrow} 商品 2。
\item 全部完成指派，最终匹配 $(1,1)+(2,3)+(3,2)=19$，与匈牙利一致。
\end{enumerate}

\needspace{4\baselineskip}\addvspace{12pt plus 3pt minus 1pt}\nobreak\textbf{行/列掩码贪心流程}

\begin{itemize}
\item 每次在整个矩阵内取全局最大值，选中后把所在行、列的元素全部掩码为 $-\infty$，重复 $N$ 次就得到匹配；该策略与 §4.2.5 路径 A、§7 谱调度离散化同属一类贪心框架。
\item 掩码不可逆，既无匈牙利算法的增广删改边、也无拍卖算法的重新竞价，故 $N\ge 3$ 时很有可能得到的是次优解。
\end{itemize}

示例：基于上面示例的 $Q$ 用贪心算法恰好得到最优值 19；但是如果换成权重矩阵 $W'=\begin{bmatrix} 8 & 7 & 7 \\ 8 & 4 & 0 \\ 8 & 0 & 1 \end{bmatrix}$（全局最优仍为 19），首轮贪心若锁定全局最大值 $(2,1)=8$，后续只能得到：
\begin{equation}
P_g' = (2,1)+(1,2)+(3,3) = 8+7+1 = 16 < 19
\end{equation}

三种方法对早期误选的补救能力对照如下：

\bigskip\noindent
\par\leftskip=2em\relax\noindent \fittab{\begin{tabular}{p{\dimexpr 0.500\linewidth-2\tabcolsep}p{\dimexpr 0.500\linewidth-2\tabcolsep}}
\toprule
第 1 步误锁定 (2,1)=8 后 & 对早期误选的补救 \\
\midrule
行/列掩码贪心 & 不可改派，锁定在 16 \\
拍卖 & 可撤销改派 \\
匈牙利 & 增广路径可删改已有匹配边 \\
\bottomrule
\end{tabular}}\par\leftskip=2em\relax{}

\bigskip

\subsubsection{与连续松弛的衔接}

§4.1--§4.2 先把 $Q$ 松弛为双随机矩阵或谱加权矩阵 $W$，然后必须投影回 $0/1$ 的 $P(t)$。该投影若采用行/列掩码类离散化（§4.2.5 路径 A、§7 谱调度），则与上文「行/列掩码贪心」属于同一类近似方法，$N\ge 3$ 时结果可能是次优、且掩码步骤不可回溯。关于权重下界需区分两种贪心：直接按 $Q$ 权重的全局贪心（附录 B.3，每次取当前最大 $Q_{ij}$ 后屏蔽其行列）所得极大匹配 $\ge$ MWM$(Q)$ 的 $1/2$，这是经典的 $1/2$-近似结论 [44]（§6.5 表 3「$\geq$50\%」）。谱调度则按 $W=Q\odot(u_1 v_1^{\mathsf{T}})$ 贪心，其 $1/2$ 界是相对 MWM$(W)$ 而非 MWM$(Q)$，故对 $\langle Q,P\rangle$ 不构成保证。两类贪心都不保证得到最优解。

\subsubsection{转入工业实现}

到这里为止，MWM 已经有了两个角色：一方面，它给出了 $\langle Q,P\rangle$ 意义下的精确参照；另一方面，它也暴露出硬件实现上的主要矛盾：每个 cycle 都要在全矩阵权重上完成全局耦合匹配。datacom switch 与片上 crossbar 调度器通常无法在每个时隙/每个 cycle 内完成「读入全矩阵权重 + 多轮全局迭代」的步骤：因为这么复杂的操作时序收敛需多拍，同时当 $N$ 较大时还需要宽权重总线与集中式状态控制，与 ns 量级的单周期 grant 实现前提不一致。

因此，工业实现通常不会追求单周期精确 MWM，而是把问题改写为更容易时序收敛的局部仲裁过程：只读 1-bit 队列非空状态，使用并行 Request--Grant--Accept 轮次满足行列互斥，再依靠跨周期指针状态模拟长期平均服务率。下文 §5 的 iSLIP 即是这条工程路线的代表。

\section{iSLIP：工业界的硬件近似}

\subsection{为什么不用 MWM？}

\bigskip\noindent
\par\leftskip=2em\relax\noindent \fittab{\begin{tabular}{p{\dimexpr 0.333\linewidth-2\tabcolsep}p{\dimexpr 0.333\linewidth-2\tabcolsep}p{\dimexpr 0.333\linewidth-2\tabcolsep}}
\toprule
约束 & Datacom 交换芯片 & 片上 crossbar arbiter（同构） \\
\midrule
时钟周期 & 约 2--4 ns & 同量级 system clock \\
规模 & 64$\times$64 等（线卡 fabric） & N$\approx$4--8 常见；N=8 已接近集中式 crossbar 的极限 \\
MWM & O(N\textsuperscript{3}) 精确匹配，单 cycle 不可完成（§4.3、附录 B.4） & 同上：不宜每周期跑完整 MWM \\
行/列掩码贪心 & N 轮比较反馈，易 Timing Violation & 长组合逻辑链同样受限 \\
迭代次数 & 每时隙 r 轮 iSLIP（datacom，取值常为 3--4） & 每 cycle r 轮 R--G--A grant 逻辑 \\
\bottomrule
\end{tabular}}\par\leftskip=2em\relax{}

\bigskip

MWM 是数学最优，但难以满足单 cycle 时序要求。iSLIP [3] 用并行 1-bit 仲裁 + 周期内少数轮 R--G--A 在统计意义上逼近 MWM，这正是 crossbar 的 RTL 主流实现方法。McKeown 与 Anderson [16] 的对比研究表明，iSLIP 在均匀（i.i.d.）流量下可达 100\% 吞吐。工业实现通常仅用 3--4 轮/ cycle。

\subsection{硬件结构：iSLIP 与两级 Round-Robin 仲裁器}

iSLIP [3] 采用并行行/列 Request--Grant--Accept：每行、每列各用一维仲裁并行工作；Grant/Accept 冲突由 Round-Robin 指针与优先级编码器解决。

每个输入端口有一个输入仲裁器，每个输出端口有一个输出仲裁器，各持有一个指针（priority pointer）。

三步协议（Request \ensuremath{\rightarrow} Grant \ensuremath{\rightarrow} Accept）：

\par{\leftskip=0pt\relax\noindent\hspace*{2em}\begin{minipage}{\dimexpr\linewidth-2em\relax}
\begin{verbatim}
第一步 Request：有报文的 VOQ 向目标输出发送 1 -bit 请求

第二步 Grant：  输出仲裁器从指针位置起顺时针扫描，向第一个有请求的输入发送 Grant

第三步 Accept： 输入仲裁器从指针位置起顺时针扫描，接受第一个收到的 Grant
\end{verbatim}
\end{minipage}\par}

Round-Robin 扫描规则：指针指向的位置是优先级起点，不是强制终点。硬件用优先级编码器（Priority Encoder）跳过无请求端口，直到找到第一个有请求的端口。

\subsection{指针更新规则（iSLIP 的核心）}

只有当 Grant 在 Accept 阶段被真正接受时，对应仲裁器的指针才前进一格；否则指针不动。

此规则带来：

\begin{itemize}
\item 空闲时：指针快速轮转，近似公平 Round-Robin；
\item 拥堵时：Accept 成功后指针错开（去同步化）；下一周期 Grant/Accept 从新指针位置开始扫描，先轮到其他输入/输出对。仍有 backlog 的 (i,j) 持续发 Request，因此在后续周期仍会被匹配，而非长期独占某一端口。
\end{itemize}

\subsection{同一周期内的多轮迭代}

iSLIP 论文中的"多轮迭代"指同一时钟周期内、指针不前进情况下的多次 Request--Grant--Accept：

\bigskip\noindent
\par\leftskip=2em\relax\noindent \fittab{\begin{tabular}{p{\dimexpr 0.500\linewidth-2\tabcolsep}p{\dimexpr 0.500\linewidth-2\tabcolsep}}
\toprule
轮次 & 作用 \\
\midrule
第 1 轮 & 高 backlog 的 (i,j) 优先匹配，对应边被锁定 \\
第 2 轮 & 剩余输入/输出继续匹配 \\
第 3 轮 & 继续填充，直到匹配数达到物理极限 \\
\bottomrule
\end{tabular}}\par\leftskip=2em\relax{}

\bigskip

全部匹配结束，报文真正发出后，此时才统一更新指针。

指针更新的精细规则：

\begin{itemize}
\item 只有在本周期第 1 轮 Accept 成功的输出侧仲裁器，才在周期末更新指针；
\item 只有在本周期第 1 轮 Accept 成功的输入侧仲裁器，才在周期末更新指针；
\item 第 2、3 轮"捡漏"匹配成功的端口，指针不动。
\end{itemize}

因此多轮迭代是在同一组指针下，对尚未匹配的端口重复 Request--Grant--Accept，以增大本周期内匹配边数（trace）。单轮 Grant\ensuremath{\rightarrow}Accept 往往只得到合法匹配集 M 的真子集。

\subsection{手算示例：单周期两轮匹配与跨周期 P̄}

下列两例共用 2$\times$2 拓扑。

示例 A：

请求掩码 R（有 backlog 则 Request；此处四个 VOQ 均非空）：
\begin{equation}
R = \begin{bmatrix} 1 & 1 \\ 1 & 1 \end{bmatrix}
\end{equation}

指针初值：g\textsubscript{1}、g\textsubscript{2} 均指向 in1；a\textsubscript{1} 指向 out1。

单周期 Grant 可先违反行约束，Grant（每列独立，只保证列和 $\leq$1）。输出 1 从 in1 扫到请求 \ensuremath{\rightarrow} Grant (1,1)；输出 2 从 in1 扫到请求 \ensuremath{\rightarrow} Grant (1,2)。Grant 矩阵 G 为：
\begin{equation}
G = \begin{bmatrix} 1 & 1 \\ 0 & 0 \end{bmatrix}
\qquad \ensuremath{\leftarrow} \text{第 1 行行和 } = 2,\ \text{尚不在合法匹配集 } \mathcal{M} \text{ 内}
\end{equation}

Accept（每行在收到的 Grant 中选至多一个），输入 1 同时收到 out1、out2 的 Grant，从 a\textsubscript{1} 指向的 out1 起接受 \ensuremath{\rightarrow} 本周期先匹配 (1,1)，得 P\textsuperscript{(}\textsuperscript{1}\textsuperscript{)}：
\begin{equation}
P^{(1)} = \begin{bmatrix} 1 & 0 \\ 0 & 0 \end{bmatrix}
\end{equation}

第 2 轮（指针本周期不动）：已匹配的输入 1 与输出 1 从后续迭代中剔除，剩余输入 2 / 输出 2 上再 Grant\ensuremath{\rightarrow}Accept，得到 (2,2)，合并为
\begin{equation}
P(t) = \begin{bmatrix} 1 & 0 \\ 0 & 1 \end{bmatrix}
\qquad \mathrm{trace} = 2
\end{equation}

示例 B：跨 10 个周期，从 P(t) 计算出 P̄

设定：Q\_11 极长（backlog 持续非零），Q\_12=0，Q\_21、Q\_22 较短。下表只记录输入 1 是否接通（即 P\_11、P\_12，取多轮合并后 P(t) 的对应项）：

\bigskip\noindent
\par\leftskip=2em\relax\noindent \fittab{\begin{tabular}{p{\dimexpr 0.250\linewidth-2\tabcolsep}p{\dimexpr 0.250\linewidth-2\tabcolsep}p{\dimexpr 0.250\linewidth-2\tabcolsep}p{\dimexpr 0.250\linewidth-2\tabcolsep}}
\toprule
t & 匹配边 & P\_11(t) & P\_12(t) \\
\midrule
1 & (1,1) & 1 & 0 \\
2 & (1,1) & 1 & 0 \\
3 & (2,2) & 0 & 0 \\
4 & (1,1) & 1 & 0 \\
5 & (1,1) & 1 & 0 \\
6 & (1,1) & 1 & 0 \\
7 & (2,1) & 0 & 0 \\
8 & (1,1) & 1 & 0 \\
9 & (1,1) & 1 & 0 \\
10 & (1,1) & 1 & 0 \\
\bottomrule
\end{tabular}}\par\leftskip=2em\relax{}

\bigskip
\begin{align}
\sum_t P_{11}(t) = 8 &\ \to\ \bar P_{11} = E[P_{11}] = 8/10 = 0.8 \nonumber \\
\sum_t P_{12}(t) = 0 &\ \to\ \bar P_{12} = 0
\end{align}

（即使指针错开， (1,1) 仍然常被轮到，故高 backlog 通道接通比例高。）

若本段时间内近似 Q\_11=1000、Q\_12=0，则 ⟨Q,P̄⟩ $\approx$ 1000$\times$0.8 = 800。取 $\lambda$\_11=0.5，则 Qᵀ$\lambda$ = 500 \textless  800，对应 §3.4.6 中 E[QᵀP] \textgreater  Qᵀ$\lambda$ 的数值版。

\subsection{iSLIP 与 Q(t) 的数学关系}

§5.2--§5.4 介绍了 iSLIP 的硬件规则（Request--Grant--Accept 加轮转指针）。本节把这些规则与 §3 的 MWM 数学衔接起来。二者之间的桥梁是一个\textbf{跨周期运行的反馈环}，而非单周期内的某个公式。下面分五步解读，随后的子节给出第 1、3、5 步的形式化数学。

\begin{enumerate}
\item \textbf{Sense（1-bit 感知）。} 每周期队列状态 Q 产生请求掩码 R\_ij = 1[Q\_ij \textgreater  0]。iSLIP 只能判断队列"是否为空"，无法获知队列"有多长"，因此单周期内不存在与 MWM 的 argmax⟨Q, P⟩ 等价的 P(t) = f(Q(t)) 形式。
\item \textbf{Decide（解耦决策）。} 输出仲裁依据自身指针给出 Grant（满足列约束），输入仲裁依据自身指针给出 Accept（满足行约束）。MWM 中行列约束同时求解的耦合 ILP，由此被拆成两次代价低且相互独立的投影，这正是 iSLIP 能够在一个周期内完成的原因。
\item \textbf{Update（反馈更新）。} 指针\textbf{仅在其 Grant 被 Accept 时前进一格}。刚被服务的 (i,j) 在下一拍让出优先级；积压但本轮未中标的 (i,j) 持续发送 Request，随着指针轮转终将再次中标。这正是使 iSLIP 区别于无权重随机匹配的动态反馈机制。
\item \textbf{Average（跨周期平均）。} 由第 3 步可知，长期积压的 (i,j) 在时间上被接通的比例更大。该比例的数学表达即为 E[P\_ij] = P̄\_ij。与 §3.4.6 中 E[P\_11] 固定在 0.5、不随 Q\_11 变化的无权重随机匹配不同，此处的 P̄\_ij 会随积压上升。
\item \textbf{Close（闭环稳定）。} 将这个"随积压上升的 P̄"代入 §3.4 的漂移条件：当 $\Lambda$ 可容许时 ⟨Q, P̄⟩ \textgreater  Qᵀ $\lambda$ 成立，队列保持有界。
\end{enumerate}

\subsubsection{第 1 步. Sense：单周期无直接代数映射}
\begin{equation}
P_{\text{MWM}}(t) = \operatorname*{argmax}_{P}\ \langle Q(t), P\rangle
\end{equation}

iSLIP 的 P(t) 只取决于：

\begin{itemize}
\item 仲裁器指针位置 g(t)、a(t)
\item 请求掩码为 1 仅依赖条件 [Q\_ij \textgreater  0]（有 backlog 则为 1，无 backlog 则为 0）
\end{itemize}

iSLIP 不感知队列长度的绝对值，只感知"有没有报文"。因此在单个时钟周期 t，不存在类似 \texttt{P(t) = f(Q(t))} 的代数公式。

\subsubsection{第 2 步. Decide：解耦耦合 ILP（Grant\ensuremath{\rightarrow}Accept）}

MWM 在合法匹配域上求解耦合 ILP（行列约束同时作用于同一 P，不能先逐行 argmax 再拼接）：
\begin{align}
\max\ & \langle Q, P\rangle = \sum_{i,j} Q_{ij}\, P_{ij} \nonumber \\
\text{s.t.}\ & \sum_j P_{ij} \le 1, \quad \forall i \quad (\text{行约束}) \nonumber \\
& \sum_i P_{ij} \le 1, \quad \forall j \quad (\text{列约束}) \nonumber \\
& P_{ij} \in \{0, 1\}
\end{align}

匈牙利、拍卖算法在该耦合 ILP 上求全局最优（O(N\textsuperscript{3})）。定义列/行可行集 $C_{\text{col}}$、$C_{\text{row}}$，$\mathcal{M} = C_{\text{col}} \cap C_{\text{row}}$。$\Pi_C(\cdot)$ 为带 RR 指针的贪心投影（非欧氏投影）。iSLIP 在 Request 掩码 $R_{ij}=\mathbf{1}[Q_{ij}>0]$ 上单轮执行
\begin{equation}
P^{(1)} = \Pi_{C_{\text{row}}}\!\big(\Pi_{C_{\text{col}}}(R)\big),
\end{equation}

Grant 对应 $\Pi_{C_{\text{col}}}(R)$（列约束，G 行和可 \textgreater 1）；Accept 对应 $\Pi_{C_{\text{row}}}(G)$（行约束，$P^{(1)}\in\mathcal{M}$）。

\subsubsection{第 3 步. Update：指针反馈}

第 1、2 步发生在单个周期内，产出一个 P(t)。把 iSLIP 从"一次性极大匹配"变成"长期追踪 MWM"的，是它的状态在\textbf{周期之间}如何变化。

回顾 §5.3 的指针规则：一个仲裁器的指针\textbf{只在它的 Grant 被 Accept 时前进一格}；若本周期待 Accept 未成功，指针不动。由此带来两个后果：

\begin{itemize}
\item \textbf{服务后让位。} 刚被服务的 (i,j)，其输入与输出仲裁器的指针都已指向 (i,j) 的下一格。下一拍扫描从这里开始，(i,j) 不再排在最前，把优先级让给其他待服务对。一个端口无法独占 crossbar。
\item \textbf{持续重试直至中标。} 积压但本轮未中标的 (i,j)（其 Grant 未被 Accept，或被掩码排除），只要 Q\_ij \textgreater  0 就每周期持续发 Request。随着其输入、输出仲裁器的指针轮转，(i,j) 终会在两侧扫描中都成为第一个待服务对而中标。积压越久，请求的周期越多，被服务的概率越大。
\end{itemize}

这就是动态反馈：\textbf{队列状态 Q 驱动请求掩码 R；R 与指针位置共同决定 P(t)；一次成功的 Accept 推动指针；新的指针位置塑造下一拍的 P(t+1)。} 跨许多周期后，服务被引向持续积压的方向，而 iSLIP 全程没有读取 Q\_ij 的数值。这就是下一步的 P̄\_ij 能够随积压上升的定性原因，也正是无权重随机匹配做不到的。

\subsubsection{第 4 步. Average：期望值与时间平均}

§3.4.6 稳定性条件里出现了 \texttt{E[QᵀP]}。这里把"期望值"和"时间平均"两个概念严格对齐。

\begin{itemize}
\item 随机变量视角
\end{itemize}

每个周期 t，调度结果 \texttt{P\_ij(t)} 是一个 0/1 随机变量：
\begin{equation}
P_{ij}(t) = \begin{cases}
1 & \text{若本周期输入 } i \text{ 成功匹配输出 } j\\
0 & \text{否则}
\end{cases}
\end{equation}

期望值定义为：
\begin{equation}
E[P_{ij}] = \lim_{T\to\infty} \frac{1}{T} \sum_{t=1}^{T} P_{ij}(t)
\end{equation}

即：在足够长的时间里，通道 (i,j) 被接通的比例。

\begin{itemize}
\item 与时间平均矩阵 P̄ 的关系
\end{itemize}

定义时间平均服务矩阵（T 个周期）：
\begin{equation}
\bar P = \frac{1}{T} \sum_{t=1}^{T} P(t)
\end{equation}

在统计的视角下，两者可作等价理解：\texttt{E[P\_ij] = P̄\_ij}。下文混用 \texttt{E[P\_ij]} 与 \texttt{P̄\_ij}，含义相同。

\begin{itemize}
\item 从 E[P\_ij] 到 E[QᵀP]
\end{itemize}

内积是对所有 (i,j) 求和，期望可逐项提出：
\begin{align}
E[Q^{\mathsf{T}}P] &= E\!\left[ \sum_{i,j} Q_{ij}\, P_{ij} \right] \nonumber \\
&= \sum_{i,j} Q_{ij}\, E[P_{ij}] \quad (\text{$Q_{ij}$ 视为当前已知常数}) \nonumber \\
&= \sum_{i,j} Q_{ij}\, \bar P_{ij} \nonumber \\
&= \langle Q, \bar P\rangle
\end{align}

物理含义：某方向 (i,j) 的队列越长，且其长期平均服务率 P̄\_ij 越高，则内积 ⟨Q, P̄⟩ 中该方向对能量削减的贡献越大。

\begin{itemize}
\item 为什么 iSLIP 能满足漂移、而随机匹配不能
\end{itemize}

在无权重随机匹配中，匹配概率期望 E[P\_11] 固定为 0.5、不随队列积压长度 Q\_11 增大，故 E[QᵀP] \textless  Qᵀ$\lambda$。与之不同，第 3 步的指针反馈使 P̄\_ij 对积压有响应：持续 backlog 的 (i,j) 在更长时间里获得更高的 P̄\_ij。长期平均下有机会满足 E[QᵀP] \textgreater  Qᵀ$\lambda$，超过固定 50/50 的 I/X 均匀分配的性能上限。

\subsubsection{第 5 步. Close：完整逻辑链}
\begin{align}
&\text{单周期：} P(t) \in M,\ P = \Pi_{C_{\text{row}}}\!\big(\Pi_{C_{\text{col}}}(\mathbf{1}[Q>0])\big)\text{；不读 } Q_{ij} \text{ 数值} \nonumber \\
&\text{周期内：多轮 Grant–Accept，指针不动} \nonumber \\
&\text{跨周期：Accept 成功后指针前进 } \to\ E[P_{ij}]=\bar P_{ij}\text{；backlog 大者 } \bar P_{ij} \text{ 更高} \nonumber \\
&\text{稳定性：} \Lambda \text{ 可容许，且 } \langle Q,\bar P\rangle > Q^{\mathsf{T}}\lambda \to \text{队列有界}
\end{align}

\subsection{MWM 与 iSLIP 对比}

本节对 MWM 与 iSLIP 逐项对照。

\bigskip\noindent
\par\leftskip=2em\relax\noindent \fittab{\begin{tabular}{p{\dimexpr 0.333\linewidth-2\tabcolsep}p{\dimexpr 0.333\linewidth-2\tabcolsep}p{\dimexpr 0.333\linewidth-2\tabcolsep}}
\toprule
特性 & MWM & iSLIP \\
\midrule
单周期计算 & 匈牙利 / 拍卖算法，O(N\textsuperscript{3})（精确组合求解） & 分布式逻辑门，O(1) 硬件延迟 \\
权重感知 & 精确读取 Q\_ij 数值 & 仅感知 Q\_ij \textgreater  0 \\
与 Q 的关系 & 每步严格 argmax⟨Q, P⟩ & 单步无映射；长期 E[P\_ij]=P̄\_ij，backlog 大的通道更高（§5.5） \\
ILP 求解 & 耦合 ILP 全局最优（匈牙利 O(N\textsuperscript{3})） & 协调解耦：Grant\ensuremath{\rightarrow}Accept 交替投影 \\
队列稳定性 / 吞吐 & 经典理论给出 100\% 吞吐结果 [4] & 多轮迭代 + 指针去同步，统计逼近 100\% 吞吐 [3] \\
稳定性判据（漂移） & \texttt{E[QᵀP] \textgreater  Qᵀ$\lambda$}（当 ‖Q‖ 大时） & 同样要求 \texttt{⟨Q, P̄⟩ \textgreater  Qᵀ$\lambda$}，但 P̄ 为时间平均矩阵 \\
工业部署 & 几乎不可能 & 广泛采用 \\
\bottomrule
\end{tabular}}\par\leftskip=2em\relax{}

\bigskip

\begin{quote}
\textbf{注（精确 vs 近似）}：精确 MWM 的逐周期实现是组合算法（匈牙利/拍卖）。Sinkhorn+BvN、谱调度（幂迭代）等是 MWM 的连续松弛/近似方法（见 §4.2、§4.1、§6.4 表 2），不等同于精确 MWM。
\end{quote}

\subsubsection{iSLIP 同族变体（仅摘要）}

McKeown iSLIP [3] 之后，DRRM（双 RR 指针 [17]，性能分析见 [18]）、FIRM（FCFS 公平 [19]）、SRR / DRDSRR（静态 RR、低延迟 MSM [20],[21]）等仍在 1-bit R--G--A 骨架上改进指针/公平性。见表 1 与文献 [17]--[21]。

\section{算法分类法（Taxonomy）}

本节把前文出现的调度算法放到同一张地图里：先按信息粒度、匹配目标、迭代次数与架构前提分类，再给出分类树、概念图和横向对比表。这样读者在进入实验前，可以先区分 MWM、iSLIP、谱调度、BvN/OT 以及其它交换架构扩展分别处在什么位置。

\subsection{分类维度}

\bigskip\noindent
\par\leftskip=2em\relax\noindent \fittab{\begin{tabular}{p{\dimexpr 0.333\linewidth-2\tabcolsep}p{\dimexpr 0.333\linewidth-2\tabcolsep}p{\dimexpr 0.333\linewidth-2\tabcolsep}}
\toprule
维度 & 含义 & 典型取值 \\
\midrule
权重感知 & 单周期内是否读取 Q\_ij 数值 & 全矩阵比较（MWM）；1-bit Request（iSLIP）；谱加权 W= Q⊙(u\textsubscript{1}v\textsubscript{1}ᵀ)（§4.2） \\
匹配目标 & 每周期优化对象 & 最大权匹配（MWM）；极大匹配（iSLIP）；双随机凸组合（BvN 时分） \\
迭代次数 & 单周期内 R--G--A 或等价轮数 & 常 3--4 轮（iSLIP）；O(log N) 轮（极大匹配，文献）；O(N\textsuperscript{3}) 离线（匈牙利） \\
跨周期记忆 & 是否保留上一周期状态 & 无（无权重随机匹配）；RR 指针（iSLIP）；上一周期匹配 M(t-1)（Tassiulas [22]）；MERGE（APSARA [24]） \\
仲裁拓扑 & 集中式 vs 分布式 & 集中式匹配（匈牙利，全局 Q）；分布式 R--G--A（iSLIP，per-port 仲裁） \\
稳定性定义 & 「100\%」指什么 & admissible 流量下队列有界（MWM [4]）；iSLIP 限均匀流量 [3]；其它易混指标见 §6.5 表 3 \\
架构前提 & 队列架构 & 纯 IQ+VOQ；CIOQ + speedup [29]；CICQ 交叉点队列 [37] \\
\bottomrule
\end{tabular}}\par\leftskip=2em\relax{}

\bigskip

\needspace{26\baselineskip}
\subsection{分类树}

输入排队 / 交叉开关调度按「信息粒度 + 匹配目标」分为五类（\ensuremath{\rightarrow} 右列为关键特性）：

\bigskip\noindent
\par\leftskip=2em\relax\noindent \fittab{\begin{tabular}{p{\dimexpr 0.333\linewidth-2\tabcolsep}p{\dimexpr 0.333\linewidth-2\tabcolsep}p{\dimexpr 0.333\linewidth-2\tabcolsep}}
\toprule
类别 & 代表算法 & 关键特性 \\
\midrule
【I】精确最优·读完整权重 Q\_ij & 图论 MWM：匈牙利 [10]、拍卖 [11] & O(N\textsuperscript{3})，100\% admissible，单周期难实现 \\
 & MWM 权重族：LQF / OCF / LPF [4],[5] & 100\% admissible，权重定义不同 \\
 & 分布式迭代 MWM：$\epsilon$-Auction / $\epsilon$-min-sum [41] & 收敛 MWM，O(n\textsuperscript{2}/$\epsilon$) 轮 \\
【II】极大匹配·0/1 近似（工业主线） & RR + 指针：iSLIP [3]、DRRM [17]、RRM [3] §II & 均匀流量 100\%；指针规则各异 \\
 & 公平性：FIRM [19] & FCFS 近似，MSM 类 \\
 & 静态/去随机 RR：SRR [20]、DRDSRR [21] & 低延迟；SRR 非均匀不稳定 \\
【III】随机化+记忆·长期平均逼近 MWM & Tassiulas [22] / 比权取大 & O(N)，100\% admissible，延迟大 \\
 & APSARA / SERENA [23],[24],[25]；随机化评估 [26] & 去随机+并行；流体模型 [24] \\
【IV】连续松弛\ensuremath{\rightarrow}合法离散 P & Sinkhorn [12] + BvN 时分 [31],[32],[35] & 离线分解，在线周期置换 \\
 & 两阶段负载均衡 BvN [33],[34] & 无中心 N\textsuperscript{2} 匹配，保序 [34] \\
 & 谱调度：幂迭代+（可选）W+离散化（§4.2） & 研究/小 N；§7 与 iSLIP 对比 \\
【V】架构扩展（调度问题形态变化） & CIOQ + speedup 2/4 [27],[28],[29],[30] & 模拟 OQ / 100\% 极大匹配 \\
 & CICQ / LIPS [36],[37],[38],[39] & 输入+输出独立调度；speedup=2 \\
 & PIFO [40]（programmable dataplane） & 与 CIOQ 理论衔接 \\
\bottomrule
\end{tabular}}\par\leftskip=2em\relax{}

\bigskip

\subsection{图示（概念）}

\needspace{10\baselineskip}
\par\medskip{\leftskip=0pt\relax\noindent\hspace*{2em}%
\begin{minipage}{\dimexpr\linewidth-2em\relax}
\centering\fitwidth{%
\begin{tikzpicture}[font=\footnotesize,>=Stealth,
  voq/.style={draw,rounded corners,inner sep=3pt,minimum width=52mm,
              minimum height=6mm,anchor=west}]
  \node[voq] (vq0) at (0,0) {[VOQ$\to$0]\,[VOQ$\to$1]\,$\cdots$\,[VOQ$\to$N$-$1]};
  \node[voq] (vq1) at (0,-10mm) {[VOQ$\to$0]\,[VOQ$\to$1]\,$\cdots$\,[VOQ$\to$N$-$1]};
  \node (vd) at (26mm,-16mm) {$\vdots$};
  \node[voq] (vqn) at (0,-22mm) {[VOQ$\to$0]\,$\cdots$\,[VOQ$\to$N$-$1]};
  \node[anchor=east] (i0) at (-8mm,0) {输入 0};
  \node[anchor=east] (i1) at (-8mm,-10mm) {输入 1};
  \node[anchor=east] (iN) at (-8mm,-22mm) {输入 N$-$1};
  \draw[->] (i0.east)--(vq0.west);
  \draw[->] (i1.east)--(vq1.west);
  \draw[->] (iN.east)--(vqn.west);
  \node[draw,fill=black!5,align=center,minimum width=16mm,minimum height=22mm]
        (cb) at (80mm,-11mm) {N$\times$N\\Crossbar};
  \coordinate (cbw0) at ([yshift=4mm]cb.west);
  \coordinate (cbw1) at (cb.west);
  \coordinate (cbwN) at ([yshift=-4mm]cb.west);
  \draw[->] (vq0.east)--(cbw0);
  \draw[->] (vq1.east)--(cbw1);
  \draw[->] (vqn.east)--(cbwN);
  \node[right=10mm of cb] (out) {输出 $0\cdots$N$-$1};
  \draw[->] (cb.east)--(out.west);
  \node[draw,dashed,align=center,below=10mm of cb] (sch) {调度器每周期选一组匹配 $P(t)$\\每行、每列至多一个 1};
  \draw[->] (sch.north)--(cb.south);
\end{tikzpicture}}

\end{minipage}\par}\medskip
\par\medskip
{\small\textbf{图 1。}N$\times$N 输入排队 + VOQ + 交叉开关。{\footnotesize 每输入端口维护 N 个 VOQ [4]，消除 HOL [1]；调度器输出合法匹配矩阵 $P(t)\in\{0,1\}^{N\times N}$。}}

\needspace{14\baselineskip}
\par\medskip{\leftskip=0pt\relax\noindent\hspace*{2em}%
\begin{minipage}{\dimexpr\linewidth-2em\relax}
\centering\fitwidth{%
\begin{tikzpicture}[font=\footnotesize,>=Stealth,
  box/.style={draw,rounded corners,align=center,inner sep=4pt}]
  \node[box,fill=black!5] (top) {队列矩阵 $Q(t)$\quad·\quad 服务矩阵 $P(t)$\quad·\quad Lyapunov $V=\lVert Q\rVert^2$};
  \node[box,below=14mm of top] (m2) {【主线 II】\\图论精确 MWM\\§4.3\\匈牙利 [10]};
  \node[box,left=10mm of m2] (m1) {【主线 I】\\线性代数/稳定性\\§2–§3, §4.1–§4.2\\MWM 导出 [4],[7]};
  \node[box,right=10mm of m2] (m3) {【主线 III】\\iSLIP 0/1 迭代\\§5\\$\bar P$ 随 backlog\\（长期平均）[3]};
  \draw[->] (top.south west) -- (m1.north);
  \draw[->] (top.south) -- (m2.north);
  \draw[->] (top.south east) -- (m3.north);
  \node[box,below=10mm of m2] (join) {§4 交汇 · Sinkhorn / BvN / 谱调度};
  \draw[->] (m1.south) -- (join.north west);
  \draw[->] (m2.south) -- (join.north);
  \draw[->] (m3.south) -- (join.north east);
\end{tikzpicture}}

\end{minipage}\par}\medskip
\par\medskip
{\small\textbf{图 2。}本文三条主线关系。}

\subsection{横向算法对比表}

下表把正文主线算法、实验对象和若干相关调度算法放在一起比较，重点区分读权重粒度、迭代预算、吞吐/稳定性语境与硬件量级。

\bigskip\noindent
\par\leftskip=2em\relax\noindent \fittab{\begin{tabular}{p{\dimexpr 0.167\linewidth-2\tabcolsep}p{\dimexpr 0.167\linewidth-2\tabcolsep}p{\dimexpr 0.167\linewidth-2\tabcolsep}p{\dimexpr 0.167\linewidth-2\tabcolsep}p{\dimexpr 0.167\linewidth-2\tabcolsep}p{\dimexpr 0.167\linewidth-2\tabcolsep}}
\toprule
算法 & 读权重 & 单周期迭代 & 吞吐/稳定 & 硬件量级 & 角色 \\
\midrule
MWM (LQF) & 全 Q\_ij & 1（若求得） & 100\% admissible [4] & O(N\textsuperscript{3}) & 理论上界 \\
匈牙利/拍卖 & 全 Q\_ij & 多轮至收敛 & 同 MWM & O(N\textsuperscript{3}) & 仿真/离线 [10],[11] \\
iSLIP & 1-bit & 可配置（常 3 轮） & 均匀高吞吐；§7 非均匀分离 & O(log N) 轮量级 & 工业基准 [3] \\
谱调度 & Q + u\textsubscript{1},v\textsubscript{1} & r 步幂迭代 & §7：贴近 MWM & O(r·N\textsuperscript{2}) & 本文 §7 \\
熵正则 OT & 全 Q\_ij & r\_sink 轮 Sinkhorn & §7：最接近 MWM 时延 & O(r\_sink·N\textsuperscript{2})+exp & 本文 §7 \\
Sinkhorn+BvN & 连续 B & 离线+周期置换 & 帧级时分 [31],[32] & O(N\textsuperscript{2})/迭代 & 松弛方法 [12] \\
Tassiulas / APSARA & 全 Q\_ij & 1（APSARA 可并行） & 100\% admissible [22],[24] & O(N)--并行 & 全局 Q、高延迟 \\
\bottomrule
\end{tabular}}\par\leftskip=2em\relax{}

\bigskip

{\small\textbf{表 2。}主线算法横向对比}

\clearpage
\subsection{复杂度与吞吐指标速查}

表中 58.6\%、50\%、100\% 三类百分比含义不同（架构 HOL 上限、极大匹配权重比下界、admissible 稳定意义），不可与后文实测吞吐--负载曲线直接对比。

\bigskip\noindent
\par\leftskip=2em\relax\noindent \fittab{\begin{tabular}{p{\dimexpr 0.200\linewidth-2\tabcolsep}p{\dimexpr 0.200\linewidth-2\tabcolsep}p{\dimexpr 0.200\linewidth-2\tabcolsep}p{\dimexpr 0.200\linewidth-2\tabcolsep}p{\dimexpr 0.200\linewidth-2\tabcolsep}}
\toprule
指标 & 数值 & 适用对象 & 含义 & 详见 \\
\midrule
HOL 饱和吞吐 & $\approx$58.6\% (=2-√2) & 输入 FIFO、无 VOQ & 架构上限 & [1] \\
极大匹配权重比 & $\geq$50\% of MWM & 单次 有权 极大匹配 & 对 MWM 的近似比下界 & §4.3 \\
MWM 稳定 & 100\% & 任意 admissible $\Lambda$（行和、列和 \textless  1） & 队列有界，定理 & [4],[7] \\
iSLIP 稳定 & 100\% & 仅均匀 i.i.d.；非均匀无一般保证（§7） & 队列有界 & [3] \\
精确 MWM（匈牙利/拍卖） & O(N\textsuperscript{3}) & 精确 MWM & 软件可接受，单周期不可实现 & §4.3、附录 B.4 \\
iSLIP 迭代 & O(log N) 轮 & 极大匹配 & 每轮 O(1) 分布式 & [3] \\
Sinkhorn & O(N\textsuperscript{2}) / 迭代 & 双随机化 & 连续松弛 & §4.1.1 \\
BvN 分解 & $O(N^{4.5})$ 等 & 离线 & 置换矩阵个数 $\leq$ N\textsuperscript{2}-2N+2 & [31] \\
\bottomrule
\end{tabular}}\par\leftskip=2em\relax{}

\bigskip

{\small\textbf{表 3。}全文易混淆数值的对照定义}

\section{对比实验：谱调度、iSLIP 与熵正则 OT（自适应温度）算法}

本章实验基于标准 IQ+VOQ 交叉开关模型，以匈牙利算法求解的最大权重匹配（MWM）为参考基准，对比 iSLIP、谱调度、熵正则最优传输（OT）三类调度算法的性能差异。其中 iSLIP 与谱调度的单周期迭代次数统一约束为 3（r=3），熵正则 OT 的行/列归一化取 r\_sink=10。

熵正则调度指：通过变换关系 $K\propto\exp(Q/\varepsilon_{\mathrm{eff}})$ 对队列矩阵进行处理（Softassign 思想 [43]；熵正则 OT / Sinkhorn 距离 [42]），结合有限轮 Sinkhorn 迭代求解得到近似双随机权重矩阵，再通过贪心取整的方法计算出离散匹配结果。实验选择熵正则调度而不是直接采用 §4.1 的连续松弛方法，是因为相较于 §4.1 对 Q 直接缩放并做 BvN 帧分解的连续松弛方法，该方法更贴近 MWM 的 ⟨Q,P⟩ 的优化目标。

\subsection{研究问题与设计原则}

实验围绕下面三个研究问题开展比较：
Q1：非均匀 admissible 流量场景下，读取完整队列信息矩阵的近似调度算法，是否优于仅采集 1-bit 队列状态的 iSLIP，性能是否更贴近 MWM 参考基准？
Q2：在上述迭代预算约束下，各算法的吞吐--$\rho$\_load、时延--$\rho$\_load 曲线的特征如何？
Q3：提升算法迭代次数能否带来有效性能收益？
实验中只替换核心调度映射 schedule(Q)\ensuremath{\rightarrow}P，其余仿真配置、流量参数与统计方式保持一致。沿用本文 §2.2 的 IQ+VOQ 模型，无队列深度上限与硬件反压，仅研究调度算法本身的性能差异。其中 MWM、谱调度、熵正则 OT 读取完整 32 位队列的数值，iSLIP 仅采集 1-bit 队列非空状态。

\bigskip\noindent
\par\leftskip=2em\relax\noindent \fittab{\begin{tabular}{p{\dimexpr 0.15\linewidth-2\tabcolsep}p{\dimexpr 0.35\linewidth-2\tabcolsep}p{\dimexpr 0.14\linewidth-2\tabcolsep}p{\dimexpr 0.20\linewidth-2\tabcolsep}p{\dimexpr 0.16\linewidth-2\tabcolsep}}
\toprule
算法 & 抽象形式 schedule(Q)\ensuremath{\rightarrow}P & 读 Q 分辨率 & 每 cycle 操作量级 & 实验角色 \\
\midrule
iSLIP（r=3） & P $\approx$ $\Pi$\_row($\Pi$\_col(1\{Q\textgreater 0\}))，R--G--A + RR 指针 & 1-bit 非空（Q\_ij\textgreater 0） & O(r·N) 1-bit RR 仲裁 & 工业成熟基准 \\
谱调度（r=3） & P = Greedy(Q ⊙ u\textsubscript{1}v\textsubscript{1}ᵀ)，幂迭代 + 贪心 & 全矩阵 + 秩-1 谱特征 & O(r·N\textsuperscript{2}) MAC + O(N\textsuperscript{2}) 贪心 & 线性代数近似（§4.2.4） \\
熵正则 OT（r\_sink=10, $\epsilon$=1） & P = Greedy(Sinkhorn(exp(Q/$\epsilon$))) & 全矩阵 + 熵正则加权 & O(r\_sink·N\textsuperscript{2}) 指数/归一化 + O(N\textsuperscript{2}) 贪心 & 熵正则 OT 方案（§7.5） \\
MWM（匈牙利） & P = Hungarian(Q)，二分图最大权匹配 & 全矩阵，精确数值 & O(N\textsuperscript{3}) 最优匹配 & 仿真参考基准 \\
\bottomrule
\end{tabular}}\par\leftskip=2em\relax{}

\bigskip

{\small\textbf{表 4。}四个调度器的抽象形式、信息粒度与每周期操作量级（同一 schedule(Q)\ensuremath{\rightarrow}P 接口）}

\needspace{26\baselineskip}
\subsection{系统模型与仿真器}

本次仿真基于自主搭建的离散事件仿真平台，统一仿真参数如下表所示。

\bigskip\noindent
\par\leftskip=2em\relax\noindent \fittab{\begin{tabular}{p{\dimexpr 0.500\linewidth-2\tabcolsep}p{\dimexpr 0.500\linewidth-2\tabcolsep}}
\toprule
项 & 规格 \\
\midrule
拓扑 & N$\times$N IQ + VOQ，单 crossbar，speedup=1，全局时钟周期同步 \\
端口规模 N & 8（核心实验规模，规模说明见下文） \\
单周期时序流程 & (1) 流量到达入队 \ensuremath{\rightarrow} (2) 调度器读取队列矩阵并输出匹配矩阵 P \ensuremath{\rightarrow} (3) 依据匹配结果出队 \\
VOQ 配置 & 无深度上限，仿真过程动态扩容，不建模硬件反压与数据包丢包 \\
时延统计定义 & 每 VOQ 维护 FIFO 时间戳，单 cell 时延 = 当前时钟周期 - 数据包队头到达周期 \\
吞吐统计定义 & 归一化吞吐 = 统计窗口输出 cell 数 / (总周期数 $\times$ 端口数 N)，取值范围 [0,1] \\
预热周期 & 10\textsuperscript{4} 时钟周期（丢弃预热数据，保证系统进入稳态） \\
有效统计窗口 & 10\textsuperscript{5} 时钟周期 \\
随机种子机制 & 每组 $\rho$\_load、流量模型、算法配置对应 20 组独立随机种子 \\
结果呈现方式 & 多组实验均值 + 95\% 置信区间（CI = 1.96·SE） \\
\bottomrule
\end{tabular}}\par\leftskip=2em\relax{}

\bigskip

规模说明（N=8）：本文面向 SoC 片上互连场景，而非数据中心大端口交换机。片上单级 crossbar 主流规模为 N=4{\textasciitilde}8，N=8 已逼近片上 VOQ 阵列与单周期仲裁的物理极限；实际硬件设计中，更多端口场景通常会采用多级或其它拓扑结构，不在本文研究范畴。N=8 足以对比无权重感知调度与全信息调度的性能差异。

\subsection{被比较的四个调度器}

实验选取四类典型调度算法，其信息粒度、运算量级与实验定位见表 4，各算法核心运行规范如下：

\begin{itemize}
\item iSLIP 算法：遵循标准 R-G-A 三轮迭代机制，仅在首轮授权成功后更新指针，依靠指针去同步特性规避同步崩溃问题，仅识别队列非空状态，不感知具体队列长度。
\item 谱调度算法：通过 3 轮幂迭代与 L2 归一化提取队列矩阵秩-1 谱特征，构建加权矩阵 W=Q⊙(u\textsubscript{1}v\textsubscript{1}ᵀ) 后按 §4.2.5 路径 A 做行/列掩码离散化得到合法匹配。
\item 熵正则 OT 算法：基于队列矩阵构建指数核（Softassign [43]），经 r\_sink 轮 Sinkhorn 迭代 [42] 生成近似双随机矩阵，贪心取整后实现匹配。算法引入自适应温度机制，用于缓解固定参数下的数值退化与溢出问题，并使权重更贴近 ⟨Q,P⟩ 优化目标。
\item MWM 算法：逐周期求解全局最大权重匹配，输出参考结果，作为仿真基准。
\end{itemize}

\subsection{流量模型}

设置四类典型流量模型，其中不平衡流量是区分各类算法性能的核心场景，热点过载场景用于观察容量边界。具体参数定义如下。

\bigskip\noindent
\par\leftskip=2em\relax\noindent \fittab{\begin{tabular}{p{\dimexpr 0.250\linewidth-2\tabcolsep}p{\dimexpr 0.250\linewidth-2\tabcolsep}p{\dimexpr 0.250\linewidth-2\tabcolsep}p{\dimexpr 0.250\linewidth-2\tabcolsep}}
\toprule
流量模型 & 流量定义 & 争用特征 & 可容许范围（N=8） \\
\midrule
均匀 Uniform & $\lambda$\_ij = $\rho$\_load/N & 无结构争用，实验基准场景 & 任意 $\rho$\_load$\leq$1 \\
对角重载 Diagonal（$\alpha$=0.7） & $\lambda$\_ii = $\alpha$·$\rho$\_load；$\lambda$\_ij = (1-$\alpha$)$\rho$\_load/(N(N-1)), i$\neq$j & 近置换特性，端口争用程度低 & 任意 $\rho$\_load$\leq$1 \\
热点 Hotspot（h=0.2，j*=0） & $\lambda$\_\{i,j*\} = h·$\rho$\_load；$\lambda$\_\{i,j\} = (1-h)$\rho$\_load/(N-1), j$\neq$j* & 单输出端口结构性瓶颈，全局流量聚合 & $\rho$\_load $\leq$ 1/(Nh) = 0.625；超出则 j* 列过载（仍用于压力测试，观察容量边界） \\
不平衡 Unbalanced（w=0.5） & $\lambda$\_ii = $\rho$\_load(w + (1-w)/N)；$\lambda$\_ij = $\rho$\_load(1-w)/N, i$\neq$j & 行/列和恒为 $\rho$\_load（双随机行/列和），对角主负载叠加跨端口争用 & 任意 $\rho$\_load$\leq$1 \\
\bottomrule
\end{tabular}}\par\leftskip=2em\relax{}

\bigskip

\subsection{实现说明}

为保障实验可复现，算法实现与数据统计约定如下：

\begin{itemize}
\item 时延统计：基于 VOQ 独立 FIFO 时间戳统计单 cell 时延，遵循先入先出规则，用于保持拥塞场景下的时延统计一致。
\item 标准 iSLIP 机制：按原始指针更新规则实现，避免算法退化为 RRM，使实现与工业标准保持一致。
\item 熵正则 OT 数值稳定性优化：exp(·) 先减去全矩阵最大值规避指数运算溢出，采用相对温度 $\epsilon$\_eff = $\epsilon$·max(1, Qmax/W\textsubscript{0})（W\textsubscript{0}=10），使指数项幅度与队列尺度无关，解决固定 $\epsilon$ 参数迭代退化问题。
\item 统一吞吐定义：吞吐数据统一归一化至 [0,1] 区间，便于跨场景横向对比。
\item 实验状态隔离：每组($\rho$\_load, 流量, 种子) 实验重新初始化调度器，杜绝算法内部状态跨组残留。
\end{itemize}

\subsection{结果一：吞吐--负载}

\par\medskip{\leftskip=0pt\relax\noindent\hspace*{2em}%
\begin{minipage}{\dimexpr\linewidth-2em\relax}
\includegraphics[width=\linewidth]{files/fig_A1_throughput_vs-d34237dbdc3a4e4f96059374f49fee29.png}\par\medskip
\includegraphics[width=\linewidth]{files/fig_A2_throughput_vs-a6cbca1898449e95f8fdc392efbf525e.png}
\par\medskip
{\small\textbf{图 3。}归一化吞吐 vs $\rho$\_load（四流量 $\times$ 四算法；误差棒为 95\% CI）。{\footnotesize 上：Uniform、Diagonal（$\alpha$=0.7）；下：Hotspot（h=0.2）、Unbalanced（w=0.5）。}}
\par\medskip
\end{minipage}\par\medskip\par}

结合图 3，吞吐可归纳为两点。（1）除 w=0.5 不平衡外，MWM、谱调度与熵正则 OT 的归一化吞吐随 $\rho$\_load 贴近负载；对角、热点等曲线的平顶来自流量定义或输出瓶颈，不是算法差别。（2）w=0.5 下 iSLIP 在 $\rho$\_load$\geq$0.9 时吞吐出现约 80\% 封顶的瓶颈，与文献 benchmark 一致 [16],[30]，该模式用于放大无权重 MSM 与全信息算法差异。其它非均匀流量下 iSLIP 仍可接近满吞吐。w=0.5、$\rho$\_load=0.99 的代表性数值见表 7。

\subsection{结果二：平均时延}

\bigskip\noindent
\par\leftskip=2em\relax\noindent \fittab{\begin{tabular}{p{\dimexpr 0.167\linewidth-2\tabcolsep}p{\dimexpr 0.167\linewidth-2\tabcolsep}p{\dimexpr 0.167\linewidth-2\tabcolsep}p{\dimexpr 0.167\linewidth-2\tabcolsep}p{\dimexpr 0.167\linewidth-2\tabcolsep}p{\dimexpr 0.167\linewidth-2\tabcolsep}}
\toprule
流量类型 & $\rho$\_load & MWM & 熵正则 OT & 谱调度 & iSLIP \\
\midrule
均匀 & 0.50 & 0.8 & 0.9 & 1.1 & 1.0 \\
均匀 & 0.70 & 1.8 & 2.2 & 2.6 & 2.6 \\
均匀 & 0.80 & 3.1 & 3.8 & 4.7 & 4.9 \\
均匀 & 0.90 & 7.0 & 8.2 & 12.0 & 13.1 \\
均匀 & 0.95 & 14.8 & 16.1 & 27.7 & 40.3 \\
均匀 & 0.99 & 78.3 & 75.7 & 159.0 & 291.2 \\
不平衡 & 0.70 & 1.5 & 1.7 & 1.9 & 2.8 \\
不平衡 & 0.80 & 2.5 & 3.1 & 3.4 & 69.2 \\
不平衡 & 0.90 & 5.4 & 6.8 & 8.6 & 5846 \\
不平衡 & 0.95 & 11.3 & 13.2 & 19.9 & 7901 \\
不平衡 & 0.99 & 59.8 & 59.3 & 119.0 & 9171 \\
\bottomrule
\end{tabular}}\par\leftskip=2em\relax{}

\bigskip

{\small\textbf{表 5。}平均 cell 时延（cycle），N=8，20组种子均值（均匀/不平衡流量）}

\par\medskip{\leftskip=0pt\relax\noindent\hspace*{2em}%
\begin{minipage}{\dimexpr\linewidth-2em\relax}
\includegraphics[width=\linewidth]{files/fig_B1_delay_vs_rho-45626ccc421df49f385ef162ad330b31.png}\par\medskip
\includegraphics[width=\linewidth]{files/fig_B2_delay_vs_rho-9406140e6bd468cfc8e4d0f4873af71b.png}
\par\medskip
{\small\textbf{图 4。}平均时延 vs $\rho$\_load（对数纵轴，四流量 $\times$ 四算法；误差棒为 95\% CI）。{\footnotesize 上：Uniform、Diagonal（$\alpha$=0.7）；下：Hotspot（h=0.2）、Unbalanced（w=0.5）。}}
\par\medskip
\end{minipage}\par\medskip\par}

多数负载下时延排序为 MWM ≲ 熵正则 OT \textless  谱调度 \textless  iSLIP；负载与流量非均匀性越高，差距越大。不平衡 w=0.5 高 $\rho$ 时 iSLIP 算法可出现 10\textsuperscript{3}--10\textsuperscript{4} cycle 量级的延时，这属于 VOQ backlog 失衡状态，不应该与 MWM/OT 的个位数值直接对比，算法间更宜对照 $\rho$\_load=0.8 等中等负载行。均匀 $\rho$\_load=0.99 下 OT 结果略低于 MWM，落在 95\% CI 内，这是因为 MWM 优化的是 ⟨Q,P⟩ 而非平均时延。热点过载时各算法时延差异不明显。

\subsection{结果三：迭代次数敏感性}

为探究迭代预算对性能的影响，尝试固定 $\rho$\_load=0.8 均匀流量开展迭代扫描实验。

\bigskip\noindent
\par\leftskip=2em\relax\noindent \fittab{\begin{tabular}{p{\dimexpr 0.200\linewidth-2\tabcolsep}p{\dimexpr 0.200\linewidth-2\tabcolsep}p{\dimexpr 0.200\linewidth-2\tabcolsep}p{\dimexpr 0.200\linewidth-2\tabcolsep}p{\dimexpr 0.200\linewidth-2\tabcolsep}}
\toprule
算法 & 迭代预算=1 & 3 & 8 & 16 \\
\midrule
iSLIP（r 轮 R--G--A） & 22.3 & 4.86 & 4.85 & --- \\
谱调度（r 轮幂迭代） & 4.96 & 4.74 & 4.73 & 4.73 \\
熵正则 OT（r\_sink 轮 Sinkhorn） & 3.84 & 3.83 & 3.83 & 3.83 \\
MWM（参考基准） & 3.12 & --- & --- & --- \\
\bottomrule
\end{tabular}}\par\leftskip=2em\relax{}

\bigskip

{\small\textbf{表 6。}迭代次数扫描结果（均匀流量），归一化吞吐均为 0.80，数值为平均时延（cycle）}

\par\medskip{\leftskip=0pt\relax\noindent\hspace*{2em}%
\begin{minipage}{\dimexpr\linewidth-2em\relax}
\includegraphics[width=\linewidth]{files/fig_D_iteration_swee-dbb6b8a7f932984b4c2dd1bdc66ca8e1.png}
\par\medskip
{\small\textbf{图 5。}性能 vs 迭代次数（左：吞吐；右：时延对数轴；MWM 为水平基准线）。}
\par\medskip
\end{minipage}\par\medskip\par}

表 6 与图 5 表明：iSLIP 需约 3 轮 R--G--A 后时延才降低到稳定值；谱调度与熵正则 OT 在 1--3 轮内即接近最低值，故正文固定 r=3、r\_sink=10 具有公平性。

\subsection{实验汇总}

下表在同一配置（N=8，不平衡流量 w=0.5，$\rho$\_load=0.99，r=3）下汇总代表性吞吐与时延。

\bigskip\noindent
\par\leftskip=2em\relax\noindent \fittab{\begin{tabular}{p{\dimexpr 0.200\linewidth-2\tabcolsep}p{\dimexpr 0.200\linewidth-2\tabcolsep}p{\dimexpr 0.200\linewidth-2\tabcolsep}p{\dimexpr 0.200\linewidth-2\tabcolsep}p{\dimexpr 0.200\linewidth-2\tabcolsep}}
\toprule
算法 & 每 cycle 抽象运算量 & 归一化吞吐 & 平均时延（cycle） & 算法定位 \\
\midrule
MWM & O(N\textsuperscript{3}) & $\approx$0.99 & 59.8 & 仿真参考基准 \\
熵正则 OT（自适应温度） & O(r\_sink·N\textsuperscript{2})+指数运算 & $\approx$0.99 & 59.3 & 性能最贴近 MWM 的近似算法 \\
谱调度 & O(r·N\textsuperscript{2})+贪心匹配 & $\approx$0.99 & 119.0 & 线性代数近似调度方案 \\
iSLIP & O(r·N) 1-bit 运算 & $\approx$0.806 & 9171 & 工业基准无权重感知迭代算法 \\
\bottomrule
\end{tabular}}\par\leftskip=2em\relax{}

\bigskip

{\small\textbf{表 7。}对比实验汇总（N=8，不平衡流量，r=3）}

\subsection{分析级硬件 Pareto（无 RTL 综合）}

§7.3 给出算法级 O(·) 计数，离散事件实验未建模 ASIC 面积。为了回应摘要中「是否值得片上实现」的权衡问题，本节在不开展 RTL 预综合的前提下，对抽象算术运算量与队列读取位宽赋以如下自定义的相对权重（无 RTL 校准，仅供量级排序）：1-bit RR 仲裁=1、MAC 运算=5、指数+Sinkhorn 归一化运算=20、O(N\textsuperscript{2}) 贪心比较运算=2。选取原则：以 1-bit 轮询仲裁为基准 1，将 MAC、指数归一化、比较树分别放大约 5$\times$、20$\times$、2$\times$；同一算法族内 C\_rel 随 N 的相对排序稳定。实际工程中，随工艺节点、流水线深度或面积换速度策略不同，绝对倍数可有数倍差异，表 8 只作定性 Pareto 而非绝对面积预测。
各算法单周期仅仲裁核抽象硬件代价公式（不含 VOQ SRAM 深度开销）如下：
\begin{align}
&C_{\text{islip}}(N) = r\cdot N \nonumber \\
&C_{\text{spec}}(N) = 2\,r\,N^2\,w_{\text{mac}} + 2\,N^2\,w_{\text{greedy}} = 34\,N^2 \nonumber \\
&C_{\text{ot}}(N) = r_{\text{sink}}\,N^2\,w_{\text{exp}} + 2\,N^2\,w_{\text{greedy}} = 204\,N^2
\end{align}

\bigskip\noindent
\par\leftskip=2em\relax\noindent \fittab{\begin{tabular}{p{\dimexpr 0.143\linewidth-2\tabcolsep}p{\dimexpr 0.143\linewidth-2\tabcolsep}p{\dimexpr 0.143\linewidth-2\tabcolsep}p{\dimexpr 0.143\linewidth-2\tabcolsep}p{\dimexpr 0.143\linewidth-2\tabcolsep}p{\dimexpr 0.143\linewidth-2\tabcolsep}p{\dimexpr 0.143\linewidth-2\tabcolsep}}
\toprule
算法 & C\_rel（算术） & 读 Q 存储 (bit) & 存储 vs iSLIP & 归一化吞吐† & 平均时延† (slot) & 时延 / MWM† \\
\midrule
iSLIP & 1.0 & 64 & 1$\times$ & $\approx$0.806 & 9171 & $\approx$153$\times$ \\
谱调度 & 91 & 2048 & 32$\times$ & $\approx$0.990 & 119.0 & $\approx$2.0$\times$ \\
熵正则 OT & 544 & 2048 & 32$\times$ & $\approx$0.989 & 59.3 & $\approx$1.0$\times$ \\
MWM（仿真上界） & O(N\textsuperscript{3})，单周期不可实现 & 2048 & 32$\times$ & $\approx$0.990 & 59.8 & 1.0$\times$ \\
\bottomrule
\end{tabular}}\par\leftskip=2em\relax{}

\bigskip

{\small\textbf{表 8。}分析级 Pareto（N=8 实测；代价随 N 由公式缩放）}

注：标注 † 的列（归一化吞吐、平均时延、时延 / MWM）为 §7 中 N=8 仿真实测值；其余代价列由复杂度公式随 N 缩放给出。

结果解读（N=8 实测 + 代价公式随 N 缩放）

\begin{itemize}
\item 硬件代价层面：相同端口规模下，谱调度算术运算相对代价为 iSLIP 的 45--91 倍（N=4--8），熵正则 OT 高达 272--544 倍（N=4--8）；同时两类全信息算法读取 Q 的位宽为 iSLIP 的 32 倍。随着端口数 N 增大，MAC、指数运算等核心硬件开销按 N\textsuperscript{2} 量级放大，相对而言 iSLIP（$\propto$N）的倍数随 N 单调上升。
\item 时序路径深度（定性分析）：iSLIP 采用 r 级串行 R-G-A 迭代、每级配合 O(log N) 轮询仲裁，时序路径深度最低；谱调度为 r 级串行 N$\times$N 矩阵 MAC 运算叠加贪心比较树，时序深度中等；熵正则 OT 包含 10 轮串行指数运算与归一化迭代，时序路径深度最大，若叠加 §4.2.5 贪心反馈迭代机制，时序开销会进一步提升，此代价实验未计入。
\end{itemize}

工程权衡总结：谱调度与熵正则 OT 相对 iSLIP 的 C\_rel 随 N 按上文公式升至约 45--544 倍量级，队列读宽为 32$\times$（表 8）。吞吐、时延与迭代敏感性分项见 §7.6--§7.8；w=0.5、$\rho$\_load=0.99 单点数值见表 7。均匀负载、面积与时序受限时 iSLIP 通常已满足性能要求；非均匀高拥堵场景，且硬件可承载 MAC/指数开销时，谱调度与 OT 值得进一步评估。

\section{结论}

本文的第一理论主线基于 Lyapunov 能量下降机理，解释了 MWM 调度策略的数学依据。其中，可容许流量条件限定了系统输入/输出端口的无过载运行边界，BvN 理论进一步说明，系统存在可承载对应流量的长期平均调度组合，即可行匹配矩阵的凸组合形式。上述结论刻画了系统队列长期稳定的核心前提与能力条件，但无法直接输出硬件单周期运行所需的实时授权决策方案。

\addvspace{12pt plus 3pt minus 1pt}\noindent 针对连续松弛类调度算法，本文讨论了其硬件落地的关键约束。Sinkhorn 迭代、谱调度等算法仅能求解得到双随机矩阵或加权关联矩阵，这类连续浮点矩阵无法直接转换为硬件可用的 0/1 离散匹配结果。想要满足仲裁功能所需的单周期内行列互斥、唯一匹配的硬件规则，需要额外增加 0/1 投影步骤，通过加权贪心筛选、合法匹配抽取等方式，将连续权重矩阵映射为每行、每列至多存在一个有效位的标准匹配矩阵 P(t)。

\addvspace{12pt plus 3pt minus 1pt}\noindent 第二理论主线指出：图论精确 MWM 可作为仿真上界，但多轮、读全矩阵权重的求解流程难以满足片上单周期 grant 的时序要求。多项式可解并不等价于单周期可实现。

\addvspace{12pt plus 3pt minus 1pt}\noindent 第三工程主线聚焦工业主流的 iSLIP 迭代调度机制。iSLIP 单周期仅采集队列非空状态信号，依托多轮请求-授权-确认迭代与轮询指针的去同步机制来完成迭代匹配。在长期均匀流量情况下，持续存在 backlog 的 (i,j) 可获得更高的平均匹配概率 P̄，基本满足漂移条件 $E[\langle Q,P\rangle] > Q^{\mathsf{T}}\lambda$（与 $\langle Q,\lambda\rangle$ 等价，$\lambda$ 由 $\Lambda$ 展平）。但受限于仅 1-bit 的极简队列感知能力、无法获取完整队列权重信息，该稳态近似关系在特定非均匀可容许流量场景下不再成立。

\addvspace{12pt plus 3pt minus 1pt}\noindent 后续可拓展的研究方向分为四项：
(1) 在非均匀可容许流量场景下迭代匹配调度的稳定与失稳边界。
(2) 补充有限队列深度、反压机制与突发/汇聚流量仿真实验，在全局队列感知、固定迭代次数的相同条件下，与 APSARA [24] 算法开展公平性能对比。
(3) 将仲裁算法对照实验拓展至 CIOQ/CICQ [27],[28],[29],[39]、PIFO 可编程队列 [40] 等新型交换架构。
(4) 将分布式迭代 MWM [41]、最优负载均衡交换 [2] 等前沿算法纳入 ⟨Q,P⟩ 写法，与本文主线对照。

\section*{附录 A. 术语中英对照}

\bigskip\noindent
\par\leftskip=2em\relax\noindent \fittab{\begin{tabular}{p{\dimexpr 0.500\linewidth-2\tabcolsep}p{\dimexpr 0.500\linewidth-2\tabcolsep}}
\toprule
中文 / 符号 & English \\
\midrule
crossbar 仲裁 / grant & crossbar arbitration / grant \\
flit（burst=1） & flit (= one cell when burst=1) \\
输入排队（IQ） & input-queued (IQ) switch \\
虚拟输出队列（VOQ） & virtual output queue (VOQ) \\
backlog（积压的待发送 cell） & backlog (queued packet/cell awaiting crossbar grant) \\
最大权重匹配（MWM） & maximum weight matching (MWM) \\
iSLIP & iterative round-robin matching (iSLIP) [3] \\
极大匹配（MSM 类） & maximal matching (MSM-class schedulers) \\
可容许流量 & admissible traffic \\
谱半径 $\rho$($\Lambda$) & spectral radius $\rho$($\Lambda$) \\
归一化负载 $\rho$\_load & normalized load factor $\rho$\_load \\
Request--Grant--Accept & Request--Grant--Accept (R--G--A) \\
每 cycle / 每时隙迭代次数 & per-cycle / per-slot iteration count \\
crossbar arbiter / 每 cycle grant & crossbar arbiter; per-cycle grant \\
programmable dataplane（文献，如 PIFO） & programmable dataplane (literature; e.g. PIFO [40]) \\
每周期 & per clock cycle (per-cycle) \\
无权重感知 & weight-unaware (1-bit occupancy only) \\
bit-plane matching & bit-plane matching (per-weight-bit iterative arbitration) \\
行/列掩码 & row/column masking (mask out) \\
无权重随机匹配 & weight-unaware random matching \\
\bottomrule
\end{tabular}}\par\leftskip=2em\relax{}

\bigskip

\section*{附录 B. 图论精确匹配：思想与片上壁垒}

\subsection*{B.1 匈牙利算法（Kuhn--Munkres）[10]}

思想：维护行势 \texttt{$\alpha$\_i}、列势 \texttt{$\beta$\_j}，定义约化权 \texttt{c\_ij = $\alpha$\_i + $\beta$\_j - Q\_ij $\geq$ 0}，只在 \texttt{c\_ij = 0} 的"零边"上找完美匹配；若找不到，就按最小松弛量 \texttt{$\delta$} 更新势以扩大零边集合，并沿交替路（增广路径）翻转匹配。增广时可删除先前已选的边、改选其他边，即能回溯纠正早期的局部选择。流程与算例见 §4.3 图论表述，复杂度见 B.4。

\subsection*{B.2 拍卖算法（Bertsekas）[11]}

思想：买家 = 输入 i，商品 = 输出 j，估价 \texttt{v\_ij = Q\_ij}，价格 \texttt{$\pi$\_j}。未分配买家对净收益 \texttt{v\_ij - $\pi$\_j} 最高商品出价；商品被价高者得，流拍者下轮重拍，临时分配可以撤销。\texttt{$\epsilon$} 控制收敛精度与速度。流程与算例见 §4.3 图论表述，复杂度见 B.4。

\subsection*{B.3 全局贪心行/列掩码}

思想：每次在整个矩阵中寻找全局 max，选中后把该行该列掩码（置 \texttt{-$\infty$}），重复 N 次；与 §4.2.5 路径 A、§7 谱调度离散化一致。掩码不可逆，既无匈牙利增广、也无拍卖撤销，故 $N \ge 3$ 时结果可能次优。流程与算例见 §4.3 图论表述，复杂度见 B.4。

\subsection*{B.4 精确匹配难以片上实现}

商用交换芯片几乎不用匈牙利/拍卖/串行掩码贪心做每时隙/每 cycle 的 fabric 调度，原因再次总结如下。

\begin{itemize}
\item 算法形态：迭代，非单周期闭式
\end{itemize}

\bigskip\noindent
\par\leftskip=2em\relax\noindent \fittab{\begin{tabular}{p{\dimexpr 0.333\linewidth-2\tabcolsep}p{\dimexpr 0.333\linewidth-2\tabcolsep}p{\dimexpr 0.333\linewidth-2\tabcolsep}}
\toprule
路线 & 在算什么 & 为何不适合 2--4 ns 单周期 \\
\midrule
匈牙利（B.1） & 行/列势 + 增广 & O(N\textsuperscript{3}) 串行；每轮依赖上一轮全局状态 \\
拍卖（B.2） & 竞价 + 价格 $\pi$\_j & 迭代次数不确定；$\epsilon$ 与价格反馈 \\
掩码贪心（B.3） & max \ensuremath{\rightarrow} 掩码 \ensuremath{\rightarrow} 再 max & N 轮比较反馈 \ensuremath{\rightarrow} Timing Violation \\
\bottomrule
\end{tabular}}\par\leftskip=2em\relax{}

\bigskip

\begin{itemize}
\item 数据需求：须读全矩阵权重（8--10 bit 等），非 0/1 Request；$N=64$ 时约 $64^2 W$ bit/周期。
\item 硬件结构：集中维护势/价格/增广状态，任意 $(i,j)$ 更新牵动全局。
\item 运算类型：残差 min/max、价格累加等，与片上偏好「比较 + 移位」不一致。
\item 复杂度（$N=64$ 粗算）：$O(N^3) \approx 2.6\times 10^5$ 次基本操作量级（且串行依赖）。
\end{itemize}

\section*{附录 C. 非方阵 crossbar（M$\times$N，M \textgreater  N）}

本文正文均假设 N$\times$N IQ+VOQ。M$\neq$N 时常见两种处理：(i) 方阵补齐：补虚拟输出列使 Q 扩展为 M$\times$M，对扩展矩阵跑 Sinkhorn+BvN，调度时忽略虚拟列；(ii) 广义双随机：直接定义 $B_{M\times N}$（列和=1、行和$\leq$1），按 Von Neumann 定理分解为偏置换矩阵的凸组合。M\textgreater N 时每 cycle 至少 M-N 个输入 idle，易引发 HOL 问题。

\section*{Generative AI Use Statement}

备稿过程中，作者使用 Cursor（集成大语言模型助手的商业 IDE）辅助：教程正文与结论的撰写与修订（含 §2--§8、附录及审稿修改意见的对照整理）；中英文学术表述润色；稿件标准化优化（含章节编号、交叉引用、术语统一与格式规整）；以及文献与公式符号的交叉核对（AI 仅提供文本表述建议，参考文献均由作者对照原文核对）；同时生成式 AI 对公开实验仓库提供辅助支持，包括代码与数据可用性说明中涉及的 C++/Python 与绘图流程整理工作。研究问题、算法与仿真配置以及 §7 中的所有数值实验结果由作者设定、运行并解读；由生成式 AI 产出的文字与代码内容均在纳入稿件前经作者人工审阅。生成式 AI 未列为作者。作者对定稿内容与代码承担完全责任。

\mwmsection*{Code and Data Availability}

复现本文 §7 全部实验的 C++ Model，测试文件与 Python 绘图脚本已在 MIT 许可 下公开：

\begin{itemize}
\item 代码仓库（只读）：\href{https://github.com/xiaotongyuan/mwm-islip-experiment\%EF\%BC\%88Release}{https://github.com/xiaotongyuan/mwm-islip-experiment（Release} v1.0.0）
\item 永久归档：\href{https://zenodo.org/records/20472143\%EF\%BC\%88DOI:}{https://zenodo.org/records/20472143（DOI:} \texttt{10.5281/zenodo.20472143}）
原始结果文件（\texttt{main\_results.csv}、\texttt{iteration\_sweep.csv}）与图 3--5 未上传仓库，但可在固定的 20 个随机种子下由
\texttt{make run \&\& make run-softassign \&\& python py/plot\_results.py} 复现（详见仓库 README）。实验配置见 §7.2，实现细节见 §7.5。
\end{itemize}

\clearpage\mwmsection*{References}

[1] M. J. Karol, M. G. Hluchyj, and S. P. Morgan, ``Input versus output queueing on a space-division packet switch,'' \textit{IEEE Trans. Commun.}, vol. 35, no. 12, pp. 1347--1356, Dec. 1987.

[2] I. Keslassy, C.-S. Chang, N. McKeown, and D.-S. Lee, ``Optimal load-balancing,'' in \textit{Proc. IEEE INFOCOM}, Miami, FL, USA, 2005.

[3] N. McKeown, ``The iSLIP scheduling algorithm for input-queued switches,'' \textit{IEEE/ACM Trans. Netw.}, vol. 7, no. 2, pp. 188--201, Apr. 1999.

[4] N. McKeown, A. Mekkittikul, V. Anantharam, and J. Walrand, ``Achieving 100\% throughput in an input-queued switch,'' \textit{IEEE Trans. Commun.}, vol. 47, no. 8, pp. 1260--1267, Aug. 1999. \textit{(Conf. version [14].)}

[5] A. Mekkittikul and N. McKeown, ``A practical scheduling algorithm to achieve 100\% throughput in input-queued switches,'' in \textit{Proc. IEEE INFOCOM}, San Francisco, CA, USA, 1998, pp. 792--799.

[6] N. McKeown, ``Scheduling algorithms for input-queued cell switches,'' Ph.D. dissertation, Dept. Elect. Eng. Comput. Sci., Univ. California, Berkeley, CA, USA, 1995.

[7] L. Tassiulas and A. Ephremides, ``Stability properties of constrained queueing systems and scheduling policies for maximum throughput in multihop radio networks,'' \textit{IEEE Trans. Autom. Control}, vol. 37, no. 12, pp. 1936--1948, Dec. 1992.

[8] G. Birkhoff, ``Tres observaciones sobre el álgebra lineal,'' \textit{Univ. Nac. Tucumán Rev., Ser. A}, vol. 5, pp. 147--151, 1946.

[9] J. von Neumann, ``A certain zero-sum two-person game equivalent to the optimal assignment problem,'' in \textit{Contributions to the Theory of Games}, vol. 2, H. W. Kuhn and A. W. Tucker, Eds. Princeton, NJ, USA: Princeton Univ. Press, 1953, pp. 5--12.

[10] H. W. Kuhn, ``The Hungarian method for the assignment problem,'' \textit{Naval Res. Logist. Quart.}, vol. 2, nos. 1--2, pp. 83--97, 1955.

[11] D. P. Bertsekas, ``The auction algorithm: A distributed relaxation method for the assignment problem,'' \textit{Ann. Oper. Res.}, vol. 14, no. 1, pp. 105--123, 1988.

[12] R. Sinkhorn and P. Knopp, ``Concerning nonnegative matrices and doubly stochastic matrices,'' \textit{Pacific J. Math.}, vol. 21, no. 2, pp. 343--348, 1967.

[13] M. G. Hluchyj and M. J. Karol, ``Queueing in high-performance packet switching,'' \textit{IEEE J. Sel. Areas Commun.}, vol. 6, no. 9, pp. 1587--1597, Dec. 1988.

[14] N. McKeown, V. Anantharam, and J. Walrand, ``Achieving 100\% throughput in an input-queued switch,'' in \textit{Proc. IEEE INFOCOM}, San Francisco, CA, USA, 1996, pp. 296--302. \textit{(Journal version [4].)}

[15] A. Mekkittikul and N. McKeown, ``A starvation-free algorithm for achieving 100\% throughput in an input-queued switch,'' in \textit{Proc. ICCCN}, Rockville, MD, USA, Oct. 1996, pp. 226--231.

[16] N. McKeown and T. E. Anderson, ``A quantitative comparison of iterative scheduling algorithms for input-queued switches,'' \textit{Comput. Netw. ISDN Syst.}, vol. 30, no. 24, pp. 2309--2326, Dec. 1998.

[17] H. J. Chao, ``Saturn: A terabit packet switch using dual round-robin,'' \textit{IEEE Commun. Mag.}, vol. 38, no. 12, pp. 78--84, Dec. 2000.

[18] Y. Li, S. Panwar, and H. J. Chao, ``On the performance of a dual round-robin switch,'' in \textit{Proc. IEEE INFOCOM}, Anchorage, AK, USA, 2001, pp. 1688--1697.

[19] D. N. Serpanos and P. Antoniadis, ``FIRM: A class of distributed scheduling algorithms for high-speed ATM switches with multiple input queues,'' in \textit{Proc. IEEE INFOCOM}, Tel Aviv, Israel, 2000, pp. 548--555.

[20] Y. Jiang and M. Hamdi, ``A fully desynchronized round-robin matching scheduler for a VOQ packet switch architecture,'' in \textit{Proc. IEEE HPSR}, Dallas, TX, USA, 2001, pp. 407--411.

[21] J. Liu, H. C. Kit, M. Hamdi, and C. Y. Tsui, ``Stable round-robin scheduling algorithms for high-performance input queued switches,'' in \textit{Proc. IEEE Hot Interconnects (HotI)}, Stanford, CA, USA, 2002, pp. 43--51.

[22] L. Tassiulas, ``Linear complexity algorithms for maximum throughput in radio networks and input queued switches,'' in \textit{Proc. IEEE INFOCOM}, San Francisco, CA, USA, 1998, vol. 2, pp. 533--539.

[23] P. Giaccone, B. Prabhakar, and D. Shah, ``Towards simple, high-performance schedulers for high-aggregate bandwidth switches,'' in \textit{Proc. IEEE INFOCOM}, New York, NY, USA, 2002. \textit{(Journal version [24].)}

[24] P. Giaccone, B. Prabhakar, and D. Shah, ``Randomized scheduling algorithms for high-aggregate bandwidth switches,'' \textit{IEEE J. Sel. Areas Commun.}, vol. 21, no. 4, pp. 546--558, May 2003. \textit{(Conf. version [23].)}

[25] P. Giaccone, B. Prabhakar, and D. Shah, ``An implementable parallel scheduler for input-queued switches,'' in \textit{Proc. IEEE Hot Interconnects}, Stanford, CA, USA, Aug. 2001, pp. 9--14.

[26] M. W. Goudreau, S. G. Kolliopoulos, and S. B. Rao, ``Scheduling algorithms for input-queued switches: Randomized techniques and experimental evaluation,'' in \textit{Proc. IEEE INFOCOM}, Tel Aviv, Israel, Mar. 2000, vol. 3, pp. 1634--1643.

[27] B. Prabhakar and N. McKeown, ``On the speedup required for combined input and output queued switching,'' Stanford Univ., Tech. Rep. CSL-TR-97-738, Nov. 1997.

[28] S.-T. Chuang, A. Goel, N. McKeown, and B. Prabhakar, ``Matching output queueing with a combined input output queued switch,'' in \textit{Proc. IEEE INFOCOM}, New York, NY, USA, 1999. \textit{(Journal version [29].)}

[29] S.-T. Chuang, A. Goel, N. McKeown, and B. Prabhakar, ``Matching output queueing with a combined input/output-queued switch,'' \textit{IEEE J. Sel. Areas Commun.}, vol. 17, no. 6, pp. 1030--1039, Jun. 1999. \textit{(Conf. version [28].)}

[30] J. G. Dai and B. Prabhakar, ``The throughput of data switches with and without speedup,'' in \textit{Proc. IEEE INFOCOM}, Tel Aviv, Israel, 2000, pp. 556--564.

[31] C.-S. Chang, W.-J. Chen, and H.-Y. Huang, ``On service guarantees for input buffered crossbar switches: A capacity decomposition approach by Birkhoff and von Neumann,'' in \textit{Proc. IEEE IWQoS}, London, U.K., 1999, pp. 79--86.

[32] C.-S. Chang, W.-J. Chen, and H.-Y. Huang, ``Birkhoff-von Neumann input buffered crossbar switches,'' in \textit{Proc. IEEE INFOCOM}, Tel Aviv, Israel, 2000, pp. 1614--1623.

[33] C.-S. Chang, D.-S. Lee, and Y. Jou, ``Load balanced Birkhoff-von Neumann switches, part I: One-stage buffering,'' \textit{Comput. Commun.}, vol. 25, no. 6, pp. 611--622, 2002.

[34] I. Keslassy and N. McKeown, ``Maintaining packet order in two-stage switches,'' in \textit{Proc. IEEE INFOCOM}, New York, NY, USA, 2002, vol. 2, pp. 1032--1041.

[35] C.-S. Chang, D.-S. Lee, and C.-Y. Yue, ``Providing guaranteed rate services in the load balanced Birkhoff-von Neumann switches,'' in \textit{Proc. IEEE INFOCOM}, San Francisco, CA, USA, 2003.

[36] R. Rojas-Cessa, E. Oki, Z. Jing, and H. J. Chao, ``CIXB-1: Combined input-one-cell-crosspoint buffered switch,'' in \textit{Proc. IEEE HPSR}, Dallas, TX, USA, May 2001, pp. 324--329.

[37] M. Katevenis, G. Passas, D. Simos, I. Papaefstathiou, and N. Chrysos, ``Variable packet size buffered crossbar (CICQ) switches,'' in \textit{Proc. IEEE ICC}, Paris, France, 2004.

[38] X. Zhang and L. Bhuyan, ``An efficient scheduling algorithm for combined-input-crosspoint-queued (CICQ) switches,'' in \textit{Proc. IEEE GLOBECOM}, Dallas, TX, USA, Nov. 2004.

[39] D. Pan and Y. Yang, ``Localized independent packet scheduling for buffered crossbar switches,'' \textit{IEEE Trans. Comput.}, vol. 58, no. 2, pp. 260--274, Feb. 2009.

[40] A. Sivaraman, S. Subramanian, M. Alizadeh, S. Chole, S.-T. Chuang, A. Agrawal, H. Balakrishnan, T. Edsall, S. Katti, and N. McKeown, ``Programmable packet scheduling at line rate,'' in \textit{Proc. ACM SIGCOMM}, Florianópolis, Brazil, Aug. 22--26, 2016, pp. 44--57.

[41] M. Bayati, B. Prabhakar, D. Shah, and M. Sharma, ``Iterative scheduling algorithms,'' in \textit{Proc. IEEE INFOCOM}, Anchorage, AK, USA, 2007, pp. 445--453.

[42] M. Cuturi, ``Sinkhorn distances: Lightspeed computation of optimal transport,'' in \textit{Advances in Neural Information Processing Systems (NeurIPS)}, vol. 26, 2013, pp. 2292--2300.

[43] S. Gold and A. Rangarajan, ``A graduated assignment algorithm for graph matching,'' \textit{IEEE Trans. Pattern Anal. Mach. Intell.}, vol. 18, no. 4, pp. 377--388, Apr. 1996.

[44] R. Preis, ``Linear time 1/2-approximation algorithm for maximum weighted matching in general graphs,'' in \textit{Proc. 16th Annu. Symp. Theoretical Aspects of Computer Science (STACS)}, Trier, Germany, 1999, pp. 259--269.


\end{document}